\documentclass[11pt,letterpaper]{article}

\usepackage{MathStyle}
\RenewDocumentCommand{\E}{}{\operatorname{\mathbb{E}}}
\usepackage{microtype}


\DeclareMathOperator{\ondisc}{ondisc}
\DeclareMathOperator*{\argmin}{arg\,min}
\newcommand{\bin}{\mathrm{bin}}
\newtheorem{proposition}{Proposition}
\crefname{appendix}{Appendix}{Appendices}
\Crefname{appendix}{Appendix}{Appendices}
\crefname{literaturetheorem}{Theorem}{Theorems}
\Crefname{literaturetheorem}{Theorem}{Theorems}
\hypersetup{pdftitle={Online Discrepancy of Sparse Random Vectors over Long Horizons},pdfauthor={Nicola Wengiel}}

\title{Online Discrepancy of Sparse Random Vectors\\
over Long Horizons}
\author{Nicola Wengiel\thanks{Email: \href{mailto:nicola.wengiel@fau.de}{nicola.wengiel@fau.de}.}}
\affil{Friedrich-Alexander-Universit\"at Erlangen-N\"urnberg\\
Department of Mathematics, Erlangen, Germany}
\date{}

\begin{document}
\maketitle

\begin{abstract}
We study online vector balancing for independent sparse random vectors: each arrival is supported on a uniformly random set of $s$ of the $n$ coordinates, with nonzero entries equal to one (binary model) or independent random signs (signed model), and an online algorithm must sign each vector upon arrival.
The prefix discrepancy is the largest $\ell_\infty$-norm of the signed sums up to the horizon $T$, the terminal discrepancy that of the sum at time $T$.
For binary inputs with $2\le s\le n/2$ and $T=\Theta(n)$, Altschuler and Tikhomirov showed that both are of order $\sqrt s+\log\log n$; they asked how the prefix discrepancy depends on the horizon and whether a matching terminal bound holds at every prescribed time.

\smallskip
Over long horizons, the two differ.
For every fixed $s\ge2$ and $c>0$, every online algorithm, even a randomized one that knows the horizon, incurs with high probability prefix discrepancy $\Omega(\log\log n+(\log T/\log n)^{\alpha_s})$ for $n\le T\le e^{cn}$, where $\alpha_s=(2^{s-1}-1)/(2^s-1)$.
For $s=2$, quadratic Greedy matches this bound; for fixed $s\ge3$, we give an online algorithm with prefix discrepancy $O(\log\log n+(\log T/\log n)^{1/2})$.
In contrast, a deterministic algorithm that knows neither the horizon nor the failure probability keeps the terminal discrepancy $O_k(\sqrt s+\log\log n)$ at each prescribed time, with failure probability $n^{-k}$ for every fixed $k$, for $2\le s\le n/2$ (binary model) and $2\le s\le n$ (signed model); for $s=2$, quadratic Greedy itself does so.
This answers their terminal question, and the same order is optimal for prefixes at all polynomial horizons $T\ge n$.

\smallskip
The lower bound uses a finite safety game realized within the input law; the upper bounds use moments of elementary symmetric polynomials and, at terminal times, a $\cosh$ potential or a translated gauge with a Foster inequality at every state.
\end{abstract}

\clearpage
\tableofcontents
\clearpage

\section{Introduction}
\label{sec:Introduction}

In online vector balancing, vectors $V_1,V_2,\ldots\in\R^n$ arrive one at a time, and an algorithm must choose a sign $x_t\in\{\pm1\}$ immediately after observing $V_t$, without knowledge of future vectors.
The aim is to keep the signed sums $\sum_{u=1}^tx_uV_u$ small in the $\ell_\infty$-norm; we call the current signed sum the \emph{state}, and the absolute values of its coordinates \emph{heights}.
Given a horizon $T$, two objectives are natural: the \emph{prefix discrepancy} and the \emph{terminal discrepancy},
\begin{equation*}
    \max_{1\le t\le T}\norm{\sum_{u=1}^tx_uV_u}_\infty\qquad\text{and}\qquad\norm{\sum_{t=1}^Tx_tV_t}_\infty.
\end{equation*}
The first measures the heights at every time up to $T$, the second only at the final time $T$.
Balancing games of this kind were introduced by Spencer \cite{Spencer77} and generalized by B\'ar\'any \cite{Barany79}.
Against an \emph{adaptive} adversary, which chooses each vector after seeing the previous signs, the discrepancy grows polynomially with the horizon: for vectors in $[-1,1]^n$, an adversary that chooses each vector orthogonal to the current signed sum forces discrepancy $\Omega(\sqrt T)$, and even for vectors with two nonzero entries in $\{\pm1\}$, such as the edge vectors $e_i-e_j$ of online carpooling, deterministic algorithms face discrepancy of order $\min\{T^{1/3},n\}$ \cite{Ajtai98,BansalCarpool26} (see also \cref{lem:DoubleCover}).
Guillen and Kobzar determined the leading order of this growth for an online Koml\'os game, in which an adaptive adversary chooses vectors of Euclidean norm at most one \cite{GuillenKobzar26}.
For vectors with exactly $s\ge3$ nonzero entries in $\{\pm1\}$, the discrepancy against an adaptive adversary also stays bounded for fixed $n$, as it does for pairs, but at a much higher level: by B\'ar\'any's width bound \cite[Section~6, Remark~(2)]{Barany79}, the smallest bound that an algorithm can maintain at all times is exactly $2^{s-2}\binom{n-1}{s-1}$, which is of order $n^{s-1}$ for fixed $s$ (\cref{prop:InfiniteValue}).
For vectors of Euclidean norm at most one that form an \emph{oblivious} sequence, fixed in advance and independent of the algorithm's random choices, randomized algorithms achieve prefix discrepancy $O(\sqrt{\log T})$ with high probability \cite{Kulkarni24}, even in linear time \cite{AdenAli26}, and this is optimal \cite{Kulkarni24}; see also \cite{Alweiss21,Liu22}.
Offline, the final signed sum of such vectors can even be kept below an absolute constant, by the recent proof of the Koml\'os conjecture \cite{GuoFangLu26,Karingula26}.
In these online worst-case models, the horizon enters polynomially or through $\sqrt{\log T}$; how it enters for random inputs is the subject of this paper.

\paragraph{Random inputs.}
When the inputs are random, the algorithm can rely on the randomness of future arrivals to counteract its accumulated imbalance.
For dense inputs, this line of work goes back to Bansal and Spencer \cite{Bansal20a}, who gave an online algorithm for vectors with independent Rademacher entries that achieves discrepancy $O(\sqrt n)$ at every prescribed time; already at $T=n$, this bound is optimal even offline.
Altschuler and Tikhomirov studied \emph{sparse} random inputs, a stochastic version of the Beck--Fiala setting, in which each vector has at most $s$ nonzero entries \cite{BeckFiala81}.
In their model, which we call the binary model, each arrival is the indicator vector of a uniformly random $s$-element subset of $[n]$.
They first showed that for $T=\Theta(n)$ the optimal prefix and terminal discrepancies are of order $\log\log n$ for $2\le s\le(\log\log n)^2/\log\log\log n$ \cite{Altschuler25}, and then, in a paper that supersedes \cite{Altschuler25}, that they are of order $\sqrt s+\log\log n$ for $2\le s\le n/2$ \cite{Altschuler26Binary}.
At such horizons the two objectives agree: their algorithm controls all prefixes, and their lower bound already holds at the terminal time.
The doubly logarithmic term, which already appears for Greedy on uniformly random edges \cite{Ajtai98}, has no offline counterpart; for fixed $s$, for instance, the offline discrepancy is less than $2s$ by the Beck--Fiala theorem.
It is reminiscent of the power of two choices in balanced allocations \cite{Azar99,Berenbrink06}.
Heuristically, it arises because the algorithm must commit to a sign before it knows which of the coordinates touched by the current arrival will be touched again; here an arrival \emph{touches} the coordinates in its support.

\paragraph{Two questions about the horizon.}
These results concern horizons comparable to the dimension.
Altschuler and Tikhomirov asked how the optimal expected prefix discrepancy depends on the horizon \cite[Section~1.4]{Altschuler25}.
In their second paper, they proved the terminal lower bound of order $\sqrt s+\log\log n$ at every prescribed time $T\ge n$ \cite[Proposition~2.5]{Altschuler26Binary}, noted that their argument does not give a matching upper bound uniformly over such times, and posed as an open problem whether, as for dense inputs \cite{Bansal20a}, such a bound holds with high probability at every time fixed in advance \cite[Section~1]{Altschuler26Binary}.

\paragraph{Why long horizons matter.}
Over a long random stream, rare patterns of arrivals occur.
For fixed $s$, a given sequence of $J$ arrivals, each touching prescribed coordinates with prescribed entries, has probability at least $n^{-O(J)}$, so with high probability it occurs as a block of consecutive arrivals among the first $T$ once $\log T$ is a large multiple of $J\log n$.
For random unit vectors and a fixed worst-case instance of offline prefix discrepancy, Bansal, Jiang, Meka, Singla and Sinha use the same argument \cite[Theorem~16]{Bansal22}.
The same holds for an adaptive strategy, since each requested arrival depends only on the past and occurs with probability $n^{-O(1)}$ conditionally on it.
Consequently, any finite adversarial strategy that forces some coordinate to height $H$ within $J$ requests from every starting state, which we call a \emph{forcing strategy}, becomes available to nature at horizons with $\log T/\log n\gtrsim J$; Bansal and Spencer observed this qualitatively for dense inputs \cite[Section~3]{Bansal20a}.
If height $H$ can be forced within about $H^{1/\alpha}$ requests, the prefix discrepancy is thus at least of order $(\log T/\log n)^\alpha$; for pairs, the cubic forcing strategies of online carpooling \cite{Ajtai98} suggest $\alpha=1/3$.
Whether the prefix discrepancy grows with $T$, and how fast, therefore depends on how quickly forcing strategies can raise the heights, and on whether an algorithm can defend against all of them simultaneously.
The natural measure of the horizon is thus the \emph{effective horizon} $\log T/\log n$.
The occurrence of a forcing strategy in the stream, which we call a \emph{forcing pattern}, produces a large height only at some time, however, and an algorithm may reduce it afterwards.
One may therefore hope that the signed sum at a prescribed time is typical, however long the stream has run; for Greedy on random edges of the complete graph, Ajtai et al.\ observed this at a typical time of a long execution \cite{Ajtai98}.
Turning this hope into a guarantee at every prescribed time requires an algorithm that recovers from every forcing pattern without knowing the horizon, and bounds that do not depend on the time.

\paragraph{Our answers.}
This paper studies the long-horizon regime $T\gg n$.
It answers the terminal question for all $T\ge n$ throughout the sparsity range of \cite{Altschuler26Binary}.
It answers the prefix question at all polynomial horizons $T\ge n$ for every sparsity in this range, and at all horizons $n\le T\le e^{cn}$ for $s=2$; for fixed $s\ge3$ and these horizons, it determines the prefix discrepancy up to the exponent of $\log T/\log n$.
As the heuristic suggests, the two answers differ: for fixed $s$, the prefix discrepancy grows with the horizon once $\log T/\log n$ exceeds a suitable power of $\log\log n$, whereas the order of the terminal discrepancy is the same at every prescribed time $T\ge n$.

\subsection{Main results}
\label{sec:MainResults}

We consider two input models.
In both, the arrivals $V_1,V_2,\ldots\in\R^n$ are independent, and each of them is supported on a uniformly random set of exactly $s$ coordinates.
In the \emph{binary model} of \cite{Altschuler25,Altschuler26Binary}, the nonzero entries are equal to one.
In the \emph{signed model}, they are independent Rademacher signs, which we call \emph{marks}.
We write $\mu_s^\bin$ and $\mu_s^\pm$ for the laws of a single arrival and $\mu_s$ for either of them.
An online algorithm chooses each sign $x_t\in\{\pm1\}$ after observing $V_1,\ldots,V_t$; it may use private randomness, that is, random bits independent of the arrivals, and may know the horizon $T$ and a target failure probability $\delta$ (see \cref{sec:Model} for the precise conventions).
Following \cite[Definition~1.9]{Altschuler25}, the optimal expected prefix discrepancy is
\begin{equation*}
    \ondisc(\mu_s,T)\colonequals\inf\E\max_{1\le t\le T}\norm{\sum_{u=1}^tx_uV_u}_\infty,
\end{equation*}
where the infimum is taken over all online algorithms.
We put
\begin{equation*}
    \alpha_s\colonequals\frac{2^{s-1}-1}{2^s-1}=\frac12-\frac1{2(2^s-1)},
    \qquad
    L\colonequals\log\frac{16n(T+1)}{\delta},
\end{equation*}
so that $\alpha_2=1/3$, $\alpha_3=3/7$, $\alpha_4=7/15$, and $\alpha_s\uparrow\frac12$.
The parameter $L$ combines the horizon and the failure probability; for $T\ge n$ and $\delta\ge T^{-C}$ with a constant $C$, it is of order $\log T$.

\paragraph{Prefix discrepancy.}
Our first result shows that, against every online algorithm, long horizons force an additional term that is polynomial in $\log T/\log n$.
For pairs, a matching upper bound holds with the same exponent; for larger supports, an upper bound of the same form holds with exponent $1/2$ in place of $\alpha_s$.

\begin{theorem}[Prefix discrepancy]
    \label{thm:Prefix}
    Fix an integer $s\ge2$.
    \begin{enumerate}[label=(\alph*)]
        \item\label{item:PrefixLower} \emph{Lower bound.}
        For every $c>0$ there are $c'=c'(s,c)>0$ and $\e_n=\e_n(s,c)\to0$ such that, in both models, for all $n\ge\max\{3,s\}$ and all integers $n\le T\le e^{cn}$, the signs $x_1,\ldots,x_T$ chosen by any online algorithm satisfy
        \begin{equation}
            \label{eq:MainLower}
            \P\bc{\max_{1\le t\le T}\norm{\sum_{u=1}^tx_uV_u}_\infty\ge c'\Big(\log\log n+\Big(\frac{\log T}{\log n}\Big)^{\alpha_s}\Big)}\ge1-\e_n.
        \end{equation}
        \item\label{item:PrefixPairs} \emph{Pairs.}
        Let $s=2$.
        Quadratic Greedy chooses $x_t\colonequals-\sgn\inp{\sum_{u<t}x_uV_u}{V_t}$, with an arbitrary fixed rule at ties.
        In both models, for all $n\ge e^{256}$, $T\ge n$ and $0<\delta<1$ with $L\le n^2$, its signs satisfy
        \begin{equation}
            \label{eq:PairsRandom}
            \P\bc{\max_{1\le t\le T}\norm{\sum_{u=1}^tx_uV_u}_\infty>C\Big(\log\log n+\Big(\frac{L}{\log n}\Big)^{1/3}\Big)}\le\delta,
        \end{equation}
        where $C$ is an absolute constant.
        \item\label{item:PrefixFixed} \emph{Fixed sparsity.}
        Let $s\ge3$ and $C_0>0$.
        There are $A_s<\infty$, depending only on $s$, and $n_0=n_0(s,C_0)$ such that, for all $n\ge n_0$, all integers $T\ge n$ and all $0<\delta<1$ with $L\le C_0n$, there is a deterministic online algorithm, depending only on $n$, $s$ and $L$, whose signs satisfy, in both models,
        \begin{equation}
            \label{eq:FixedSparsity}
            \P\bc{\max_{1\le t\le T}\norm{\sum_{u=1}^tx_uV_u}_\infty>A_s\Big(\log\log n+\Big(\frac{L}{\log n}\Big)^{1/2}\Big)}\le\delta.
        \end{equation}
    \end{enumerate}
\end{theorem}

In part~\ref{item:PrefixLower}, the doubly logarithmic term is the baseline of Altschuler and Tikhomirov, which we extend to the signed model in \cref{prop:Baseline}; the horizon term is new.
The rule in part~\ref{item:PrefixPairs} is deterministic and parameter-free; in particular, it depends neither on $T$ nor on $\delta$.
On every sequence of vectors with exactly two nonzero entries in $\{\pm1\}$, the prefix discrepancy of quadratic Greedy up to time $T$ is at most $\min\{2n,4(2T)^{1/3}\}$ (\cref{lem:DoubleCover}); this is a direct consequence of the cube-root bound of Bansal, Prabhu, Singla and Sundaram for online carpooling \cite[Lemma~3.1]{BansalCarpool26}.
Part~\ref{item:PrefixPairs} is a long-horizon counterpart of this cube-root phenomenon, which they also established for random arrivals at linear horizons \cite[Theorems~1.2 and~1.3]{BansalCarpool26}: over random arrivals, the exponent $1/3$ persists, but the number of arrivals is replaced by the effective horizon $\log T/\log n$.
The rule in part~\ref{item:PrefixFixed} is a weighted gradient rule: it chooses the sign that does not increase the sum of the heights of the touched coordinates, each weighted according to its current height; the weights grow exponentially, but slowly, up to a \emph{knee} and linearly beyond it, and only the position of the knee depends on $L$.
Restarting this rule with geometrically increasing values of $L$, whenever the signed sum since the last restart exceeds a threshold that grows with $L$, gives a single deterministic rule, independent of $T$ and $\delta$, with the same guarantee up to the constant (\cref{prop:Restart}).
The proof gives $A_s=O(s^2\log(es))$, and in the range $L\ge\sqrt n$ the threshold in \eqref{eq:FixedSparsity} can even be replaced by $O(\sqrt{sL/\log n})$; we state part~\ref{item:PrefixFixed} for fixed $s$, since these bounds lose powers of $s$ compared with the dependence expected for growing sparsity (\cref{q:Joint}).
Combining part~\ref{item:PrefixLower} with parts~\ref{item:PrefixPairs} and~\ref{item:PrefixFixed} for $\delta=T^{-2}$, we obtain the following answer to the prefix question of Altschuler and Tikhomirov in our range of horizons; for $s\ge3$, the lower and upper bounds have different horizon exponents, $\alpha_s$ and $\frac12$, the optimal exponent remains open, and we conjecture that it is $\frac12$ (\cref{conj:SquareRoot}).

\begin{corollary}[Expected prefix discrepancy]
    \label{cor:ExpectedPrefix}
    Fix $c>0$.
    Uniformly for $n\ge3$ and integers $n\le T\le e^{cn}$,
    \begin{equation*}
        \ondisc(\mu_2,T)\asymp_c\log\log n+\Big(\frac{\log T}{\log n}\Big)^{1/3},
    \end{equation*}
    and, for each fixed $s\ge3$, uniformly for $n\ge s$ and integers $n\le T\le e^{cn}$,
    \begin{equation*}
        \log\log n+\Big(\frac{\log T}{\log n}\Big)^{\alpha_s}
        \lesssim_{s,c}\ondisc(\mu_s,T)\lesssim_{s,c}
        \log\log n+\Big(\frac{\log T}{\log n}\Big)^{1/2}.
    \end{equation*}
\end{corollary}

\paragraph{Terminal discrepancy.}
The next theorem shows that an algorithm can recover from the forcing patterns behind \cref{thm:Prefix}\ref{item:PrefixLower} at every prescribed time, with constants that do not depend on that time, without knowledge of the horizon, and with polynomially small failure probability.

\begin{theorem}[Terminal discrepancy at every prescribed time]
    \label{thm:Terminal}
    There is a deterministic online algorithm, depending only on $n$ and $s$, with the following guarantee.
    For every fixed $k\ge1$ there are constants $C_k$ and $n_k$ such that, for all $n\ge n_k$, in the binary model with $2\le s\le n/2$ and in the signed model with $2\le s\le n$, and for every prescribed time $T\in\N$, the signs chosen by the algorithm satisfy
    \begin{equation}
        \label{eq:Terminal}
        \P\bc{\norm{\sum_{t=1}^Tx_tV_t}_\infty>C_k\big(\sqrt s+\log\log n\big)}\le n^{-k}.
    \end{equation}
\end{theorem}

The bound holds at every prescribed time $T$; by \cref{thm:Prefix}\ref{item:PrefixLower}, for fixed $s$ no bound of this order can hold simultaneously for all times up to a long horizon.
The process is never restarted, and the algorithm uses no information about the horizon or the failure probability, so the bounds for all $T$ and all $k$ concern one and the same sequence of signs.
Together with the terminal lower bound of Altschuler and Tikhomirov, which we extend to the signed model in \cref{prop:Baseline}, \cref{thm:Terminal} shows that the terminal discrepancy is of order $\sqrt s+\log\log n$ at every prescribed time $T\ge n$, in the binary model with $2\le s\le n/2$, the range considered in \cite{Altschuler26Binary}, and in the signed model with $2\le s\le n$.
Here the upper bound holds with failure probability $n^{-k}$ for every fixed $k$, and the lower bound with probability $1-o(1)$.
This answers their terminal question.
For $s=1$, and for $s=n$ in the binary model, a simple alternating rule keeps the prefix discrepancy equal to one at all times, which is optimal, and its terminal discrepancy is at most one (\cref{rem:SizeOne}); binary inputs with $n/2<s<n$ are not covered by \cref{thm:Terminal}.

The algorithm of \cref{thm:Terminal} is one of two deterministic rules, selected before the first arrival according to $n$ and $s$.
For $s>\log\log(en)$, it is a \emph{gauge rule} (\cref{sec:Gauge}): it follows the gradient of a potential built from the profile of Altschuler and Tikhomirov \cite[Section~3.1]{Altschuler26Binary}, an inverse-square profile with an exponential continuation; we translate this profile with the state, and, like their rule, ours gives priority to a small set of exceptional coordinates.
For $s\le\log\log(en)$ and $n\ge2^{40}s^{3/2}$, which holds once $n$ exceeds an absolute constant, it is a \emph{hybrid rule} (\cref{sec:CoshHybrid}) that follows the gradient of the potential $y\mapsto\sum_i\cosh(\omega y_i)$ at the current state, with $\omega\asymp s^{-1/2}$, and gives priority to a small population of high coordinates; in all remaining cases, it is the gauge rule.
For $s=2$, quadratic Greedy itself satisfies \eqref{eq:Terminal} (\cref{cor:GreedyTerminal}); thus, for pairs, a single parameter-free rule attains the optimal order of the terminal discrepancy at every time $T\ge n$ and, for $n\le T\le e^{cn}$, of the prefix discrepancy.

In expectation, \cref{thm:Terminal}, together with a bound, uniform in $T$, on the second moment of the terminal discrepancy of the same rule (\cref{sec:ExpectedTerminal}), and the terminal lower bound give the following.

\begin{corollary}[Expected terminal discrepancy]
    \label{cor:ExpectedTerminal}
    Consider the binary model with $2\le s\le n/2$ and the signed model with $2\le s\le n$.
    There is an absolute constant $n_0$ such that, uniformly for $n\ge n_0$ and integers $T\ge n$,
    \begin{equation*}
        \inf\E\norm{\sum_{t=1}^Tx_tV_t}_\infty\asymp\sqrt s+\log\log n,
    \end{equation*}
    where the infimum is taken over all online algorithms.
    The upper bound is achieved, for every $T$, by one and the same algorithm, namely that of \cref{thm:Terminal}.
\end{corollary}

Since the algorithm of \cref{thm:Terminal} does not depend on $k$, we may apply \cref{thm:Terminal} at every time $t\le T$ with a larger exponent; for $T\le n^\beta$, a union bound over these times turns the terminal bound into a prefix bound.
The terminal lower bound of \cref{prop:Baseline} also bounds the prefix discrepancy from below, so this determines the optimal order at all polynomial horizons.

\begin{corollary}[Polynomial horizons]
    \label{cor:PolynomialPrefix}
    Fix $\beta\ge1$, and consider the binary model with $2\le s\le n/2$ and the signed model with $2\le s\le n$.
    Uniformly for $n\ge3$ and integers $n\le T\le n^\beta$,
    \begin{equation*}
        \ondisc(\mu_s,T)\asymp_\beta\sqrt s+\log\log n.
    \end{equation*}
    More precisely, for every fixed $k\ge1$ there are constants $C_{\beta,k}$ and $n_{\beta,k}$ such that, for all $n\ge n_{\beta,k}$ and all integers $1\le T\le n^\beta$, the signs chosen by the algorithm of \cref{thm:Terminal} satisfy
    \begin{equation}
        \label{eq:PolynomialPrefix}
        \P\bc{\max_{1\le t\le T}\norm{\sum_{u=1}^tx_uV_u}_\infty>C_{\beta,k}\big(\sqrt s+\log\log n\big)}\le n^{-k}.
    \end{equation}
\end{corollary}

Thus the order $\sqrt s+\log\log n$, which Altschuler and Tikhomirov established for $T=\Theta(n)$, persists for the prefix discrepancy at every polynomial horizon, for every sparsity in their range, and it is attained by a single rule that does not know the horizon; restarting an algorithm designed for $T=\Theta(n)$ on consecutive blocks of fixed length would only give a bound growing linearly in $T/n$.
For fixed $s$, the order persists as long as $\log T\lesssim\log n\,(\log\log n)^2$, by \cref{thm:Prefix}\ref{item:PrefixPairs} and~\ref{item:PrefixFixed}, and it fails once $\log T\gg\log n\,(\log\log n)^{1/\alpha_s}$, by \cref{thm:Prefix}\ref{item:PrefixLower}; for $s=2$, the first range extends to $\log T\lesssim\log n\,(\log\log n)^3$, so the two thresholds coincide (\cref{sec:ExpectedPrefix}).
For sparsity growing with $n$, \cref{cor:PolynomialPrefix} determines $\ondisc(\mu_s,T)$ at all polynomial horizons.
At longer horizons, the joint dependence on $s$ and $T$ is open (\cref{q:Joint}): \cref{rem:GrowingSparsity} gives a lower bound for growing $s$; for $s\le\log^2(en)$, the gauge rule keeps the order $\sqrt s+\log\log n$ in expectation as long as $\log T\le\sqrt s\min\{s,\log(en)\}$ (\cref{prop:GrowingPrefix}); and the generic bounds of \cref{rem:GenericBenchmarks}, based on the result of Hmadi \cite{Hmadi26}, are of the optimal order $\sqrt s$ once $s\gtrsim\log T\,(\log\log T)^2$.

\paragraph{Prefixes and the effective horizon.}
For fixed $s\ge2$ and $n\le T\le e^{cn}$, our bounds express the horizon dependence of the optimal expected prefix discrepancy through the effective horizon $\log T/\log n$.
This dependence is far below the polynomial dependence on $T$ for adaptive adversaries, and also below the generic oblivious bound $O(\sqrt{s\log(T/\delta)})$ \cite{Kulkarni24,AdenAli26}, which gives $O(\sqrt n)$ at $T=\floor{e^{cn}}$; there, the optimal prefix order lies between $(n/\log n)^{\alpha_s}$ and $(n/\log n)^{1/2}$, and it equals $(n/\log n)^{1/3}$ for $s=2$.
Beyond exponential horizons, the optimal prefix discrepancy converges as $T\to\infty$, and its limit is of order $n^{s-1}$, as for an adaptive adversary; in the signed model with $s\ge3$, it equals the adaptive value $2^{s-2}\binom{n-1}{s-1}$ (\cref{cor:BeyondExponential}).
Random inputs thus eventually force discrepancy of the adaptive order, but only at horizons with $\log T\gg n$, since for $T\le e^{cn}$ the optimal prefix discrepancy is at most of order $(n/\log n)^{1/2}$.

\paragraph{Terminal recovery.}
At terminal times, the horizon does not matter.
For fixed $s\ge2$ and $T=\floor{e^{cn}}$, \cref{cor:ExpectedPrefix,cor:ExpectedTerminal} show that the optimal expected terminal discrepancy is of order $\log\log n$, whereas the optimal expected prefix discrepancy is at least of order $(n/\log n)^{\alpha_s}$.
Thus the two objectives, which agree up to constants for $T=\Theta(n)$ \cite{Altschuler26Binary}, differ there by a factor of order at least $(n/\log n)^{\alpha_s}/\log\log n$.
The rule of \cref{thm:Terminal} cannot prevent the forcing patterns of \cref{thm:Prefix}\ref{item:PrefixLower}, but it recovers from them: the analysis at time $T$ involves only a window of length $n\,\mathrm{polylog}(n)$ before $T$, together with moment bounds that hold uniformly in time.

\paragraph{Further results.}
The lower bound extends to all horizons $T\ge n$ for pairs, to growing sparsity and to random edge orientations, and it gives lower bounds for the terminal radius at failure probabilities between $e^{-n}$ and a fixed negative power of $n$ (\cref{sec:ProofUniversalLower}); for pairs, quadratic Greedy satisfies a weaker bound of order $\log n+(\log(T/\delta))^{1/3}$ at all horizons and failure probabilities (\cref{prop:PairsAllParameters}).
In the appendices, we show that quadratic Greedy survives at least a quadratic number of requests in the finite game, so that for fixed $s$ no forcing argument based on the game gives an exponent larger than $1/2$, and that the radix construction has the optimal exponent for our width certificate with static coefficients (\cref{sec:MethodLimits}).
We also show that, in the signed model with every fixed $s\ge3$, quadratic Greedy with ties broken towards $+1$ does not beat the square-root order (\cref{sec:SpecificRules}).

\subsection{Comparison with previous work}
\label{sec:Comparison}

Related results differ in the input model, the discrepancy objective, and the dependence on the horizon.
For background on discrepancy theory, including prefix and online discrepancy, we refer to the monograph of Bansal and Nikolov \cite{BansalNikolov26}.
\Cref{tab:Comparison} summarizes the most closely related online guarantees.

\paragraph{Sparse random inputs at linear horizons.}
For the binary model, Altschuler and Tikhomirov proved the lower bound $\Omega(\sqrt s+\log\log n)$ for the terminal discrepancy at every prescribed time $T\ge n$, and gave an efficient algorithm with prefix discrepancy $O(\sqrt s+\log\log n)$ for $T=n$ \cite{Altschuler25,Altschuler26Binary}.
Their upper bound extends to $T=\Theta(n)$ by restarting on consecutive blocks of at most $n$ arrivals, with a constant depending on $T/n$ \cite[Remark~1.2]{Altschuler26Binary}.
At $T=\Theta(n)$, the $\sqrt s$ term is necessary even offline \cite{MacRury23}, whereas the $\log\log n$ term comes from an isolation argument that uses fresh supports \cite[Section~1.2.1 and Lemma~2.3]{Altschuler26Binary}.
Compared with their results, we add the horizon term to the lower bound, control all prefixes up to exponential horizons for fixed $s$, show that their order persists for the prefix discrepancy at all polynomial horizons (\cref{cor:PolynomialPrefix}), and match their terminal lower bound at every prescribed time, without restarts.

\paragraph{Relation to earlier work on the terminal bound.}
The gauge rule of \cref{thm:Terminal} builds on the algorithm of Altschuler and Tikhomirov: its profile is a continuous version of their potential, which consists of an inverse-square profile and an exponential continuation \cite[Section~3.1]{Altschuler26Binary}, its curvature estimates follow theirs \cite[Lemmas~3.3 and~3.5]{Altschuler26Binary}, its one-step service estimates use their finite-population sampling inequality \cite[Lemma~3.4]{Altschuler26Binary}, and, like their rule, it gives priority to exceptional coordinates, although our exceptional set depends only on the current state.
The analysis at a prescribed time follows the scheme of our earlier work \cite{Wengiel26}: a drift inequality at every state gives moments that are bounded uniformly in time, and, for a rare-activation law, a maximal inequality on a deterministic window before $T$ and the clearance of the high levels within this window give a terminal bound \cite[Lemma~27 and Appendix~B.2--B.4]{Wengiel26}.
New here are the translation of the profile with the state, which yields such an inequality for the profile of Altschuler and Tikhomirov; the estimates for exact supports in both models that this inequality and the clearance require; the control of each level by moments of elementary symmetric polynomials; and a bootstrap that couples a large exponential moment of the gauge with these moments up to a common exit time.
Unlike the window argument of \cite[Appendix~B]{Wengiel26}, which gives a fixed failure probability, this bootstrap makes the failure probability polynomially small at the sharp radius.

\begin{table}[t]
    \centering
    \caption{Selected online guarantees from the literature, in our notation. Prefix bounds hold simultaneously up to the horizon; terminal bounds concern a prescribed time.}
    \label{tab:Comparison}
    \begingroup
    \renewcommand{\arraystretch}{1.15}
    \begin{tabularx}{\textwidth}{@{}>{\raggedright\arraybackslash}p{0.24\textwidth}>{\raggedright\arraybackslash}p{0.25\textwidth}>{\raggedright\arraybackslash}X@{}}
        \toprule
        \textbf{Work} & \textbf{Input and objective} & \textbf{Guarantee and scope}\\
        \midrule
        Kulkarni--Reis--Rothvoss \cite{Kulkarni24}; Aden-Ali \cite{AdenAli26} & Fixed vectors with $\norm{v_t}_2\le1$; oblivious; prefix & $O(\sqrt{\log(T/\delta)})$; gives $O(\sqrt{s\log(T/\delta)})$ for our inputs.\\
        Altschuler--Tikhomirov \cite{Altschuler26}; Hmadi \cite{Hmadi26} & Fixed vectors in $[-1,1]^n$, $s$-sparse (or with $\norm{v_t}_2\le\sqrt s$ in \cite{Hmadi26}); prefix & $O(\sqrt s)$ once $s$ exceeds $\log T$ by a doubly logarithmic factor.\\
        Bansal--Prabhu--Singla--Sundaram \cite{BansalCarpool26} & Graph edges $e_i-e_j$; prefix & Greedy: $O(\min\{T^{1/3},n\})$ adversarially. Random edges of a $\Delta$-regular graph, $T=O(n)$: $O((\log n)^{1/3})$, and $O((\log n/\log\Delta)^{1/3}+\log\log n)$ if $\Delta=\Omega(\log^3n)$.\\
        Bansal--Spencer \cite{Bansal20a} & Rademacher vectors, $s=n$; terminal & $O(\sqrt n)$ at every prescribed time.\\
        Altschuler--Tikhomirov \cite{Altschuler26Binary} & Binary model; prefix upper, terminal lower bound & $\Theta(\sqrt s+\log\log n)$ for $T=\Theta(n)$, $2\le s\le n/2$; lower bound at every $T\ge n$.\\
        \bottomrule
    \end{tabularx}
    \endgroup
\end{table}

\paragraph{Oblivious and stochastic online balancing.}
For oblivious sequences of vectors of Euclidean norm at most one, the self-balancing walk of Alweiss, Liu and Sawhney \cite{Alweiss21} gives prefix discrepancy $O(\log(nT))$, and a Gaussian fixed-point walk of Liu, Sah and Sawhney gives the order $\sqrt{\log(nT)}$ for partial colorings and a relaxed set of signs \cite{Liu22}; Kulkarni, Reis and Rothvoss \cite{Kulkarni24} obtained the optimal bound $O(\sqrt{\log(T/\delta)})$ with failure probability $\delta$, together with a matching lower bound for a randomized oblivious adversary in two dimensions, and Aden-Ali \cite{AdenAli26} achieved this bound in linear time.
Applied to $V_t/\sqrt s$, these results give $O(\sqrt{s\log(T/\delta)})$ prefix discrepancy for our inputs (\cref{rem:GenericBenchmarks}), and $O(\sqrt{s\log(n/\delta)})$ at every prescribed terminal time \cite[Corollary~3]{Kulkarni24}.
For $s$-sparse vectors in $[-1,1]^n$, Altschuler and Tikhomirov \cite{Altschuler26} and Hmadi \cite{Hmadi26} obtained prefix discrepancy $O(\sqrt s)$ once $s$ exceeds $\log T$ by a doubly logarithmic factor, namely for $s\gtrsim\log T\,(\log\log T)^{2+\eta}$ with a fixed $\eta>0$ in \cite{Altschuler26} and for $s\gtrsim\log T\,(\log\log T)^2$ in \cite{Hmadi26}; Hmadi's result applies more generally to vectors with Euclidean norm at most $\sqrt s$ and entries at most one in absolute value.
For \iid arrivals from general distributions, Bansal, Jiang, Singla and Sinha \cite{Bansal20b} and Bansal, Jiang, Meka, Singla and Sinha \cite{Bansal21} obtained polylogarithmic dependence on $T$.
These results apply to much more general inputs; for sparse random inputs they leave a dependence on $\log T$ rather than $\log T/\log n$, and they do not provide lower bounds for sparse random inputs.
The lower bound of \cite{Kulkarni24} uses a specific input distribution in two dimensions and does not apply to \iid uniform supports.
Offline, prefix discrepancy is the signed-series problem behind the Steinitz lemma: Bansal, Jiang, Meka, Singla and Sinha showed that random perturbations of the input reduce the dependence on $T$ in Banaszczyk's prefix bound $O(\sqrt{\log n+\log T})$ \cite{Banaszczyk12} to $O(\sqrt{\log n+\log\log T})$ \cite{Bansal22}, and Russo and Wei recently reduced an online version of the Steinitz problem to online vector balancing \cite{RussoWei26}.

\paragraph{Terminal guarantees at every prescribed time.}
For dense Rademacher inputs, Bansal and Spencer \cite{Bansal20a} obtained terminal discrepancy $O(\sqrt n)$ at every prescribed time; \cref{thm:Terminal} extends this to sparse inputs at the scale $\sqrt s+\log\log n$ and, for $s=n$ in the signed model, recovers their order, with failure probability $n^{-k}$ instead of an exponentially small one.
In earlier work \cite{Wengiel26}, we studied random vectors with independent coordinates, including Bernoulli-masked inputs, in which each coordinate is retained independently with a fixed probability, and obtained terminal guarantees that hold at every prescribed time without knowledge of the horizon.
There, too, a doubly logarithmic factor appears: for a law with rare activations, the optimal order at a fixed failure probability is $\Theta(\sqrt n\log\log n)$ at all sufficiently late prescribed times.
The present paper shows that, for sparse inputs, the terminal discrepancy has the same order at every prescribed time $T\ge n$ and that this order is attained without knowledge of the horizon, whereas for fixed $s$ the order of the prefix discrepancy grows with the horizon (\cref{thm:Prefix}\ref{item:PrefixLower}), even though the upper bounds of \cref{thm:Prefix} are attained by rules that do not know the horizon: quadratic Greedy for $s=2$ and, up to the constant, the restart rule of \cref{prop:Restart} for $s\ge3$.

\paragraph{Online carpooling.}
For $s=2$, our problem is close to online carpooling, or online edge orientation, in which each arrival is an edge $e_i-e_j$ of a graph.
Ajtai, Aspnes, Naor, Rabani, Schulman and Waarts \cite{Ajtai98} showed, building on a chip game of Anderson, Lov\'asz, Shor, Spencer, Tardos and Winograd \cite{Anderson89}, that at most $h^3$ requests suffice to force unfairness $h$, which is half the discrepancy, against any deterministic algorithm whenever $h\le\frac12\ceil{(n-1)/2}$, and that for uniformly random edges of the complete graph, Greedy has expected unfairness of order $\log\log n$ at a typical time of a long execution.
Gupta, Krishnaswamy, Kumar and Singla \cite{Gupta20} obtained polylogarithmic bounds for random edges of general graphs.
Bansal, Prabhu, Singla and Sundaram \cite{BansalCarpool26} proved that Greedy achieves $O(\min\{T^{1/3},n\})$ for adversarial arrivals.
For $T=O(n)$ uniformly random edges of a $\Delta$-regular graph, they obtained the bound $O((\log n)^{1/3})$, and $O((\log n/\log\Delta)^{1/3}+\log\log n)$ when $\Delta=\Omega(\log^3n)$; on the complete graph this is $O(\log\log n)$.
Their lower bound $\Omega((\log n/\log\Delta)^{1/3})$ for $n$ random arrivals with $\Delta=\Omega(\log n)$, against every online algorithm, matches the cube-root term of their upper bound.
It uses a disjoint union of cliques of size $\Delta+1$ and prescribes in each of them a forcing sequence of cubic length that adapts to the current heights; since the cliques are independent, the sequence is realized in one of them with high probability \cite[Theorem~4.3]{BansalCarpool26}.
Our lower bounds use the same principle, which already underlies the lower bound of Ajtai et al.\ against an oblivious adversary \cite[Theorem~3.2]{Ajtai98}: a forcing sequence that is unlikely in a single attempt but succeeds in one of many attempts, here with consecutive blocks of time in place of disjoint groups of vertices.
Our proof for pairs uses their deterministic chip-game estimates, the cube-root bound and the spreadness lemma \cite[Lemmas~3.1 and~3.3]{BansalCarpool26}, as its only structural input (\cref{thm:ChipGame}).
The stochastic analysis is different: our arrivals carry random marks or are binary, the reduction of \cref{lem:DoubleCover} to a chip game with two chips per coordinate produces two correlated chip moves per arrival, and the horizon is exponential rather than linear in $n$.

\paragraph{Offline discrepancy of worst-case sparse inputs.}
Spencer showed that any $n$ vectors in $[-1,1]^n$ admit signs with discrepancy $O(\sqrt n)$ \cite[Theorem~1]{Spencer85}, and Beck and Fiala showed that vectors in $[-1,1]^n$ with at most $s$ nonzero entries admit discrepancy less than $2s$ \cite{BeckFiala81}.
Their conjecture that $O(\sqrt s)$ suffices was approached through Banaszczyk's bound $O(\sqrt{s\log T})$ for $T$ vectors \cite{Banaszczyk98}, the bound $O(\sqrt s)$ for $s\ge\log^2T$ of Bansal and Jiang \cite{Bansal26}, and its extension to $s\ge(\log T)^{1+o(1)}$ by the online algorithm of Altschuler and Tikhomirov \cite{Altschuler26}.
It was recently proved, as a consequence of the Koml\'os conjecture, by Guo, Fang and Lu \cite[Theorem~1.1 and Corollary~1.2]{GuoFangLu26} (see also \cite{Karingula26}), and polynomial-time algorithms followed \cite{GuoFangLu26b,Li26,AnandChen26}; in the meantime, Altschuler gave an efficient algorithm with discrepancy $O(\sqrt s\,(1+\log^*T))$ when no coordinate is used by more than $e^{O(\sqrt s)}$ of the vectors \cite{Altschuler26c}.
Thus, offline, the final signed sum of any sequence of vectors in $[-1,1]^n$ with at most $s$ nonzero entries can be kept below $C\sqrt s$, for an absolute constant $C$ and at every horizon.

\paragraph{Offline discrepancy of sparse random set systems.}
Offline, the discrepancy of sparse random set systems has been studied extensively \cite{Ezra19,BansalMeka20,Potukuchi18,Hoberg19,Franks20,MacRury23}.
With many more columns (arrivals) than rows (coordinates), it becomes bounded: Franks and Saks \cite{Franks20} showed that the binary model with $T\ge Cn^3\log^2n$ columns has offline discrepancy at most two with high probability; see also \cite{Ezra19,Hoberg19,Altschuler22}.
Online, by contrast, the prefix discrepancy eventually grows with the horizon for fixed $s$, and already the terminal discrepancy is at least of order $\sqrt s+\log\log n$ \cite{Altschuler26Binary}; by \cref{thm:Terminal}, the optimal terminal discrepancy is of this order at every prescribed time $T\ge n$.

\subsection{Organization}
\label{sec:Organization}

\Cref{sec:TechnicalOverview} explains the main ideas: for the lower bound, a finite safety game and a width certificate; for the prefix upper bounds, moments of elementary symmetric polynomials, which control the number of high coordinates over the entire horizon; and for the terminal bound, a translated gauge with layers cleared in deterministic windows and a bootstrap inside a controlled height region, or, for $s\le\log\log(en)$, a $\cosh$ potential with priority for high coordinates.
\Cref{sec:Preliminaries} fixes the model, establishes the baseline lower bound at every prescribed time $T\ge n$ in both models, and collects the moment tools shared by the upper bounds.
\Cref{sec:LowerBounds} proves \cref{thm:Prefix}\ref{item:PrefixLower}, \cref{sec:Pairs,sec:FixedSparsity} prove parts~\ref{item:PrefixPairs} and~\ref{item:PrefixFixed}, and \cref{sec:Terminal} proves \cref{thm:Terminal} and its counterpart for quadratic Greedy at $s=2$.
\Cref{sec:Consequences} derives the corollaries, removes the dependence of the rule of part~\ref{item:PrefixFixed} on the horizon by restarts, extends the prefix bound at polynomial horizons to longer horizons for growing sparsity, and discusses benchmarks and complements; \cref{sec:OpenQuestions} collects open questions and a conjecture.
The appendices treat the limits of the lower-bound method (\cref{sec:MethodLimits}) and the forcing of quadratic Greedy (\cref{sec:SpecificRules}).

\section{Technical overview}
\label{sec:TechnicalOverview}

This section explains the main ideas behind \cref{thm:Prefix,thm:Terminal}.
The lower bound rests on a deterministic game and on an exact realization of its moves within the input law, each with a probability that is polynomial in $1/n$ (\cref{sec:OverviewLower}).
The prefix upper bounds control two quantities over the entire horizon: the number of coordinates above a moderate level, by moments of elementary symmetric polynomials that all our upper bounds share (\cref{sec:OverviewPopulation}), and an energy above that level (\cref{sec:OverviewPairs,sec:OverviewFixed}).
For $s>\log\log(en)$, the terminal upper bound controls, at every state, a level that moves with the state, the gauge, and clears the coordinates far above this level within a deterministic window before the prescribed time; for smaller $s$, a hybrid rule with a $\cosh$ potential is used (\cref{sec:OverviewTerminal}).
Both input models are handled by the same arguments; how the two models are related differs between the three upper bounds and is explained in each subsection.

We write $S_t\colonequals\sum_{u=1}^tx_uV_u$ for the signed sum after $t$ arrivals, $S_i$ for the $i$th coordinate of the current signed sum, and $D_T\colonequals\max_{0\le t\le T}\norm{S_t}_\infty$ for the prefix discrepancy up to time $T$.
Apart from $S$, $T$, $L$, $D_T$ and $\alpha_s$, the symbols in this section are local to each subsection and follow the notation of the section in which the corresponding argument is carried out.

\subsection{Lower bounds: a safety game and a width certificate}
\label{sec:OverviewLower}

The lower bound of \cref{thm:Prefix}\ref{item:PrefixLower} has three ingredients: a reduction from long horizons to short forcing strategies, a finite safety game in which such strategies live, and a width certificate with mixed-radix coefficients that bounds their forcing time.

\paragraph{From forcing to long horizons.}
Track a pool of $m$ coordinates, and suppose that, from every state in which all pool heights are below $H$, an adversary who chooses the support and the marks of the next arrival can force some height $H$ within $J$ arrivals, whatever signs the algorithm chooses; we call $J$ the \emph{forcing time}.
Each arrival prescribed by the adversary occurs with probability at least $(s/(2en))^s$, conditionally on the entire past.
Hence, within a block of $J$ consecutive arrivals, height $H$ is forced with conditional probability at least $n^{-O_s(J)}$, and $\floor{T/J}$ disjoint blocks give
\begin{equation*}
    \P(D_T<H)\le\exp\big\{-\floor{T/J}\,n^{-O_s(J)}\big\}
\end{equation*}
(\cref{prop:BlockBound}).
This bound holds against every online algorithm, including randomized ones that know $T$, and the argument does not require the state to be reset between blocks.
If the forcing time grows like $J\asymp H^\beta$, the height $H\asymp(\log T/\log n)^{1/\beta}$ is forced with high probability.
The task is thus to find adversarial strategies with a small exponent $\beta$ and a pool small enough to fit into $[n]$.

\paragraph{The safety game.}
These strategies live in a finite game.
Its state $z\in\{-R,\ldots,R\}^m$ represents the pool; the adversary requests a direction $v\in\{-1,0,1\}^m$ with at most $r\le s$ nonzero coordinates, and the \emph{controller}, that is, the algorithm, moves the state to $z+v$ or $z-v$.
The \emph{safety sets} $Q_j$, obtained by iterating an operator of B\'ar\'any \cite{Barany79}, consist of the states from which the controller can keep the state in the cube during $j$ further requests; thus $Q_J=\emptyset$ means that an exit from the cube can be forced within $J$ requests.
Whether this happens for some $J$ is decided by a width bound of B\'ar\'any: it does if and only if $2R$ is smaller than the number of pairs $\pm v$ of directions whose support contains a given coordinate (\cref{prop:InfiniteValue}).
For random inputs, what matters is the size of $J$.
Both input models realize this game exactly, with cube radius $R\asymp H$ (\cref{lem:Realization}); in the binary model, the differences of paired coordinates act as virtual signed coordinates.

\paragraph{A width certificate.}
The idea is to follow a family of linear scores of the safe states: every score that fails to decrease from $Q_j$ to $Q_{j+1}$ forces a kink in a convex function whose slope increases by at most $2R$ in total, so most scores decrease and the safety sets become empty.
To show $Q_J=\emptyset$, we distinguish the first pool coordinate, fix integer coefficients $a$ on the pool with $a_1=0$, and follow the support function $g_j(t)=\max_{z\in Q_j}\inp{a+te_1}{z}$ of $Q_j$ along the line of normals $a+te_1$.
Its slopes are first coordinates of states, so they lie in $[-R,R]$, and $g_j$ is convex.
For a direction $v$ with $v_1=1$, the \emph{root} $t_v=-\inp av$ is the value of $t$ at which the normal $a+te_1$ is orthogonal to $v$, and we call $g_j(t_v)$ the \emph{score} of $v$.
If the score does not decrease from $Q_j$ to $Q_{j+1}$, a maximizer and its safe successor $z\pm v$ are both maximizers at $t_v$, with first coordinates differing by one; this produces a kink of size at least one in $g_j$ at $t_v$.
Hence, among $N$ distinct roots, all but $2R$ scores decrease by at least one in every step from $Q_j$ to $Q_{j+1}$.
Since the scores are bounded by $R$ times the $\ell_1$-norms of the normals, the safety sets die after about $R\cdot(\norm a_1+\max\abs{t_v})$ steps (\cref{lem:WidthCertificate}).

\paragraph{Mixed radices.}
For directions with $v_1=1$ and at most $k$ further nonzero coordinates, roots are signed sums of at most $k$ coefficients.
We need at least $4R$ consecutive roots with a small coefficient mass $\norm a_1$.
For $k=1$, the roots are the coefficients themselves, so $\norm a_1\asymp R^2$ and $J\asymp R^3$.
For larger $k$, we organize the coefficients by mixed-radix digits (\cref{prop:Radix}), an $\ell_1$-version of the recursive additive bases of Bandeira, Ferber and Kwan \cite[Lemma~3.1]{Bandeira17}, who control the sum of squares instead.
Optimizing the radices gives $\norm a_1\asymp R^{\eta_k}$ with $\eta_k=2^k/(2^k-1)$, a pool of size $O(kR^{\eta_k/2})$, and forcing time $J\asymp R^{1+\eta_k}$ (\cref{prop:OptimizedRadices}).
For $k=s-1$, the forcing time has exponent $\beta=1+\eta_{s-1}$, and the forced height has exponent $1/\beta=\alpha_s$.

Two limits apply.
A counting argument shows that no static choice of coefficients does better for this certificate, as long as it is evaluated with the budget of \cref{lem:WidthCertificate} (\cref{prop:StaticBarrier}).
More fundamentally, quadratic Greedy survives at least a quadratic number of requests in the game (\cref{prop:QuadraticSurvival}), so for fixed $s$ no forcing argument based on the game can give an exponent larger than $1/2$ (\cref{sec:MethodLimits}).

\subsection{Population control by elementary symmetric moments}
\label{sec:OverviewPopulation}

All our upper bounds control the \emph{population} at a level $b$, the number of coordinates of height at least $b$; two coordinates \emph{meet} when they are selected by the same arrival.
The tool is a family of moments of elementary symmetric polynomials, which bound the probability that $r$ given coordinates are simultaneously high without assuming any independence between coordinates (\cref{sec:Tools}).

\paragraph{Triangular recursions.}
For weights $Y_i$ that increase with the height, the elementary symmetric polynomials $e_r(Y)$ obey a \emph{triangular recursion}: after one arrival, the conditional expectation of $e_r(Y)$ is at most a contraction of $e_r(Y)$ plus a multiple of $e_{r-1}(Y)$.
Such a recursion keeps all these moments bounded uniformly in time, and hence it bounds the population at a moderate level at every time (\cref{lem:TriangularRecursion}).

\paragraph{Layers.}
Moments of $\vartheta^{(\abs{S_i}-b)_+}-1$, multiplied by an indicator that vanishes once the population at a lower level becomes too large, then show that the population at the next level drops from $q$ to roughly $sq^2/n$, as in the analysis of two-choice allocations \cite{Azar99,Berenbrink06} and in the doubly exponential level estimates of Altschuler and Tikhomirov \cite[Section~3.5]{Altschuler26Binary}; after $O(\log\log n)$ levels it is of order $L$ (\cref{lem:LocalClearance,lem:LayerIteration}).
These layer estimates need only one property of the rule: if exactly one of the selected coordinates is high, a \emph{lone high coordinate}, the rule moves it inward.
Each population claim fails with probability at most $(T+1)e^{-L}$, and there are only $O(\log\log n)$ claims.

\subsection{Upper bound for pairs: an energy witness and population control}
\label{sec:OverviewPairs}

For $s=2$, quadratic Greedy is analyzed through an energy above a level and the population at that level.

\paragraph{A deterministic energy witness.}
Quadratic Greedy can be viewed as the Greedy algorithm of online carpooling on a doubled set of chips.
If each coordinate is represented by two chips at positions $S_i$ and $-S_i$, every arrival becomes two Greedy moves of chips (\cref{lem:DoubleCover}).
The spreadness lemma of Bansal, Prabhu, Singla and Sundaram \cite[Lemma~3.3]{BansalCarpool26}, combined with Jensen's inequality as in their proof of \cite[Lemma~3.1]{BansalCarpool26}, then shows the following: if a height first reaches $b+u$, then at some earlier time the excess energy $F_b=\sum_i(\abs{S_i}-b)_+^2$ is at least $u^3/32$.

\paragraph{Excursions of the excess energy.}
Above a level $b$, the energy $F_b$ can increase only when two coordinates of height at least $b$ meet with equal heights, which has probability $O(q^2/n^2)$ if the population at $b$ is at most $q$.
It decreases whenever a coordinate with positive excess meets a coordinate below $b$, which has probability of order $m/n$, where $m$ is the number of coordinates with positive excess.
For $F_b\gtrsim u^3$ while all heights are below $b+u$, we have $m\gtrsim u$, and $z^{F_b}$ is a supermartingale with $z\asymp\sqrt{nu}/q$.
An excursion to energy $u^3/32$ therefore has probability $\exp(-\Omega(u^3\log z))$.
With a suitable choice of $q$ for $L\le\sqrt n$ and for $L>\sqrt n$, we have $\log z\gtrsim\log n$, and this gives $u\asymp(L/\log n)^{1/3}$ (\cref{lem:CubicExcursions}).
This is where the exponent $1/3$ arises: the witness requires energy of order $u^3$, and each unit of energy reduces the probability by a factor $n^{-\Omega(1)}$.

\paragraph{Population control.}
It remains to show that the population at a level $b$ that fits into the bound is at most $q$ at all times up to $T$.
For $L\le\sqrt n$, the triangular recursion with $Y_i=2^{\abs{S_i}}$ bounds the population at a constant level by a small fraction of $n$ (\cref{lem:GeometricMoments}), and the layers of \cref{sec:OverviewPopulation} reduce it to $q$ at a level $b=O(\log\log n+u)$.
For $L>\sqrt n$, a global quadratic moment bounds the population above a level of order $\log n$, which is smaller than $(L/\log n)^{1/3}$ in this range (\cref{lem:GlobalQuadratic}).

\paragraph{Binary inputs.}
In the binary model, the outcome of an arrival is deterministic given the signs of the two selected coordinates; its deviation from the signed average is a correction with positive semidefinite kernels, such as $\min(Y_i,Y_j)$ for $Y_i=2^{\abs{S_i}}$, so the binary model inherits the signed recursions with slightly worse constants.

\subsection{Upper bound for fixed sparsity: flat weights with a linear tail}
\label{sec:OverviewFixed}

For $s\ge3$, a coordinate of maximal height is no longer guaranteed to move inward under quadratic Greedy.
The rule of \cref{sec:FixedSparsity} is the exact gradient rule for the potential $\Psi(S)=\sum_i\psi(\abs{S_i})$, whose profile $\psi$ has centered differences $a_h$, the weights, which grow exponentially with a small rate $\rho\asymp s^{-2}$ up to a \emph{knee} $B$ and linearly above it; in particular, $\psi$ grows quadratically above the knee.
Quadratic Greedy is no better: in the signed model, with ties broken towards $+1$, it can be driven, with high probability, to height $\Omega_s(\sqrt{\log T/\log n})$ for every fixed $s\ge3$ and $C_s\log n\le\log T\le n^2$ (\cref{prop:GreedyThreeTouches}).
Three properties drive the proof: population control, which holds for every $L$, and two excursion bounds, a tilted supermartingale for $L\ge\sqrt n$ and a squared hinge above a layered knee for smaller $L$.
Both models are handled by fresh orientations.

\paragraph{Population control.}
For $Y_i=e^{\rho\abs{S_i}}$ and degrees $r$ up to a small multiple $\e_sn$ of $n$, the elementary symmetric polynomials $e_r(Y)$ contract in expectation as long as the population above a level $b\le B$ is small (\cref{lem:StoppedMoments}).
The key estimate compares the actual decision with a fictitious decision based on the weights $Y_i$, which would be optimal for the monomials of degree $r$ up to a quadratic error (\cref{lem:NewtonService}).
The support-averaged anti-concentration inequality of \cref{lem:ClippedAntiConcentration} then gives an inward drift of order $\sqrt s/n$ per unit weight.
Coordinates above the knee, where the actual weights deviate from $Y_i$, are handled by comparing the linear tail with the exponential core.

\paragraph{A tilted supermartingale.}
Above the knee, the weights grow linearly with slope $\kappa\asymp s^{-1}$, so the curvature of $\psi$ is at most $\kappa$ and one arrival increases $\Psi$ by at most a small constant, whereas $\psi(B+j)\ge\kappa j^2/2$.
An exponential tilt $e^{\eta\Psi}$ with $\eta\asymp\log n$ is a supermartingale as long as few coordinates are high (\cref{lem:TiltedDrift}).
Reaching height $2B$ requires energy of order $B^2$ and therefore has probability $n^{-\Omega_s(B^2)}$; with $B\asymp\sqrt{L/\log n}$ this handles horizons with $L\ge\sqrt n$.

\paragraph{Layers and a squared hinge.}
For smaller $L$, the knee is placed by the layer iteration of \cref{lem:LayerIteration}, with gaps between consecutive levels that make a lone high coordinate dominate the flat core.
Above the knee, a lone high coordinate moves inward because its linear weight dominates, so a squared hinge slightly above the knee can only increase when two high coordinates meet.
Its excursions to level $u^2$ have probability $n^{-\Omega_s(u^2)}$ (\cref{lem:HingeExcursions}), which gives $u\asymp\sqrt{L/\log n}$.

\paragraph{Fresh orientations.}
To treat both models by one proof, we condition on a full fresh vector of marks in $\{\pm1\}^n$, an \emph{orientation}, of which only the restriction to the support is revealed (\cref{lem:FreshOrientations}).
All one-step estimates are proved for arbitrary fixed orientations and averaged over the uniform support only.
The binary model corresponds to the orientation with all marks equal to one; there only the support average is available, which is why the support-averaged anti-concentration bound is needed (\cref{rem:BinaryTransfer}).

\subsection{Terminal discrepancy: a translated gauge}
\label{sec:OverviewTerminal}

A terminal bound must show that the heights at a fixed time $T$ are of the typical order, however long the stream has run, and it cannot use a union bound over the past.
We therefore look for a functional $\Phi$ of the state that satisfies a \emph{Foster inequality} \cite{Foster53,Hajek82}, a drift inequality of the form $\E[\Phi(S')\mid S]\le(1-\gamma)\Phi(S)+C$ for the state $S'$ after one further arrival, at \emph{every} state $S$, so that its moments stay bounded uniformly in time.
The argument has three ranges of $s$.
For $s\ge(\log(en))^{2/3}$, the Foster inequality and a local clearance argument give the sharp radius; \cref{sec:Terminal} treats $s\le(\log(en))^2$ and $s>(\log(en))^2$ separately, with a moderate and an absolute exponent.
For $\log\log(en)<s<(\log(en))^{2/3}$, the same argument gives only a coarse bound, which is upgraded by a bootstrap inside a controlled height region.
For $s\le\log\log(en)$, a different, hybrid rule is used.

\paragraph{The gauge and the rule.}
Let $f$ be the inverse-square profile of Altschuler and Tikhomirov, $f(u)=(\rho^2/(\rho^2-u^2))^2$ with $\rho\asymp\sqrt s$, with an exponential continuation beyond $\rho-O(1)$ \cite[Section~3.1]{Altschuler26Binary}.
The \emph{gauge} $B(S)$ is the unique real level $b$ with $\sum_if((\abs{S_i}-b)_+)=4n$; it moves with the state, it is convex and $1$-Lipschitz in $\ell_\infty$, and $\norm S_\infty\le B(S)+O(\sqrt s+\log n)$.
Coordinates whose excess over the gauge, $\abs{S_i}-B(S)$, exceeds a threshold $h_*=O(\sqrt s+\log\log n)$ are called \emph{exceptional}; there are at most $n/(s\,\mathrm{polylog}(n))$ of them.
As in the rule of Altschuler and Tikhomirov \cite[Algorithm~3.2]{Altschuler26Binary}, the highest selected exceptional coordinate is moved inward; if no selected coordinate is exceptional, the rule follows the gradient of the gauge, which is proportional to the gradient of $\sum_if((\abs{S_i}-b)_+)$ at the fixed level $b=B(S)$.

\paragraph{A Foster inequality at every state.}
A second-order expansion of $B$ along the update, together with the finite-population sampling inequality of Altschuler and Tikhomirov \cite[Lemma~3.4]{Altschuler26Binary}, shows that, in the binary model, $B$ decreases in expectation by at least order $\sqrt s/n$ whenever $B>1$ (\cref{lem:GaugeDrift}), and that it increases in expectation by at most $O(s/n)$ at every state (\cref{lem:GaugeRemainder}).
This yields, for every state $S$ and the state $S'$ after one further arrival,
\begin{equation*}
    \E\big[e^{\theta B(S')_+}\bigm|S\big]\le\Big(1-\frac{c\min\{\theta\sqrt s,s\}}n\Big)e^{\theta B(S)_+}+\frac{C\theta e^{2\theta}s}n,
\end{equation*}
first for an absolute $\theta$ and then, for $s\le(\log(en))^2$, for the \emph{moderate exponent} $\theta=c_0\min\{s,\log(n/s)\}$ (\cref{prop:GaugeFoster,prop:ModerateFoster}).
Since the inequality holds at every state, it transfers to the signed model by flipping the signs of the coordinates of the state according to the fresh marks (\cref{lem:MarkConjugation}), and it controls the gauge at every fixed time and on every deterministic window.

\paragraph{Local clearance in a deterministic window.}
As long as $B\le B_0$, at most $n/(s\,\mathrm{polylog}(n))$ coordinates exceed $B_0+h_*$, and above this level a lone high coordinate moves inward.
The layer iteration of \cref{sec:OverviewPopulation}, now started from arbitrary states at deterministic times before $T$, reduces the population after a burn-in of $(n/s)\,\mathrm{polylog}(n)$ steps per level, so the argument involves only a window of length $(n/s)\,\mathrm{polylog}(n)$ before $T$ and never conditions on the future; we call this the \emph{composition} argument (\cref{prop:Composition}).
With the moderate exponent, the gauge exceeds $O(1+\log n/s)$ in this window only with probability $n^{-k}$, which gives the \emph{coarse terminal bound} $O(\sqrt s+\log\log n+\log n/s)$ with failure probability $n^{-k}$ (\cref{prop:CoarseTerminal}); for $s\ge(\log(en))^2$, the absolute exponent and the bound $\norm S_\infty\le B(S)+O(\sqrt s+\log n)$ already suffice.
This is sharp for $s\ge(\log(en))^{2/3}$, since then $\log n/s\le\sqrt s$.

\paragraph{A bootstrap inside a controlled height region.}
For $\log\log(en)<s<(\log(en))^{2/3}$, the coarse bound places all heights below $H_0=O(\log n/\log\log n)$ at a deterministic time before $T$, with failure probability $n^{-k}$.
While all heights stay below a barrier of the same order and the gauge stays below $H_0+2$, one arrival moves the gauge by at most $n^{-3/4}$, so the exponential moment of the gauge with the large exponent $n^{1/4}$ satisfies a Foster inequality (\cref{lem:HighDrift}); this moment keeps the gauge below its barrier, and a local moment of degree one keeps all heights below theirs.
Both moments are taken on the event that the common first exit time from this region has not occurred before time $t$, so they include the values at the exit time, and neither estimate conditions on the event established by the other.
After a clearance time of $O(n\log n)$ steps, the gauge stays below $2$ until $T$, and the composition at gauge level $2$ gives the sharp radius with failure probability $n^{-k}$ (\cref{prop:Bootstrap}).

\paragraph{Very small sparsity.}
For $s\le\log\log(en)$ and $n$ larger than an absolute constant, \cref{thm:Terminal} uses a hybrid rule that follows the gradient of the hyperbolic cosine potential $\sum_i\cosh(\omega S_i)$ of Spencer \cite{Spencer77}, with $\omega\asymp s^{-1/2}$, and, like the gauge rule, gives priority to a small population of high coordinates.
The moments of the elementary symmetric polynomials of its weights, of all degrees up to order $n/s^{3/2}$, satisfy a triangular recursion at every state, and the layer iteration gives the radius $\sqrt s\log(s+1)+\log\log n=O(\log\log n)$ with failure probability $n^{-k}$ (\cref{sec:CoshHybrid}).
For $s=2$, quadratic Greedy has the same guarantee: the geometric moments of \cref{sec:GeometricMoments} bound its population and its heights at every fixed time, and \cref{prop:Composition} applies directly (\cref{cor:GreedyTerminal}).

\section{Preliminaries}
\label{sec:Preliminaries}

This section fixes the model and the notation, records an anti-concentration estimate for random supports and signs, establishes the baseline lower bound $\Omega(\sqrt s+\log\log n)$ in both models at every prescribed time $T\ge n$, for the ranges of $s$ in \cref{prop:Baseline}, extending the lower bound of Altschuler and Tikhomirov to the signed model, and collects the moment tools shared by the upper bounds.

\subsection{Model and conventions}
\label{sec:Model}

Throughout, $n\ge2$ and $1\le s\le n$ are integers.
The \emph{support} $A$ of an arrival is a uniformly random $s$-element subset of $[n]$, which we call a \emph{uniform $s$-subset}.
In the \emph{binary model} the arrival is $V=\mathbf 1_A$.
In the \emph{signed model} it is $V=\sum_{i\in A}\e_ie_i$, where $e_i$ denotes the $i$th unit vector and the \emph{marks} $\e_i$ are independent Rademacher signs that are also independent of $A$.
We write $\mu_s^\bin$ and $\mu_s^\pm$ for the laws in the binary and the signed model, respectively, and $\mu_s$ for either of them.
The arrivals $V_1,V_2,\ldots$ are independent with law $\mu_s$, and $A_t$ denotes the support of $V_t$.

An \emph{online algorithm} chooses $x_t\in\{\pm1\}$ as a measurable function of $V_1,\ldots,V_t$ and of an auxiliary random variable $\Theta_0$, independent of all arrivals, which represents private randomness.
The algorithm may depend on $n$, $s$, the horizon $T$, and a target failure probability $\delta$; it is \emph{deterministic} if it does not use $\Theta_0$.
We put
\begin{equation}
    \label{eq:SignedSums}
    S_0\colonequals0,\qquad S_t\colonequals\sum_{u=1}^tx_uV_u,\qquad D_T\colonequals\max_{0\le t\le T}\norm{S_t}_\infty.
\end{equation}
We call $D_T$ the \emph{prefix discrepancy} up to time $T$ and $\norm{S_T}_\infty$ the \emph{terminal discrepancy} at time $T$.
Let $\Fc_t\colonequals\sigma(\Theta_0,V_1,\ldots,V_t)$.
Then $x_t$ is $\Fc_t$-measurable, and $V_{t+1}$ is independent of $\Fc_t$.
We call $S_t$ the \emph{pre-state} of the arrival $V_{t+1}$, and one-step estimates at a pre-state are conditional on $\Fc_t$.
Following \cite[Definition~1.9]{Altschuler25}, the optimal expected prefix discrepancy is
\begin{equation}
    \label{eq:Ondisc}
    \ondisc(\mu_s,T)\colonequals\inf\E D_T,
\end{equation}
where the infimum is taken over all online algorithms.

Our universal lower bounds hold uniformly over online algorithms in this generality; in particular, they apply to randomized algorithms that know the horizon.
We call deterministic online algorithms \emph{rules}.
The upper bounds are proved for explicit rules, and the forcing result of \cref{sec:SpecificRules} concerns only the rule considered there.
When a rule depends on $T$ or $\delta$, its parameters are fixed before the first arrival and never change during the run; we state explicitly when a rule is independent of the horizon.
Stopping times appear only in the analysis and never modify the algorithm, except in the restart rule of \cref{def:RestartRule}, which starts a new epoch, with a new rule and a reset local state, when the local state crosses a threshold.

We write $S_i(t)$ and $V_t(i)$ for the $i$th coordinates of $S_t$ and $V_t$, and $h_i(t)\colonequals\abs{S_i(t)}$ for the \emph{height} of coordinate $i$ at time $t$; when the time is clear, we write $S_i$ and $h_i$.
An arrival \emph{selects}, or \emph{touches}, the coordinates in its support, and the signed arrival moves a selected coordinate \emph{inward} if it decreases its height and \emph{outward} if it increases it.
Since $S_0=0$ and all entries of the arrivals lie in $\{-1,0,1\}$, all heights are integers and change by at most one per arrival.
Hence $D_T\le T$, and, for every integer $H\ge1$, at the first time at which some height is at least $H$, this height equals $H$.

Logarithms are natural unless a base is indicated.
We write $f\lesssim g$ or $f=O(g)$ if $f\le Cg$ for an absolute constant $C$, $f=\Omega(g)$ if $g=O(f)$, and $f\asymp g$ or $f=\Theta(g)$ if $f\lesssim g\lesssim f$; subscripts such as $\lesssim_s$ or $O_s$ indicate that the constant may depend on the displayed parameters.
We write $f=o(g)$, or $g\gg f$, if $f/g\to0$ as $n\to\infty$.
The phrase \emph{with probability $1-o(1)$} means that the failure probability is bounded by a quantity that tends to zero as $n\to\infty$ and does not depend on the online algorithm or on any other parameter in the stated range.
For $I\subseteq[n]$ and $y\in\R^n$, the vector $y_{-I}$ is obtained by deleting the coordinates in $I$, and $y_I$ is the restriction of $y$ to $I$; we write $y_{-i}\colonequals y_{-\{i\}}$.

For a fixed set $I\subseteq[n]$ of size $k$ and distinct $i,j\in[n]$, a uniform $s$-subset $A$ satisfies
\begin{equation}
    \label{eq:Inclusion}
    \P(I\subseteq A)=\frac{(s)_k}{(n)_k}\le\Big(\frac sn\Big)^k,\qquad\P(i\in A)=\frac sn,\qquad\P(\{i,j\}\subseteq A)=\frac{s(s-1)}{n(n-1)},
\end{equation}
where $(a)_k\colonequals a(a-1)\cdots(a-k+1)$.

\begin{remark}[Supports of size one]
    \label{rem:SizeOne}
    For $s=1$, let $i$ be the coordinate selected at time $t$.
    The rule that chooses $x_t$ with $x_tV_t(i)=-\sgn S_i(t-1)$ whenever $S_i(t-1)\ne0$, and $x_t\colonequals1$ otherwise, keeps every height in $\{0,1\}$.
    Hence $D_T=1$ for every $T\ge1$ in both models, which is optimal, and the terminal discrepancy is at most one.
    The same holds in the binary model with $s=n$, where all coordinates are equal, by alternating the sign $x_t$.
    These endpoints are excluded from the constructions below.
\end{remark}

\subsection{Anti-concentration for random supports and signs}
\label{sec:Sampling}

The following anti-concentration estimate is the main place where the randomness of the supports enters the one-step drift estimates of the upper bounds, through \cref{lem:NewtonService} and in \cref{sec:FixedSparsity,sec:Terminal}.
Its first three inequalities hold for arbitrary fixed coefficients and use no randomness of signs; \eqref{eq:SignedAC} is the corresponding estimate for random signs on a fixed set.
Inequality~\eqref{eq:UnclippedAC} is, with the better constant $1/4$, the first step of the proof of the sampling estimate of Altschuler and Tikhomirov \cite[Lemma~3.4]{Altschuler26Binary}.
Our proof uses their construction of the support from a random matching and replaces their combination of Jensen's and Khintchine's inequalities by the Paley--Zygmund inequality, which also yields the clipped forms \eqref{eq:ClippedAC} and \eqref{eq:ExponentialAC} needed in the drift estimates.

\begin{lemma}[Clipped anti-concentration]
    \label{lem:ClippedAntiConcentration}
    Let $2s\le n$, let $A$ be a uniform $s$-subset of $[n]$, and let $z\in\R^n$.
    Then
    \begin{align}
        \label{eq:ClippedAC}
        \E\min\Big\{\Big|\sum_{i\in A}z_i\Big|,1\Big\}&\ge\frac1{144}\,\E\min\{\norm{z_A}_2,1\},\\
        \label{eq:UnclippedAC}
        \E\Big|\sum_{i\in A}z_i\Big|&\ge\frac1{144}\,\E\norm{z_A}_2,\\
        \label{eq:ExponentialAC}
        \E\Big[1-e^{-\abs{\sum_{i\in A}z_i}}\Big]&\ge\frac1{288}\,\E\min\{\norm{z_A}_2,1\}.
    \end{align}
    Moreover, for every fixed set $A'\subseteq[n]$ and independent Rademacher signs $(\e_i)$,
    \begin{equation}
        \label{eq:SignedAC}
        \E_\e\Big[1-e^{-\abs{\sum_{i\in A'}\e_iz_i}}\Big]\ge\frac1{32}\min\{\norm{z_{A'}}_2,1\}.
    \end{equation}
\end{lemma}

\begin{proof}
    We generate $A$ in three steps: we choose a uniform $2s$-subset $B\subseteq[n]$, then a uniform perfect matching $M$ of $B$, and finally, independently, one endpoint of each matched pair.
    The resulting set $A$ is a uniform $s$-subset of $[n]$.
    Condition on $B$ and $M$; up to \eqref{eq:ConditionalAC}, probabilities and expectations are conditional on them.
    Then
    \begin{equation*}
        W\colonequals\sum_{i\in A}z_i=\mu+\sum_{\{i,j\}\in M}\e_{ij}\frac{z_i-z_j}2,\qquad\mu\colonequals\frac12\sum_{i\in B}z_i,
    \end{equation*}
    with independent fair signs $\e_{ij}$.
    Put $Q\colonequals\sum_{i\in B}z_i^2$ and $\sigma^2\colonequals\frac14\sum_{\{i,j\}\in M}(z_i-z_j)^2$.
    Then $\sigma^2\le Q/2$, $\E W^2=\mu^2+\sigma^2$, and a direct expansion gives $\E W^4\le3(\mu^2+\sigma^2)^2$.
    The Paley--Zygmund inequality applied to $W^2$ yields $\P\{W^2\ge(\mu^2+\sigma^2)/2\}\ge1/12$, and therefore
    \begin{equation}
        \label{eq:ConditionalAC}
        \E\big[\min\{\abs W,1\}\bigm|B,M\big]\ge\frac1{12\sqrt2}\min\Big\{\sqrt{\mu^2+\sigma^2},1\Big\}.
    \end{equation}
    If $\mu^2\ge Q/8$, the right side is at least $\min\{\sqrt Q,1\}/48$.
    Otherwise $(\sum_{i\in B}z_i)^2<Q/2$.
    Each pair of $B$ is matched with probability $1/(2s-1)$, so
    \begin{equation*}
        \E\big[\sigma^2\bigm|B\big]=\frac{2sQ-(\sum_{i\in B}z_i)^2}{4(2s-1)}\ge\frac Q4.
    \end{equation*}
    Since $\sigma^2\le Q/2$, the event $\{\sigma^2\ge Q/8\}$ has conditional probability at least $1/3$ given $B$, and on this event \eqref{eq:ConditionalAC} is again at least $\min\{\sqrt Q,1\}/48$.
    In both cases $\E[\min\{\abs W,1\}\mid B]\ge\min\{\sqrt Q,1\}/144$.
    Since $A\subseteq B$, we have $\norm{z_A}_2\le\sqrt Q$, and averaging over $B$ proves \eqref{eq:ClippedAC}.
    Both sides of \eqref{eq:UnclippedAC} are homogeneous of degree one in $z$; after scaling so that $\sum_i\abs{z_i}\le1$, neither side is clipped in \eqref{eq:ClippedAC}, which proves \eqref{eq:UnclippedAC}.
    Finally, $1-e^{-x}\ge\frac12\min\{x,1\}$ for $x\ge0$ gives \eqref{eq:ExponentialAC}.

    For \eqref{eq:SignedAC}, put $W\colonequals\sum_{i\in A'}\e_iz_i$ and $Q\colonequals\norm{z_{A'}}_2^2$.
    Then $\E W^2=Q$ and $\E W^4\le3Q^2$, so the Paley--Zygmund inequality gives $\P\{\abs W\ge\sqrt Q/2\}\ge(3/4)^2/3=3/16$.
    Hence the left side of \eqref{eq:SignedAC} is at least $\frac3{16}\cdot\frac12\min\{\sqrt Q/2,1\}\ge\frac3{64}\min\{\sqrt Q,1\}\ge\frac1{32}\min\{\sqrt Q,1\}$.
\end{proof}

\subsection{The baseline lower bound}
\label{sec:Baseline}

For the binary model, the following lower bound is due to Altschuler and Tikhomirov.

\begin{literaturetheorem}[{\cite[Theorem~1.1 and Proposition~2.5]{Altschuler26Binary}}]
    \label{thm:ATLower}
    There is an absolute constant $c>0$ such that, in the binary model with $2\le s\le n/2$, every online algorithm satisfies, for every prescribed $T\ge n$,
    \begin{equation*}
        \P\bc{\norm{S_T}_\infty\ge c(\sqrt s+\log\log n)}=1-o(1).
    \end{equation*}
\end{literaturetheorem}

The failure probability in \cref{thm:ATLower} satisfies our convention in \cref{sec:Model}.
Indeed, the proof of \cite[Proposition~2.5]{Altschuler26Binary} conditions on the history up to time $T-n$ and bounds the conditional failure probability, for $s\le(\log\log n)^2$, by $O(n^{-0.8})+n^{-1/2+o(1)}$, using the isolation lemma below, and, for larger $s$, by the square root of the probability that $2n$ independent arrivals admit signs whose signed sum has $\ell_\infty$-norm below $c_1\sqrt s$, where $c_1>0$ is an absolute constant.
By the offline estimate of MacRury, Masa\v{r}\'{\i}k, Pai and P\'erez-Gim\'enez \cite[Theorem~1.2]{MacRury23}, the latter probability tends to zero along every sequence $s_n\to\infty$ with $s_n\le n/2$, and hence uniformly in $(\log\log n)^2<s\le n/2$.
These bounds depend neither on the algorithm nor on $T$.
Since $D_T\ge\norm{S_T}_\infty$, the theorem also bounds the prefix discrepancy.
The isolation lemma is the following statement about supports alone.

\begin{literaturetheorem}[Isolation lemma {\cite[Lemma~2.3]{Altschuler26Binary}}]
    \label{thm:Isolation}
    Let $2\le s\le(\log\log n)^2$, and let $m$ and $r$ be integers in the interval $[n^{0.99},2n/\log n]$.
    Let $I,J\subseteq[n]$ be disjoint sets of size $r$, and let $A_1,\ldots,A_m$ be independent uniform $s$-subsets of $[n]$.
    With probability $1-O(n^{-0.9})$, the following hold:
    \begin{enumerate}[label=(\roman*)]
        \item at least $r/2$ elements of $I$, and at least $r/2$ elements of $J$, lie in none of the sets $A_k$;
        \item at least $m(r/n)^2/100$ indices $k$ are \emph{isolated}: $A_k$ contains exactly one element of $I$ and exactly one element of $J$, and neither of these two elements lies in any other $A_{k'}$, $k'\ne k$.
    \end{enumerate}
\end{literaturetheorem}

In \cite[Lemma~2.3]{Altschuler26Binary}, where the sparsity is denoted by $d$ and the set size by $s$, the sets $I,J$ may be measurable with respect to the past of the process, and $A_1,\ldots,A_m$ are the supports of the next $m$ arrivals, which we call a \emph{block}.
Since the conclusion concerns only the supports in the block, which are independent of the past, the version for deterministic sets stated here implies the conditional version with respect to the filtration $(\Fc_t)$ of \cref{sec:Model}, in both models: for every deterministic time $t$ and all disjoint $\Fc_t$-measurable sets $I,J$ of size $r$, the conclusion holds for the supports $A_{t+1},\ldots,A_{t+m}$ with conditional probability $1-O(n^{-0.9})$ given $\Fc_t$.

The following proposition transfers \cref{thm:ATLower} to the signed model.
Part~\ref{item:BaselineLogLog} is the isolation argument of Altschuler and Tikhomirov \cite[Lemma~2.2 and Propositions~2.4 and~2.5]{Altschuler26Binary}, who also run it on the last $n$ arrivals from an arbitrary history; we make explicit that it is insensitive to the marks, and we treat the state at the start of the final blocks by a different case distinction.
Part~\ref{item:BaselineSqrt} is an offline statement: even with hindsight, no choice of signs for the last $n$ arrivals keeps all coordinates below $\sqrt s/32$.

\begin{proposition}[Baseline lower bound]
    \label{prop:Baseline}
    There are absolute constants $c>0$, $C$ and $n_0$ such that the following holds for every $n\ge n_0$, every prescribed time $T\ge n$ and every online algorithm.
    \begin{enumerate}[label=(\roman*)]
        \item\label{item:BaselineLogLog} In both models with $2\le s\le(\log\log n)^2$, $\P\{\norm{S_T}_\infty\ge c\log\log n\}\ge1-Cn^{-0.8}$.
        \item\label{item:BaselineSqrt} In the signed model with $128\le s\le n$, with probability at least $1-10e^{-s/8}-2^{-0.8n}$, every $y\in\{\pm1\}^n$ satisfies
        \begin{equation*}
            \Big\|S_{T-n}+\sum_{u=1}^ny_uV_{T-n+u}\Big\|_\infty\ge\frac{\sqrt s}{32}.
        \end{equation*}
        \item\label{item:BaselineCombined} $\P\{\norm{S_T}_\infty\ge c(\sqrt s+\log\log n)\}=1-o(1)$ in the signed model for $2\le s\le n$ and, using \cref{thm:ATLower}, in the binary model for $2\le s\le n/2$.
    \end{enumerate}
\end{proposition}

\begin{proof}
    \emph{Part~\ref{item:BaselineLogLog}.}
    Put $q_*\colonequals\floor{\log\log n/(2\log4)}$, $m\colonequals\ceil{n/\log n}$, and $r_j\colonequals\ceil{n(\log n)^{-4^j}}$ for $0\le j\le q_*$.
    Since $4^{q_*}\le\sqrt{\log n}$, we have $r_j\ge n\exp(-\sqrt{\log n}\log\log n)\ge n^{0.99}$ for large $n$, and all of $m,r_0,\ldots,r_{q_*}$ lie in $[n^{0.99},2n/\log n]$.
    Put $t_0\colonequals T-q_*m$; since $q_*m\le n\le T$, this is a nonnegative deterministic time.
    The last $q_*m$ arrivals are divided into $q_*$ consecutive blocks of length $m$, the $j$-th block consisting of the arrivals $t_0+(j-1)m+1,\ldots,t_0+jm$.

    Consider first the state $S_{t_0}$, and suppose that every integer is taken as a value $S_i(t_0)$ by fewer than $2r_0$ coordinates.
    Then fewer than $(2q_*-1)\cdot2r_0\le4q_*r_0\le n/2$ coordinates satisfy $\abs{S_i(t_0)}<q_*$.
    The last $q_*m$ arrivals select at most $sq_*m\le(\log\log n)^3\cdot2n/\log n<n/2$ coordinates.
    Hence some coordinate with $\abs{S_i(t_0)}\ge q_*$ is not selected after time $t_0$, and in this case $\norm{S_T}_\infty\ge q_*$.

    Otherwise, choose $\Fc_{t_0}$-measurably an integer $v$ taken by at least $2r_0$ coordinates and two disjoint sets $I_0,J_0$ of size $r_0$ of such coordinates, and put $\lambda_0^+\colonequals\lambda_0^-\colonequals v$.
    We show by induction on $j$ that, outside an event of probability $O(jn^{-0.9})+je^{-n^{0.96}/16}$ conditionally on $\Fc_{t_0}$, at the end of block $j$ there are disjoint $\Fc_{t_0+jm}$-measurable sets $I_j,J_j$ of size $r_j$ and integers $\lambda_j^-\le\lambda_j^+$ with $\lambda_j^+-\lambda_j^-\ge j$, such that $S_i(t_0+jm)=\lambda_j^+$ for $i\in I_j$ and $S_i(t_0+jm)=\lambda_j^-$ for $i\in J_j$.
    For $j=0$ this holds by construction.

    Suppose the claim holds for $j-1$, and apply \cref{thm:Isolation}, conditionally on $\Fc_{t_0+(j-1)m}$, to $I_{j-1},J_{j-1}$ and the supports of block $j$.
    Outside an event of conditional probability $O(n^{-0.9})$, at least $r_{j-1}/2$ coordinates of each set are not selected in the block, and there are $N\ge m(r_{j-1}/n)^2/100$ isolated arrivals.
    Consider an isolated arrival at time $t$, with coordinates $i\in I_{j-1}$ and $\ell\in J_{j-1}$.
    We call it \emph{agreeing} if $V_t$ has equal entries at $i$ and $\ell$; in the binary model every arrival is agreeing.
    On an agreeing isolated arrival the update adds the same sign $y=x_tV_t(i)\in\{\pm1\}$ to $S_i$ and $S_\ell$, and since $i$ and $\ell$ are selected by no other arrival of the block, at the end of the block either $S_i=\lambda_{j-1}^++1$ or $S_\ell=\lambda_{j-1}^--1$, whatever the algorithm does.
    In the signed model, conditionally on $\Fc_{t_0+(j-1)m}$ and on all supports of the block, the isolated arrivals are agreeing independently with probability $1/2$; by Chernoff's bound, at least $N/4$ of them are agreeing, except with probability $e^{-N/16}$.
    Hence, in both models, at least $N/8$ coordinates of one of the two sets have moved away from the other level by one, that is, up for $I_{j-1}$ or down for $J_{j-1}$.
    Suppose this happens for $I_{j-1}$.
    Let $I_j$ consist of $r_j$ such coordinates and $J_j$ of $r_j$ coordinates of $J_{j-1}$ that are not selected in the block, both chosen $\Fc_{t_0+jm}$-measurably, and put $\lambda_j^+\colonequals\lambda_{j-1}^++1$, $\lambda_j^-\colonequals\lambda_{j-1}^-$; the other case is symmetric.
    These choices are possible because $r_{j-1}/2\ge r_j$ and
    \begin{equation*}
        \frac N8\ge\frac n{800\log n}\,(\log n)^{-2\cdot4^{j-1}}\ge\frac{2n}{(\log n)^{4^j}}\ge r_j,
    \end{equation*}
    where the middle inequality is $(\log n)^{2\cdot4^{j-1}-1}\ge1600$, valid for large $n$.
    Moreover $N\ge n^{1-o(1)}\ge n^{0.96}$, so the Chernoff error is at most $e^{-n^{0.96}/16}$.

    After $q_*$ blocks, at time $T$, the two levels satisfy $\lambda_{q_*}^+-\lambda_{q_*}^-\ge q_*$, so one of them has absolute value at least $q_*/2$.
    Hence $\norm{S_T}_\infty\ge q_*/2\ge c\log\log n$ outside an event of probability $O(q_*n^{-0.9})+q_*e^{-n^{0.96}/16}\le Cn^{-0.8}$.

    \emph{Part~\ref{item:BaselineSqrt}.}
    We first secure a good configuration of the last $n$ supports $A_{T-n+1},\ldots,A_T$, which are independent of $\Fc_{T-n}$.
    The number $d_i\colonequals\abs{\{u\le n:i\in A_{T-n+u}\}}$ of these supports that contain coordinate $i$ has distribution $\Bin(n,s/n)$, so $\P\{d_i<s/2\}\le e^{-s/8}$, and by Markov's inequality the set $G\colonequals\{i:d_i\ge s/2\}$ has $\abs G\ge0.9n$ except with probability $10e^{-s/8}$.
    Now condition on $\Fc_{T-n}$ and on supports with $\abs G\ge0.9n$; the marks of the last $n$ arrivals remain independent Rademacher signs.
    Fix $y\in\{\pm1\}^n$.
    Conditionally, the coordinates $R_i\colonequals S_i(T-n)+\sum_{u\le n}y_uV_{T-n+u}(i)$ are independent, and $R_i-S_i(T-n)$ is a sum of $d_i$ independent Rademacher variables.
    Every point probability of such a sum is at most $\binom{d_i}{\floor{d_i/2}}2^{-d_i}\le d_i^{-1/2}$, so, for $i\in G$,
    \begin{equation*}
        \P\Big\{\abs{R_i}\le\frac{\sqrt s}{32}\Big\}\le\frac{\sqrt s/16+1}{\sqrt{d_i}}\le\frac{\sqrt2}{16}+\sqrt{\frac2s}\le\frac14,
    \end{equation*}
    since $s\ge128$.
    By independence, all coordinates in $G$ satisfy $\abs{R_i}\le\sqrt s/32$ with probability at most $4^{-0.9n}$, and a union bound over the $2^n$ vectors $y$ gives the bound $2^{-0.8n}$.

    \emph{Part~\ref{item:BaselineCombined}.}
    Decreasing $c$ keeps part~\ref{item:BaselineLogLog} valid; we may therefore assume that it holds with $2c$ in place of $c$, that $2c\le1/32$, and that $c$ is at most the constant of \cref{thm:ATLower}.
    We also take $n_0$ so large that $(\log\log n)^2\ge128$ for $n\ge n_0$.
    If $s\le(\log\log n)^2$, then $\log\log n\ge\frac12(\sqrt s+\log\log n)$, and part~\ref{item:BaselineLogLog} applies.
    If $s>(\log\log n)^2$, then $s\ge128$ and $\sqrt s\ge\frac12(\sqrt s+\log\log n)$; in the signed model, part~\ref{item:BaselineSqrt} applies to the signs that the algorithm chooses for the last $n$ arrivals, with failure probability at most $10e^{-(\log\log n)^2/8}+2^{-0.8n}$, and in the binary model we use \cref{thm:ATLower}.
\end{proof}

Thus, as already observed in \cite[Remark~1.2]{Altschuler26Binary}, the doubly logarithmic lower bound is not a horizon effect: it is present at every time $T\ge n$, in particular already at $T=n$.

\subsection{Moment tools for the upper bounds}
\label{sec:Tools}

The upper bounds of \cref{sec:Pairs,sec:FixedSparsity,sec:Terminal} share four tools: a triangular recursion for moments of elementary symmetric polynomials, a comparison between the weights of such a moment and the weights used by a rule, a clearance estimate for the population above a level together with its iteration over levels, and an excursion bound for energy functionals.
The triangular recursion and the clearance estimate are used in all three upper bounds, the comparison of weights in \cref{sec:FixedSparsity,sec:Terminal}, and the excursion bound in \cref{sec:Pairs,sec:FixedSparsity}.
Throughout, we write $p\colonequals s/n$, and we call $N_b(S)\colonequals\abs{\{i:h_i\ge b\}}$ the \emph{population} at level $b$ of a state $S$ with heights $h_i=\abs{S_i}$.
All four tools apply to both models.
Besides their explicit hypotheses, such as a drift inequality for the tracked moments and, in the clearance estimate, the isolating property of the rule defined below, they use only the uniform law of the supports and the fact that the heights are integers and change by at most one per arrival.
We use exact sampling without replacement throughout.

\paragraph{Elementary symmetric polynomials.}
For $y\in[0,\infty)^n$, let $e_r(y)$ denote the elementary symmetric polynomial of degree $r$ in the coordinates of $y$, with $e_0\colonequals1$ and $e_r\colonequals0$ for $r<0$ or $r>n$; the same notation is used for vectors with fewer coordinates, such as $y_{-i}$.
We use the identities
\begin{equation}
    \label{eq:ElementaryIdentities}
    \sum_{i=1}^ny_ie_{r-1}(y_{-i})=re_r(y),\qquad\sum_{i=1}^ne_{r-1}(y_{-i})=(n-r+1)e_{r-1}(y).
\end{equation}
The first identity counts each monomial of degree $r$ once for each of its $r$ variables; the second counts each monomial of degree $r-1$ once for each of the $n-r+1$ variables it does not contain.
The generating polynomial $\prod_i(1+zy_i)=\sum_re_r(y)z^r$ is a polynomial in $z$ with nonnegative coefficients.
An inequality between such polynomials is meant \emph{coefficientwise}; it is preserved under multiplication by polynomials with nonnegative coefficients.

\paragraph{Triangular recursions.}
The first tool turns a one-step drift inequality in which each degree is fed only by the degree below into a bound on all moments, uniformly in time, and hence into a bound on the number of high coordinates.

\begin{lemma}[Triangular recursion]
    \label{lem:TriangularRecursion}
    Let $r_*\ge1$, let $\mathsf u,\mathsf v>0$ with $\mathsf ur_*\le1$, and put $\mathsf M\colonequals\max\{1,\mathsf v/\mathsf u\}$.
    Let $X_0\equiv1$, let $X_1(t),\ldots,X_{r_*}(t)\ge0$ be adapted processes with $X_r(0)\le\binom nr$, and let $\Ac_t\in\Fc_t$ be events with $\Ac_{t+1}\subseteq\Ac_t$; we interpret $\Ac_{-1}$ as the sure event.
    Suppose that, for all $t\ge0$ and $1\le r\le r_*$,
    \begin{equation}
        \label{eq:TriangularDrift}
        \E\big[X_r(t+1)\bigm|\Fc_t\big]\le(1-\mathsf ur)X_r(t)+\mathsf v(n-r+1)X_{r-1}(t)\qquad\text{on }\Ac_t.
    \end{equation}
    Then $\E[X_r(t)\mathbf1_{\Ac_{t-1}}]\le\mathsf M^r\binom nr$ for all $t\ge0$ and $0\le r\le r_*$.
    If, moreover, $X_r=e_r(\varphi(h_1),\ldots,\varphi(h_n))$ for a nondecreasing function $\varphi\ge1$ of the height, then, for all $b\ge0$, $t\ge0$ and $1\le r\le r_*$,
    \begin{equation}
        \label{eq:TriangularPopulation}
        \P\big(N_b(S_t)\ge r,\ \Ac_{t-1}\big)\le\Big(\frac{e\mathsf Mn}{r\varphi(b)}\Big)^r.
    \end{equation}
\end{lemma}

\begin{proof}
    Put $K_r\colonequals\mathsf M^r\binom nr$.
    Since $(n-r+1)\binom n{r-1}=r\binom nr$ and $\mathsf v\le\mathsf u\mathsf M$, we have $(1-\mathsf ur)K_r+\mathsf v(n-r+1)K_{r-1}\le K_r$.
    We show $\E[X_r(t)\mathbf1_{\Ac_t}]\le\E[X_r(t)\mathbf1_{\Ac_{t-1}}]\le K_r$ by induction on $t$, simultaneously for all $r$; the first inequality holds because $\Ac_t\subseteq\Ac_{t-1}$.
    For $t=0$ it holds because $X_r(0)\le\binom nr\le K_r$.
    Since $\Ac_t\in\Fc_t$ and $1-\mathsf ur\ge0$, \eqref{eq:TriangularDrift} and the induction hypothesis give
    \begin{equation*}
        \E\big[X_r(t+1)\mathbf1_{\Ac_t}\big]\le(1-\mathsf ur)\E\big[X_r(t)\mathbf1_{\Ac_t}\big]+\mathsf v(n-r+1)\E\big[X_{r-1}(t)\mathbf1_{\Ac_t}\big]\le K_r.
    \end{equation*}
    If $N_b(S_t)\ge r$, then $X_r(t)$ contains a monomial with $r$ factors at least $\varphi(b)$, and all its monomials are nonnegative; Markov's inequality and $\binom nr\le(en/r)^r$ prove \eqref{eq:TriangularPopulation}.
\end{proof}

\paragraph{Weights of a moment.}
For $Y\in[1,\infty)^n$ and $1\le r<n$, put
\begin{equation}
    \label{eq:NewtonWeights}
    \chi_i\colonequals\frac{Y_ie_{r-1}(Y_{-i})}{e_r(Y)},\qquad\zeta_i\colonequals\frac{Y_ie_{r-2}(Y_{-i})}{e_{r-1}(Y)},\qquad v_i\colonequals\frac{Y_ie_{r-1}(Y)}{e_r(Y)},
\end{equation}
with $\zeta_i\colonequals0$ for $r=1$.
Thus $\chi_i$ is the probability that $i$ belongs to a random $r$-set $I$ chosen with probability proportional to $\prod_{j\in I}Y_j$, while the weights $v_i$ are proportional to $Y_i$; since $e_r$ is affine in each variable, replacing $Y_i$ by $\theta Y_i$ multiplies $e_r(Y)$ by $1+(\theta-1)\chi_i$.
A rule that follows the gradient of $\sum_iY_i$ uses weights proportional to $v_i$, whereas the moment $e_r(Y)$ responds to the weights $\chi_i$.
The following lemma shows that choosing the sign according to the weights of the rule instead of those of the moment costs only a quadratic error, and it bounds the resulting service to the moment from below for a uniformly random support.

\begin{lemma}[Weights of a moment]
    \label{lem:NewtonService}
    \leavevmode
    \begin{enumerate}[label=(\roman*)]
        \item\label{item:NewtonBasic} $\sum_i\chi_i=r$, $0\le\zeta_i\le\chi_i<1$, $v_i=\chi_i/(1-\zeta_i)$ and $\chi_i/Y_i\le r/n$.
        If $Y_i\ge Y_j$, then $\chi_i\ge\chi_j$, $v_i/\chi_i\ge v_j/\chi_j$ and $e_{r-1}(Y_{-i})\le e_{r-1}(Y_{-j})$, and $\chi_i>\chi_j$ if $Y_i>Y_j$.
        \item\label{item:NewtonPointwise} Let $\mathsf k:[1,\infty)\to[0,1]$ be nondecreasing, put $\pi_i\colonequals\mathsf k(Y_i)\chi_i$ and $\hat w_i\colonequals\mathsf k(Y_i)v_i$, and let $0\le\e_0\le\frac14$ be such that $\chi_i\le\pi_i+\e_0$ whenever $\pi_i>0$.
        For every finite set $A$ and signs $(\sigma_i)_{i\in A}$, every sign $f\in\{\pm1\}$ of the sum $\sum_A\sigma_i\hat w_i$, that is, every $f$ with $f\sum_A\sigma_i\hat w_i\ge0$, satisfies
        \begin{equation}
            \label{eq:PointwiseService}
            f\sum_A\sigma_i\pi_i\ge\Big|\sum_A\sigma_i\pi_i\Big|-16\sum_{\{i,j\}\subseteq A}\pi_i\pi_j-4\e_0\sum_A\pi_i.
        \end{equation}
        \item\label{item:NewtonAveraged} In the setting of \ref{item:NewtonPointwise}, let $A$ be a uniform $s$-subset of a set $\mathsf C\subseteq[n]$ with $\abs{\mathsf C}\ge2s$, let the signs $\sigma_i$ be fixed, and let $f=f(A)$ be chosen as in \ref{item:NewtonPointwise} for every $A$.
        Then
        \begin{equation}
            \label{eq:AveragedService}
            \E\Big[f\sum_A\sigma_i\pi_i\Big]\ge\Big[\frac{\sqrt s}{144\abs{\mathsf C}}-\frac{8s(s-1)r}{\abs{\mathsf C}(\abs{\mathsf C}-1)}-\frac{4\e_0s}{\abs{\mathsf C}}\Big]\sum_{i\in\mathsf C}\pi_i.
        \end{equation}
    \end{enumerate}
\end{lemma}

In the applications, $x=-f$ is the sign chosen by the rule, and a selected coordinate $i$ with $\pi_i>0$ moves inward if and only if $\sigma_i=f$; thus $f\sum_A\sigma_i\pi_i$ is the inward movement of the selected coordinates, weighted by $\pi_i$.

\begin{proof}
    \ref{item:NewtonBasic}
    The identity $\sum_i\chi_i=r$ is \eqref{eq:ElementaryIdentities}.
    With $U\colonequals e_{r-2}(Y_{-i})$, $V\colonequals e_{r-1}(Y_{-i})$ and $W\colonequals e_r(Y_{-i})$, we have $\chi_i=Y_iV/(W+Y_iV)$ and $\zeta_i=Y_iU/(V+Y_iU)$, and the inequality $UW\le V^2$ for consecutive elementary symmetric functions, a weak form of Newton's inequality \cite[Section~2.22, Theorem~53]{Hardy52}, gives $\zeta_i\le\chi_i$; moreover $W>0$ since $r<n$.
    Since $e_{r-1}(Y)=V+Y_iU$, we have $v_i/\chi_i=e_{r-1}(Y)/V=1/(1-\zeta_i)$.
    Counting extensions of $(r-1)$-sets and using $Y_j\ge1$ gives $W\ge(n-r)V/r$, hence $e_r(Y)\ge W+V\ge(n/r)V$ and $\chi_i/Y_i=V/e_r(Y)\le r/n$.
    Finally, $\chi_i-\chi_j=(Y_i-Y_j)e_{r-1}(Y_{-\{i,j\}})/e_r(Y)$ with $e_{r-1}(Y_{-\{i,j\}})\ge1$, and $e_{r-1}(Y_{-j})-e_{r-1}(Y_{-i})=(Y_i-Y_j)e_{r-2}(Y_{-\{i,j\}})$; since $v_i/\chi_i=e_{r-1}(Y)/e_{r-1}(Y_{-i})$, the latter identity also gives the monotonicity of $v_i/\chi_i$.

    \ref{item:NewtonPointwise}
    By \ref{item:NewtonBasic}, a coordinate with larger $Y_i$ has a weakly larger $\pi_i$ and, where $\pi_i>0$, a weakly larger ratio $\hat w_i/\pi_i=v_i/\chi_i$.
    Moreover, if $\pi_\ell>0$ is maximal among the $\pi_i$, $i\in A$, then $Y_\ell\ge Y_i$ for every $i\in A$ with $\pi_i>0$, because $Y_i>Y_\ell$ would give $\pi_i>\pi_\ell$.
    We call $\lvert\sum_A\sigma_i\pi_i\rvert-f\sum_A\sigma_i\pi_i$ the \emph{loss}; it is at most $2\sum_A\pi_i$, and it vanishes unless $f\sum_A\sigma_i\pi_i<0$.
    Suppose first that $\pi_\ell\ge\sum_{A\setminus\{\ell\}}\pi_i$ for some $\ell\in A$; we may assume $\pi_\ell>0$, since otherwise all $\pi_i$ and $\hat w_i$ vanish.
    Then $\hat w_i\le\pi_i\hat w_\ell/\pi_\ell$ for $i\in A$, hence $\hat w_\ell\ge\sum_{A\setminus\{\ell\}}\hat w_i$.
    If $\sum_A\sigma_i\hat w_i\ne0$, both sums have the sign $\sigma_\ell$, or the $\pi$-sum vanishes; if $\sum_A\sigma_i\hat w_i=0$, equality holds throughout, which forces $\sum_A\sigma_i\pi_i=0$.
    In both cases the loss vanishes.
    Otherwise put $\pi_{\max}\colonequals\max_A\pi_i$ and $\pi_{\mathrm{rest}}\colonequals\sum_A\pi_i-\pi_{\max}>\pi_{\max}$; then $\sum_A\pi_i^2\le\pi_{\max}(\pi_{\max}+\pi_{\mathrm{rest}})\le2\pi_{\max}\pi_{\mathrm{rest}}\le2\sum_{\{i,j\}\subseteq A}\pi_i\pi_j$.
    If $f\sum_A\sigma_i\pi_i<0$, then the two sums have opposite signs or the $\hat w$-sum vanishes, so $\lvert\sum_A\sigma_i\pi_i\rvert\le\sum_A\abs{\hat w_i-\pi_i}$, and the loss is at most $2\sum_A\abs{\hat w_i-\pi_i}$.
    If $\chi_i\le\frac12$ for every $i\in A$ with $\pi_i>0$, then $\hat w_i-\pi_i=\pi_i\zeta_i/(1-\zeta_i)\le2\pi_i\chi_i\le2\pi_i^2+2\e_0\pi_i$, and the loss is at most $8\sum_{\{i,j\}\subseteq A}\pi_i\pi_j+4\e_0\sum_A\pi_i$.
    Otherwise some $\pi_i\ge\frac12-\e_0\ge\frac14$, so $\pi_{\max}\ge\frac14$, and the bound $2\sum_A\pi_i$ on the loss gives at most $4\pi_{\mathrm{rest}}\le16\pi_{\max}\pi_{\mathrm{rest}}$.
    This proves \eqref{eq:PointwiseService}.

    \ref{item:NewtonAveraged}
    By \eqref{eq:UnclippedAC} on $\mathsf C$ and $\norm{\pi_A}_2\ge\norm{\pi_A}_1/\sqrt s$, we have $\E\lvert\sum_A\sigma_i\pi_i\rvert\ge\frac{\sqrt s}{144\abs{\mathsf C}}\sum_{\mathsf C}\pi_i$.
    By \eqref{eq:Inclusion} with $[n]$ replaced by $\mathsf C$, every pair of $\mathsf C$ is selected with probability $s(s-1)/(\abs{\mathsf C}(\abs{\mathsf C}-1))$, and $\sum_{\{i,j\}\subseteq\mathsf C}\pi_i\pi_j\le\frac r2\sum_{\mathsf C}\pi_i$ because $\sum_i\pi_i\le\sum_i\chi_i=r$.
    Averaging \eqref{eq:PointwiseService} proves \eqref{eq:AveragedService}.
\end{proof}

\paragraph{Clearance of layers.}
The third tool bounds the population above a level for rules that move a lone high coordinate inward.
Consider a rule that chooses the sign as a function of the current state and the arrival; the rules of \cref{sec:FixedSparsity,sec:Terminal} are of this form, and \cref{sec:Pairs} explains why this tool also applies to quadratic Greedy with an arbitrary rule at ties.
For integers $b\ge1$ and $g\ge0$, we call a state $S$ \emph{$(b,g)$-isolating} for such a rule if, for every possible arrival in which exactly one selected coordinate has height at least $b$ and this coordinate has height at least $b+g$, the rule moves this coordinate inward.
For $g=0$, this means that a selected coordinate that is the only one of height at least $b$ moves inward.
The name refers to a lone high coordinate in the sense of \cref{sec:OverviewPopulation}; it is a property of the rule and unrelated to the isolated arrivals of \cref{thm:Isolation}, which concern the supports.
Put
\begin{equation*}
    C_{\mathrm{cl}}\colonequals2^{20},\qquad A_{\mathrm{cl}}\colonequals256\,C_{\mathrm{cl}},\qquad c_*\colonequals\log(32/e).
\end{equation*}

\begin{lemma}[Local clearance]
    \label{lem:LocalClearance}
    Fix such a rule and integers $b\ge1$, $g\ge0$ and $1\le q\le n/(4C_{\mathrm{cl}}s)$, and put
    \begin{equation*}
        \vartheta\colonequals(C_{\mathrm{cl}}pq)^{-1/2}\ge2,\qquad W_i(S)\colonequals\vartheta^{(\abs{S_i}-b-g)_+}-1,\qquad\gamma\colonequals p/4.
    \end{equation*}
    For a state $S$, let $S'$ denote the state after one further arrival, signed by the rule.
    \begin{enumerate}[label=(\roman*)]
        \item\label{item:LocalDrift} At every $(b,g)$-isolating state $S\in\Z^n$ with $N_b(S)\le q$, and for $1\le r\le q$,
        \begin{equation}
            \label{eq:LocalDrift}
            \E\big[e_r(W(S'))\bigm|S\big]-e_r(W(S))\le-\gamma re_r(W(S))+\sum_{j=1}^r\mathsf a_je_{r-j}(W(S)),
        \end{equation}
        where $\mathsf a_1\colonequals2\vartheta^2(pq)^2$ and $\mathsf a_j\colonequals2(\vartheta pq)^j/j!$ for $j\ge2$.
        \item\label{item:LocalMoment} Let $(S_t)$ be generated by the rule, let $t_0$ be a deterministic time, let $G\in\Fc_{t_0}$ be an event on which $\sum_iW_i(S_{t_0})\le M$, and let $\kappa\ge t_0$ be a stopping time such that, for $t_0\le t<\kappa$, the state $S_t$ is $(b,g)$-isolating and satisfies $N_b(S_t)\le q$.
        Put $\Ac_t\colonequals G\cap\{\kappa>t\}$.
        Then, for all $t\ge0$ and $0\le r\le q$,
        \begin{equation}
            \label{eq:LocalMoment}
            \E\big[e_r(W(S_{t_0+t}))\mathbf1_{\Ac_{t_0+t}}\big]\le\frac{\mu_t^r}{r!},\qquad\mu_t\colonequals2q+M(1-\gamma/2)^t.
        \end{equation}
        Here $t$ counts the steps after $t_0$.
        \item\label{item:LocalPopulation} In the setting of \ref{item:LocalMoment}, if $Me^{-\gamma t/2}\le q$, then for all integers $u\ge1$ and $1\le r\le q$,
        \begin{equation}
            \label{eq:LocalPopulation}
            \P\big(N_{b+g+u}(S_{t_0+t})\ge r,\ \Ac_{t_0+t}\big)\le\Big(\frac{8eq}{r\vartheta^u}\Big)^r.
        \end{equation}
    \end{enumerate}
    If the process starts from the zero state at time $t_0=0$, one may take $M=0$; then the \emph{burn-in}, the number of steps $t$ needed for $Me^{-\gamma t/2}\le q$, vanishes.
\end{lemma}

\begin{proof}
    Fix a $(b,g)$-isolating state $S$ with $N_b(S)\le q$, let $\mathsf H\colonequals\{i:h_i\ge b\}$, put $\lambda_q\colonequals pq\le1/(4C_{\mathrm{cl}})$, and let $A$ be the support of the next arrival.
    We first prove \eqref{eq:LocalDrift}.
    Fix a monomial $\prod_{i\in I}W_i$ of $e_r(W(S))$ that is positive at $S$, an \emph{old positive monomial}, with $\abs I=r$; every $i\in I$ has height at least $b+g+1$, so $I\subseteq\mathsf H$.
    Write $\mathsf J\colonequals\abs{A\cap I}$ and $\mathsf K\colonequals\abs{A\cap\mathsf H}$.
    If $\mathsf K=0$, the factors do not change.
    The event $\{\mathsf K=1,\mathsf J=1\}$ has probability at least $rp(1-2\lambda_q)$, by conditioning on the selected coordinate of $I$ and a union bound for the other coordinates of $\mathsf H$.
    On this event the selected coordinate of $I$ is the only selected coordinate of height at least $b$, and its height is at least $b+g$; since $S$ is $(b,g)$-isolating, the rule moves it inward, and the monomial is multiplied by at most $1/\vartheta$.
    If $\mathsf K\ge2$, every selected coordinate of $\mathsf H$ satisfies $W_i'\le\vartheta W_i+(\vartheta-1)$, and we first account for the multiplicative part, which multiplies the monomial by at most $\vartheta^{\mathsf J}$.

    Conditionally on the event that $d$ specified coordinates are selected, the other $s-d$ selected coordinates form a uniform $(s-d)$-subset of the remaining $n-d$ coordinates, so by \eqref{eq:Inclusion} every $k$-set of remaining coordinates is included with probability $(s-d)_k/(n-d)_k\le p^k$; hence, for any set $I'$ of remaining coordinates,
    \begin{equation}
        \label{eq:LocalInclusion}
        \E\big[\vartheta^{\abs{A\cap I'}}\bigm|\text{specified selected coordinates}\big]\le(1+p(\vartheta-1))^{\abs{I'}}\le e^{\vartheta p\abs{I'}}.
    \end{equation}
    Using $\vartheta^{\mathsf J}-1\le(\vartheta-1)\mathsf J\vartheta^{\mathsf J-1}$, sum first over a selected $i\in I$ and then over another selected $j\in\mathsf H\setminus\{i\}$; their joint inclusion probability is at most $p^2$, and if $j\in I$ its old factor contributes one extra $\vartheta$.
    By \eqref{eq:LocalInclusion},
    \begin{equation*}
        \E\big[(\vartheta^{\mathsf J}-1)\mathbf1\{\mathsf K\ge2\}\big]\le\vartheta(\vartheta-1)rqp^2e^{\vartheta pr}\le2\vartheta^2rp\lambda_q=\frac{2rp}{C_{\mathrm{cl}}},
    \end{equation*}
    since $\vartheta pq=\sqrt{\lambda_q/C_{\mathrm{cl}}}\le1/(2C_{\mathrm{cl}})$.
    The net coefficient of the old monomial is therefore at most $-(1-\vartheta^{-1})rp(1-2\lambda_q)+2rp/C_{\mathrm{cl}}\le-rp/4$.

    It remains to account for the additive parts.
    A coordinate with $W_i=0$ can acquire $W_i'>0$ only if its height is $b+g$ and it moves outward; such a coordinate belongs to $\mathsf H$, and if it is the only selected coordinate of $\mathsf H$, the rule moves it inward.
    Hence there are no additive parts on $\{\mathsf K\le1\}$.
    On $\{\mathsf K\ge2\}$, expand $e_r$ and choose $j$ additive factors $(\vartheta-1)$ on coordinates disjoint from an old-factor set $I$.
    For $j\ge2$, the new coordinates have joint inclusion probability at most $p^j$, the conditional growth of the old factors is at most $e^{\vartheta p\abs I}\le2$, and there are at most $q^j/j!$ choices; this gives $\mathsf a_je_{r-j}$.
    For $j=1$, a second selected coordinate of $\mathsf H$ is necessary.
    Summing over the new coordinate and this additional coordinate gives a factor $q^2p^2$; if the latter belongs to the old-factor set, its multiplier contributes at most $\vartheta$; and the remaining old factors grow by at most a factor $e^{\vartheta p\abs I}\le2$, as before.
    The coefficient is consequently at most $2\vartheta(\vartheta-1)p^2q^2\le\mathsf a_1$.
    This expansion is valid even when an old positive coordinate supplies an additive factor, and terms with $j>s$ may be retained as positive upper bounds.
    This proves \eqref{eq:LocalDrift}, without any independence assumption on the coordinates that newly reach height $b+g+1$.

    To compare the additive terms with the contraction, let $\mu\ge2q$; division by $\gamma r\mu^r/r!$ gives
    \begin{equation*}
        \frac{\sum_{j=1}^r\mathsf a_j\mu^{r-j}/(r-j)!}{\gamma r\mu^r/r!}\le\frac4{C_{\mathrm{cl}}}+\frac8{pr}\sum_{j\ge2}\frac{(\vartheta pr/2)^j}{j!}\le\frac4{C_{\mathrm{cl}}}+\vartheta^2pr\,e^{\vartheta pr/2}\le\frac6{C_{\mathrm{cl}}}<\frac14.
    \end{equation*}
    The induction for \eqref{eq:LocalMoment} starts from $e_r(W)\le(\sum_iW_i)^r/r!\le M^r/r!$ on $G$.
    All coefficients $1-\gamma r$ in \eqref{eq:LocalDrift} are nonnegative.
    Since $\Ac_{t_0+t+1}\subseteq\Ac_{t_0+t}\in\Fc_{t_0+t}$ and the state at time $t_0+t$ satisfies the hypotheses of \ref{item:LocalDrift} on $\Ac_{t_0+t}$, \eqref{eq:LocalDrift} and the induction hypothesis give $\E[e_r(W(S_{t_0+t+1}))\mathbf1_{\Ac_{t_0+t}}]\le(1-3\gamma r/4)\mu_t^r/r!$, whereas $\mu_{t+1}^r\ge(1-\gamma r/2)\mu_t^r$.
    This proves \eqref{eq:LocalMoment}, simultaneously in all degrees; degree zero is bounded by one.
    After the burn-in, $\mu_t\le3q$.
    If at least $r$ coordinates have height at least $b+g+u$, then $e_r(W)\ge(\vartheta^u-1)^r\ge(\vartheta^u/2)^r$, since $W_i$ is nondecreasing in the height, and Markov's inequality with $r!\ge(r/e)^r$ proves \eqref{eq:LocalPopulation}.
\end{proof}

The estimate is never conditioned on future survival: it bounds the moment of the state at a deterministic time, multiplied by an adapted survival indicator.
Iterating it over $O(\log\log n)$ levels gives the layer structure behind all our upper bounds, in the spirit of the analysis of two-choice allocations \cite{Azar99,Berenbrink06} and of the doubly exponential level estimates of Altschuler and Tikhomirov \cite[Section~3.5]{Altschuler26Binary}.
The layer iteration runs \cref{lem:LocalClearance} at levels $b_0<b_1<\cdots<b_J$, each started at a deterministic time after a burn-in, and shows that the population above successive levels drops doubly exponentially, from $q_0$ to at most $2R$ within at most $\log_2\log_2n+3$ levels.
For a process started from the zero state at time zero, as in the prefix bounds of \cref{sec:Pairs,sec:FixedSparsity,sec:GrowingPrefix}, one may take $t_0=\cdots=t_{J+1}=0$ and $M=0$ in the lemma, since all local variables vanish at time zero; then \eqref{eq:LayerFailure} controls all levels over the entire horizon $[0,T]$.
The general starting times are needed for the terminal analysis in \cref{sec:Terminal}.

\begin{lemma}[Layer iteration]
    \label{lem:LayerIteration}
    Let $g\ge0$, $R\ge1$ and $q_0$ be integers with $2R<q_0\le n/(8A_{\mathrm{cl}}s)$.
    Define the \emph{cutoffs} $q_{j+1}\colonequals\ceil{\max\{A_{\mathrm{cl}}sq_j^2/n,R\}}$ for $j\ge0$, until the first index $J$ with $q_J\le2R$, and put $\vartheta_j\colonequals(C_{\mathrm{cl}}pq_j)^{-1/2}$.
    Then $q_{j+1}\le q_j/2$ for $j<J$, $J\le\log_2\log_2n+3$, and $2\le\vartheta_j\le\sqrt n$.

    Let $(S_t)$ be generated by a rule as above, fix an integer $b_0\ge1$, put $b_j\colonequals b_0+j(g+2)$, and let $\Gc_t\in\Fc_t$ be events on which $N_{b_0}(S_t)\le q_0$ and $S_t$ is $(b_j,g)$-isolating for every $j\le J$.
    Let $T\ge0$, $M\ge0$ and deterministic times $t_0\le t_1\le\cdots\le t_{J+1}\le T$ be such that $t_{j+1}-t_j\ge(8n/s)\log(1+M)$ for $j\le J$, and such that $\sum_i\vartheta_j^{(h_i(t_j)-b_j-g)_+}\le n+M$ on $\Gc_{t_j}$ for every $j\le J$.
    Let $\Vc$ be the event that $\Gc_t$ holds for all $t\in[t_0,T]$.
    Then
    \begin{equation}
        \label{eq:LayerFailure}
        \P\Big(\Vc\cap\big\{N_{b_{j+1}}(S_t)>q_{j+1}\text{ for some }j<J\text{ and }t\in[t_{j+1},T]\big\}\Big)\le J(T-t_0+1)e^{-c_*R}.
    \end{equation}
    Moreover, let $\Lc_t$ be the event that $N_{b_j}(S_u)\le q_j$ for all $1\le j\le J$ and $u\in[t_j,t]$.
    Then, for all integers $d\ge1$ and $1\le r\le q_J$, and every deterministic $t\in[t_{J+1},T]$,
    \begin{equation}
        \label{eq:LayerFinal}
        \P\big(\Vc\cap\Lc_t\cap\{N_{b_J+g+d}(S_t)\ge r\}\big)\le\Big(\frac{8eq_J}{r\vartheta_J^d}\Big)^r.
    \end{equation}
\end{lemma}

\begin{proof}
    For $j<J$ we have $q_j>2R$, hence $R<q_j/2$ and $A_{\mathrm{cl}}sq_j/n\le A_{\mathrm{cl}}sq_0/n\le1/8$, so $q_{j+1}\le\max\{q_j/8+1,q_j/2\}=q_j/2$, as $q_j\ge3$.
    If $q_{j+1}>2R$, the maximum defining $q_{j+1}$ is its first term, which then exceeds one, so $q_{j+1}\le2A_{\mathrm{cl}}sq_j^2/n$; thus $z_j\colonequals2A_{\mathrm{cl}}sq_j/n$ satisfies $z_{j+1}\le z_j^2$ as long as $j+1<J$, and $z_0\le1/4$.
    For $j<J$, this gives $4/n\le z_j\le4^{-2^j}$, whence $J\le\log_2\log_2n+3$.
    Moreover $C_{\mathrm{cl}}pq_j\le C_{\mathrm{cl}}sq_0/n\le1/2048$ and $q_j\ge1$, so $2\le\vartheta_j\le\sqrt n$, and every cutoff satisfies the hypothesis of \cref{lem:LocalClearance}.

    For $0\le j\le J$, apply \cref{lem:LocalClearance}\ref{item:LocalMoment} at level $b_j$, with gap $g$ and cutoff $q_j$, from the time $t_j$, with $G\colonequals\Gc_{t_j}$ and with $\kappa$ the first time $t\ge t_j$ at which $\Gc_t$ fails or $N_{b_i}(S_t)>q_i$ for some $1\le i\le j$ with $t\ge t_i$; we call the resulting moments the \emph{local process} of level $j$ and $\kappa$ its \emph{killing time}.
    Its hypotheses hold before the killing time, and for $t\ge t_{j+1}$ we have $Me^{-\gamma(t-t_j)/2}\le M/(1+M)\le q_j$, so \eqref{eq:LocalPopulation} applies at every deterministic time $t\in[t_{j+1},T]$.
    For $j<J$, take $u=2$ and $r=q_{j+1}$; since $b_j+g+2=b_{j+1}$ and $q_{j+1}\ge A_{\mathrm{cl}}sq_j^2/n$, the base of \eqref{eq:LocalPopulation} is
    \begin{equation*}
        \frac{8eq_j}{q_{j+1}\vartheta_j^2}=\frac{8eC_{\mathrm{cl}}sq_j^2}{nq_{j+1}}\le\frac{8eC_{\mathrm{cl}}}{A_{\mathrm{cl}}}=\frac e{32},
    \end{equation*}
    and the probability is at most $(e/32)^{q_{j+1}}\le e^{-c_*R}$.

    Suppose that $\Vc$ holds and that some condition $N_{b_{j+1}}(S_t)\le q_{j+1}$ with $t\ge t_{j+1}$ fails.
    Consider the first time $t$ at which this happens, and the smallest index $j+1$ failing at that time.
    The local process of level $j$ has not been killed at time $t$, since $\Gc$ holds on $[t_0,T]$ and no condition of a level $i\le j$ has failed up to time $t$.
    Hence the failure is one of the events bounded above, and a union bound over the at most $J(T-t_0+1)$ pairs of a level and a time proves \eqref{eq:LayerFailure}.
    On $\Vc\cap\Lc_t$, the local process of level $J$ is alive at time $t$, and \eqref{eq:LocalPopulation} with $u=d$ proves \eqref{eq:LayerFinal}.
\end{proof}

\paragraph{Excursions.}
The last tool bounds the probability that an energy functional $\Phi_t$, for which $z^{\Phi_t}$ is a supermartingale above a level $E_-$, reaches a higher level $E_+$ before time $T$; it is in the spirit of the drift analysis of Hajek \cite{Hajek82}.

\begin{lemma}[Excursions]
    \label{lem:Excursions}
    Let $\Phi_t\ge0$ be an adapted process with $\Phi_0=0$ and $\Phi_{t+1}-\Phi_t\le K_\Phi$, let $z>1$ and $0\le E_-<E_+$, and let $\Ac_t\in\Fc_t$ be events such that
    \begin{equation*}
        \E\big[z^{\Phi_{t+1}-\Phi_t}\bigm|\Fc_t\big]\le1\qquad\text{on }\Ac_t\cap\{\Phi_t>E_-\}.
    \end{equation*}
    Then, for every integer $T\ge1$,
    \begin{equation}
        \label{eq:Excursions}
        \P\big(\Phi_t\ge E_+\text{ for some }t\le T\text{, and }\Ac_u\text{ holds for all }u<t\big)\le Tz^{E_-+K_\Phi-E_+}.
    \end{equation}
\end{lemma}

\begin{proof}
    Call $t\le T$ an \emph{entrance} if $\Phi_t>E_-\ge\Phi_{t-1}$; then $\Phi_t\le E_-+K_\Phi$.
    From an entrance $t$, stop at the first time $\kappa\ge t$ at which $\Phi_\kappa\le E_-$, $\Phi_\kappa\ge E_+$, $\Ac_\kappa$ fails, or $\kappa=T$.
    Up to $\kappa$, the process $z^{\Phi}$ is a nonnegative supermartingale, so, conditionally on $\Fc_t$, the probability that $\Phi_\kappa\ge E_+$ is at most $z^{E_-+K_\Phi-E_+}$.
    On the event in \eqref{eq:Excursions}, let $t'$ be the first time with $\Phi_{t'}\ge E_+$; since $\Phi_0=0$, the last entrance $t\le t'$ exists, and the excursion started at $t$ is stopped at $t'$ with $\Phi_{t'}\ge E_+$.
    A union bound over the at most $T$ entrances proves \eqref{eq:Excursions}.
\end{proof}

\section{Universal lower bounds over long horizons}
\label{sec:LowerBounds}

This section proves \cref{thm:Prefix}\ref{item:PrefixLower}.
The argument has three parts.
A finite safety game separates the deterministic content: an adversary that controls the next request can force a large height within a bounded number $J$ of requests (\cref{sec:SafetyGame,sec:Radix}).
Each request is then realized by the actual random input with conditional probability at least $n^{-O(s)}$, uniformly over the past (\cref{sec:Realization}).
Finally, repeated attempts over consecutive blocks of length $J$, each of which succeeds with a conditional probability that is bounded below uniformly over the past, convert the forcing time into a lower bound at long horizons (\cref{prop:BlockBound,sec:ProofUniversalLower}).
Along the way, \cref{prop:InfiniteValue} determines the radii of the cube for which forcing is possible at all; this yields the order of the optimal prefix discrepancy as $T\to\infty$ (\cref{cor:BeyondExponential}).

The game and its forcing strategy go back to B\'ar\'any \cite{Barany79}.
What is new is threefold: a quantitative width certificate that bounds the forcing time; an $\ell_1$-version of the recursive additive bases of \cite[Lemma~3.1]{Bandeira17}, with constants uniform in the number of touches and a controlled pool; and an exact realization of every request within both random input laws.
Realizing an adaptive forcing sequence by random arrivals, which gives lower bounds against every online algorithm, is also the idea behind the lower bounds of Ajtai et al.\ against an oblivious adversary \cite[Theorem~3.2]{Ajtai98} and of Bansal, Prabhu, Singla and Sundaram for random edges \cite[Theorem~4.3]{BansalCarpool26}.
For a fixed instance, the same repetition over consecutive blocks gives the conditional lower bound of Bansal, Jiang, Meka, Singla and Sinha for the offline prefix discrepancy of random unit vectors, where one block is close to a given worst-case configuration \cite[Theorem~16]{Bansal22}.

Throughout this section the inputs follow either model of \cref{sec:Model}, and the online algorithm is arbitrary; it may use private randomness and know the horizon.

\subsection{A finite safety game and a width certificate}
\label{sec:SafetyGame}

Fix integers $m,R,r\ge1$.
The \emph{alphabet} and the \emph{safe cube} are
\begin{equation*}
    \Vc_{m,r}\colonequals\big\{v\in\{-1,0,1\}^m:1\le\abs{\supp v}\le r\big\},\qquad Q_0\colonequals\{-R,\ldots,R\}^m,
\end{equation*}
and the \emph{safety sets} are defined recursively by
\begin{equation}
    \label{eq:SafetySets}
    Q_{j+1}\colonequals\big\{z\in Q_j:\ z+v\in Q_j\text{ or }z-v\in Q_j\text{ for every }v\in\Vc_{m,r}\big\}.
\end{equation}
In the underlying game, an adversary requests a direction $v\in\Vc_{m,r}$, and a controller, which stands for the algorithm, moves the state $z$ to $z+v$ or $z-v$.
Thus $Q_j$ is the set of states from which the controller can keep the state in $Q_0$ during the next $j$ requests, whatever they are.
If $Q_J=\emptyset$, we call $J$ a \emph{forcing time}.
The sets $Q_j$ are nested, invariant under permutations and sign changes of the coordinates, and in particular symmetric, $Q_j=-Q_j$.
The recursion \eqref{eq:SafetySets} iterates the operator $X\mapsto X\cap\bigcap_v((X+v)\cup(X-v))$ of B\'ar\'any \cite[Section~3]{Barany79}, and the following lemma, including the bound on the number of steps, is his strategy for the first player, our adversary, in the proof of \cite[Theorem~1]{Barany79}; we include the short proof to state it in our notation.

\begin{lemma}[Forcing from emptiness]
    \label{lem:SafetyForcing}
    Suppose that $Q_J=\emptyset$.
    Then there is a map $F:Q_0\to\Vc_{m,r}$ with the following property.
    For every $z_0\in Q_0$ and every sequence of signs $x_0,x_1,\ldots\in\{\pm1\}$, the sequence $z_{\ell+1}\colonequals z_\ell+x_\ell F(z_\ell)$, continued as long as $z_\ell\in Q_0$, satisfies $z_\ell\notin Q_0$ for some $\ell\le J$.
\end{lemma}

\begin{proof}
    For $z\in Q_0$ let $d(z)\colonequals\max\{j:z\in Q_j\}\le J-1$ be its \emph{depth}.
    Since $z\notin Q_{d(z)+1}$, there is $v\in\Vc_{m,r}$ such that neither $z+v$ nor $z-v$ lies in $Q_{d(z)}$; let $F(z)$ be such a direction.
    Then each successor either leaves $Q_0$ or has depth at most $d(z)-1$.
    Hence the state leaves $Q_0$ after at most $d(z_0)+1\le J$ steps.
\end{proof}

The lemma reduces forcing to proving $Q_J=\emptyset$.
The recursion \eqref{eq:SafetySets} and \cref{lem:SafetyForcing} apply verbatim to every finite alphabet $\Vc\subset\Z^m\setminus\{0\}$ with $\Vc=-\Vc$ in place of $\Vc_{m,r}$; we then speak of the safety sets of $\Vc$.

Whether a forcing time exists at all is decided by closed sets in the sense of B\'ar\'any.
A set $C\subseteq\Z^m$ is \emph{$\Vc$-closed} if $z+v\in C$ or $z-v\in C$ for all $z\in C$ and $v\in\Vc$ \cite[Section~2]{Barany79}.
The finite nested sets $Q_j$ stabilize at the largest $\Vc$-closed subset of $Q_0$ \cite[Section~3]{Barany79}: their final value is $\Vc$-closed, and by induction on $j$, every $\Vc$-closed subset of $Q_0$ lies in every $Q_j$.
Hence $Q_J=\emptyset$ for some $J$ if and only if $Q_0$ contains no nonempty $\Vc$-closed set.
Let $\Vc^+\subseteq\Vc$ contain one vector from each pair $\pm v$, and suppose that no two vectors of $\Vc^+$ are collinear.
By B\'ar\'any's width bound \cite[Section~6, Remark~(2)]{Barany79}, every nonempty bounded $\Vc$-closed set $C$ satisfies
\begin{equation}
    \label{eq:BaranyWidth}
    \max_{z\in C}\inp uz-\min_{z\in C}\inp uz\ge\sum_{v\in\Vc^+}\abs{\inp uv}\qquad\text{for all }u\in\R^m.
\end{equation}
Conversely, the set of subset sums of $\Vc^+$ is $\Vc$-closed \cite[Section~3]{Barany79}.
The controller keeps the state in this set, starting from $0$, by the \emph{parity rule}: it moves the state by $+v$ at the odd-numbered requests from the pair $\pm v$, where $v\in\Vc^+$, and by $-v$ at the even-numbered ones.
The state is then the sum of those $v\in\Vc^+$ whose pair has been requested an odd number of times.
For our alphabets, the two bounds match.

\begin{proposition}[Infinite-horizon value]
    \label{prop:InfiniteValue}
    Let $1\le r\le m$, and let $\Vc$ be the alphabet $\Vc_{m,r}$ or, if $r\ge3$, the alphabet $\Vc^{=}_{m,r}$ of all vectors in $\{-1,0,1\}^m$ with exactly $r$ nonzero coordinates.
    Let $\Delta$ be the number of pairs $\pm v$ in $\Vc$ with $v_1\ne0$, that is,
    \begin{equation*}
        \Delta=\sum_{j=1}^r2^{j-1}\binom{m-1}{j-1}\quad\text{for }\Vc_{m,r},\qquad\Delta=2^{r-1}\binom{m-1}{r-1}\quad\text{for }\Vc^{=}_{m,r}.
    \end{equation*}
    Then the safety sets of $\Vc$ satisfy $Q_J=\emptyset$ for some $J$ if and only if $2R<\Delta$.
    If $2R\ge\Delta$, the parity rule, for a suitable choice of $\Vc^+$, keeps the state in $Q_0$ forever, starting from $0$.
\end{proposition}

\begin{proof}
    Two vectors in $\{-1,0,1\}^m$ are collinear only if they agree up to sign, so \eqref{eq:BaranyWidth} applies for every choice of $\Vc^+$.
    Since $\Vc$ is invariant under permutations of the coordinates, each coordinate lies in the supports of exactly $\Delta$ vectors of $\Vc^+$.
    If $2R<\Delta$, then \eqref{eq:BaranyWidth} with $u=e_1$ shows that every nonempty bounded $\Vc$-closed set has width larger than $2R$ in the direction $e_1$; hence $Q_0$ contains no such set, and $Q_J=\emptyset$ for some $J$.

    If $2R\ge\Delta$, it suffices to choose $\Vc^+$ such that, at each coordinate $i$, the numbers of $v\in\Vc^+$ with $v_i=1$ and with $v_i=-1$ differ by at most one.
    Then every subset sum $z$ of $\Vc^+$ satisfies $\norm z_\infty\le\ceil{\Delta/2}\le R$, so the parity rule keeps the state in $Q_0$, and in particular $Q_J\ne\emptyset$ for all $J$.
    For a support $\{a_1<\cdots<a_j\}$ with $j\ge3$, take the vectors
    \begin{equation*}
        \sigma_2\sigma_3\Big(e_{a_1}+\sum_{\ell=2}^j\sigma_\ell e_{a_\ell}\Big),\qquad(\sigma_2,\ldots,\sigma_j)\in\{\pm1\}^{j-1},
    \end{equation*}
    one from each pair with this support.
    Their entries at $a_1$, $a_2$, $a_3$ and at $a_\ell$ with $\ell\ge4$ are $\sigma_2\sigma_3$, $\sigma_3$, $\sigma_2$ and $\sigma_2\sigma_3\sigma_\ell$, and each of these equals $1$ for exactly half of the sign vectors.
    Hence, at each coordinate of the support, these vectors have as many entries $1$ as entries $-1$, which proves the claim for $\Vc^{=}_{m,r}$.
    For $\Vc_{m,1}$, take $\Vc^+=\{e_i:i\in[m]\}$; each coordinate then lies in the support of exactly one of these vectors.
    For $\Vc_{m,r}$ with $r\ge2$, take in addition the vectors $e_i$ for $i\in[m]$ and, for each support $\{a,b\}$ with $a<b$, the vectors $(-1)^a(e_b+e_a)$ and $(-1)^a(e_b-e_a)$.
    At coordinate $i$, the vectors with support $\{i,b\}$ for some $b>i$ are balanced, and the others contribute $1+2\sum_{a<i}(-1)^a\in\{1,-1\}$ to the difference of the two numbers.
\end{proof}

Thus $\ceil{\Delta/2}$ is the smallest radius that the controller can maintain forever against an adaptive adversary; for $\Vc^{=}_{m,r}$ it equals $2^{r-2}\binom{m-1}{r-1}$, which is of order $m^{r-1}$ for fixed $r$.
For $r=m$, the alphabet $\Vc_{m,m}$ consists of all nonzero vectors in $\{-1,0,1\}^m$, and $\ceil{\Delta/2}=(3^{m-1}+1)/2$ is the value that B\'ar\'any computed for this alphabet \cite[Section~6, Remark~(2)]{Barany79}.
The proposition does not bound the forcing time, which is what matters for random inputs (\cref{prop:BlockBound}).
Our width certificate bounds it, using support functions along a line.
Fix $a\in\Z^m$ with $a_1=0$.
For a direction $v$ with $v_1=1$, the number $t_v\colonequals-\inp av$ is its \emph{root}, and $w_v\colonequals a+t_ve_1$ is its \emph{normal}; by construction $\inp{w_v}v=0$.

\begin{lemma}[Selected-root width certificate]
    \label{lem:WidthCertificate}
    Let $a\in\Z^m$ with $a_1=0$, and let $v^{(1)},\ldots,v^{(N)}\in\Vc_{m,r}$ be directions with first coordinate $1$ and pairwise distinct roots $t_1,\ldots,t_N$.
    Put $w_\nu\colonequals a+t_\nu e_1$ and $W\colonequals\sum_{\nu=1}^N\norm{w_\nu}_1$.
    If $N>2R$, then $Q_J=\emptyset$ for
    \begin{equation}
        \label{eq:WidthForcingTime}
        J\colonequals\floor{\frac{2RW}{N-2R}}+1.
    \end{equation}
    The same holds for the safety sets of any alphabet that contains $v^{(1)},\ldots,v^{(N)}$.
\end{lemma}

\begin{proof}
    For nonempty $Q_j$ consider the support function along the line $t\mapsto a+te_1$,
    \begin{equation*}
        g_j(t)\colonequals\max_{z\in Q_j}\inp{a+te_1}z.
    \end{equation*}
    It is convex and piecewise linear, and its slopes are first coordinates of maximizers; hence they lie in $[-R,R]$, and the total increase of the slope over the real line is at most $2R$.
    The \emph{scores} $g_j(t_\nu)$ are integers, they are nonincreasing in $j$, and $\abs{g_j(t_\nu)}\le R\norm{w_\nu}_1$.

    Suppose that $Q_{j+1}\ne\emptyset$ and that $g_{j+1}(t_\nu)=g_j(t_\nu)$ for some $\nu$.
    Let $z\in Q_{j+1}$ maximize $\inp{w_\nu}\cdot$ over $Q_{j+1}$; then $z$ also maximizes it over $Q_j$.
    By \eqref{eq:SafetySets}, $z'\colonequals z+xv^{(\nu)}\in Q_j$ for some $x\in\{\pm1\}$, and $\inp{w_\nu}{z'}=\inp{w_\nu}z$ because $\inp{w_\nu}{v^{(\nu)}}=0$.
    Thus $z$ and $z'$ are maximizers of $g_j$ at $t_\nu$ whose first coordinates differ by one, and the right derivative of $g_j$ at $t_\nu$ exceeds the left derivative by at least one.
    Since the roots are distinct, at most $2R$ indices $\nu$ have an unchanged score.
    All other scores decrease by at least one, so
    \begin{equation*}
        \sum_{\nu=1}^Ng_{j+1}(t_\nu)\le\sum_{\nu=1}^Ng_j(t_\nu)-(N-2R).
    \end{equation*}
    If $Q_J$ were nonempty, the sum of the scores would decrease $J$ times by at least $N-2R$ while staying in $[-RW,RW]$, contradicting \eqref{eq:WidthForcingTime}.
    The argument uses only the selected directions, which gives the last assertion.
\end{proof}

Qualitatively, the lemma also follows from the width bound \eqref{eq:BaranyWidth}, applied to the alphabet $\Vc'$ of the vectors $\pm v^{(\nu)}$, whose safety sets contain those of every larger alphabet.
The selected directions are pairwise non-collinear and have first coordinate one, so every nonempty bounded $\Vc'$-closed set would have width at least $N>2R$ in the direction $e_1$, whereas the cube has width $2R$.
What we need is the explicit forcing time \eqref{eq:WidthForcingTime}.

The lemma is useful when many distinct roots can be produced with a small coefficient mass $\norm a_1$: each normal has $\ell_1$-norm $\norm a_1+\abs{t_\nu}$, and at least $4R$ roots make the denominator in \eqref{eq:WidthForcingTime} comparable to $N$.
We call $2RW$, the length of the a priori interval $[-RW,RW]$ for the sum of the scores, the \emph{budget} of the certificate; the actual range of the scores may be smaller.
A direction with first coordinate one and $k$ further nonzero coordinates, that is, with \emph{touch count} $\abs{\supp v}=k+1$, has a root that is a signed sum of at most $k$ coefficients.
The next subsection chooses the coefficients so that these sums cover a long interval of integers.

\subsection{Mixed-radix coefficients}
\label{sec:Radix}

The coefficients below are grouped into levels and form a mixed-radix additive basis: every integer in a long interval is one coefficient of the top level minus at most one coefficient from each lower level.
Such bases go back to Rohrbach \cite{Rohrbach37}.
Bandeira, Ferber and Kwan \cite[Lemma~3.1]{Bandeira17} chose the level sizes recursively to bound the smallest sum of squares of an additive basis of order $h$ of $[n]$ by $n^{2\cdot3^h/(3^h-1)}$, up to a factor depending on $h$.
Our cost is instead the $\ell_1$-norm of the coefficients, which enters the budget of \cref{lem:WidthCertificate}, and the same balancing of the levels gives the exponent $\eta_k=2^k/(2^k-1)$ in \cref{prop:OptimizedRadices}.
In addition, we need three things: constants that do not depend on $k$, which matters for growing sparsity (\cref{rem:GrowingSparsity}); a bound on the number $m$ of tracked coordinates, the \emph{pool}, so that the pool fits into $[n]$; and the realization of all requests within the exact input laws (\cref{sec:Realization}).

\begin{proposition}[Radix construction]
    \label{prop:Radix}
    Let $k,R\ge1$ and let $q_1,\ldots,q_k$ be positive integers.
    Put $P_0\colonequals1$ and $P_i\colonequals\prod_{j\le i}q_j$, and suppose that $P_k\ge4R$.
    Set
    \begin{align}
        \label{eq:RadixPool}
        m&\colonequals2+\sum_{i=1}^k(q_i-1),\\
        \label{eq:RadixMass}
        \Lambda&\colonequals\frac12\sum_{i<k}P_{i-1}q_i(q_i-1)+\frac12P_{k-1}q_k(q_k+1),\\
        \label{eq:RadixTime}
        J&\colonequals4R(\Lambda+P_k)+1.
    \end{align}
    Then the safety sets of the alphabet $\Vc_{m,k+1}$ satisfy $Q_J=\emptyset$.
\end{proposition}

\begin{proof}
    Give the first coordinate the coefficient zero, and give the remaining $m-1$ coordinates, called \emph{levers}, distinct positive coefficients, grouped into \emph{levels}: level $i<k$ consists of $P_{i-1},2P_{i-1},\ldots,(q_i-1)P_{i-1}$, and level $k$ of $P_{k-1},2P_{k-1},\ldots,q_kP_{k-1}$.
    A level may be empty.
    Every coefficient of a lower level $i$ is smaller than $P_i$, and $P_i$ is at most every coefficient of a later nonempty level, so these $m-1$ coefficients are distinct, and their sum is $\Lambda$.
    Write $[c]$ for the lever with coefficient $c$, and let $a$ be the resulting coefficient vector.

    Let $1\le t\le P_k$ and $b\colonequals\ceil{t/P_{k-1}}\in\{1,\ldots,q_k\}$.
    The mixed-radix expansion of $bP_{k-1}-t\in[0,P_{k-1})$ reads
    \begin{equation*}
        bP_{k-1}-t=\sum_{i<k}u_iP_{i-1},\qquad0\le u_i<q_i.
    \end{equation*}
    The direction $v\colonequals e_1+e_{[bP_{k-1}]}-\sum_{i<k,\,u_i>0}e_{[u_iP_{i-1}]}$ has at most $k+1$ nonzero coordinates and satisfies $\inp av=t$, so its root is $-t$.
    Reversing all lever signs, with first coordinate still $+1$, gives root $t$, and $e_1$ has root $0$.
    This yields $N=2P_k+1$ directions with the distinct roots $-P_k,\ldots,P_k$.
    Each normal has $\ell_1$-norm $\Lambda+\abs t\le\Lambda+P_k$, so $W\le N(\Lambda+P_k)$.
    Since $P_k\ge4R$, we have $N-2R\ge N/2$, and \cref{lem:WidthCertificate} gives $Q_{J'}=\emptyset$ for some $J'\le4R(\Lambda+P_k)+1=J$; hence $Q_J=\emptyset$, since the safety sets are nested.
\end{proof}

Any radices satisfying the hypotheses may be used.
The trivial choice $q_1=\cdots=q_{k-1}=1$, $q_k=4R$ gives $\Lambda\asymp R^2$ and $J\asymp R^3$; this is the cubic forcing time behind the case $s=2$.
The next choice balances the levels; for each fixed $k$, it is optimal for this certificate up to a factor depending only on $k$ (\cref{prop:StaticBarrier}).

\begin{proposition}[Optimized radices]
    \label{prop:OptimizedRadices}
    Let $k,R\ge1$ with $2^k\le4R$, and put
    \begin{equation*}
        \eta_k\colonequals\frac{2^k}{2^k-1},\qquad\beta_k\colonequals1+\eta_k=2+\frac1{2^k-1},\qquad\gamma_k\colonequals\frac{\eta_k}2.
    \end{equation*}
    There are radices in \cref{prop:Radix} such that
    \begin{equation}
        \label{eq:OptimizedRadices}
        4R\le P_k\le8R,\qquad \Lambda\le112R^{\eta_k},\qquad J\le481R^{\beta_k},\qquad m\le2+2k(4R)^{\gamma_k}.
    \end{equation}
\end{proposition}

\begin{proof}
    Put $Z\colonequals4R$,
    \begin{equation*}
        A\colonequals\Big(\frac Z{2^{k-1+2^{-k}}}\Big)^{\eta_k},\qquad\Pi_i\colonequals2^{i-1+2^{-i}}A^{1-2^{-i}}\qquad(0\le i\le k).
    \end{equation*}
    Since $\eta_k(1-2^{-k})=1$, we have $\Pi_0=1$ and $\Pi_k=Z$, and $Z\ge2^k$ is equivalent to $A\ge2$.
    The ideal ratios $q_i^*\colonequals\Pi_i/\Pi_{i-1}$ satisfy
    \begin{equation}
        \label{eq:IdealRatios}
        q_i^*=2\Big(\frac A2\Big)^{2^{-i}}\ge2,\qquad\frac{\Pi_i^2}{2\Pi_{i-1}}=2^{i-1}A.
    \end{equation}
    Round the cumulative products: $P_0\colonequals1$, $q_i\colonequals\ceil{\Pi_i/P_{i-1}}$, and $P_i\colonequals P_{i-1}q_i$.
    By induction, $\Pi_i\le P_i\le2\Pi_i$; indeed $P_{i-1}\le2\Pi_{i-1}\le\Pi_i$ and $P_i<\Pi_i+P_{i-1}\le2\Pi_i$.
    In particular $Z\le P_k\le2Z$.

    For a lower level, $q_i-1\le\Pi_i/P_{i-1}$ and $q_i\le\Pi_i/P_{i-1}+1$, so $\frac12P_{i-1}q_i(q_i-1)\le\Pi_i^2/(2P_{i-1})+\Pi_i/2\le2^{i-1}A+\Pi_i/2$.
    For the last level, $\frac12P_{k-1}q_k(q_k+1)=P_k^2/(2P_{k-1})+P_k/2\le2^{k+1}A+\Pi_k$.
    By \eqref{eq:IdealRatios}, $\sum_{i<k}\Pi_i\le\Pi_k=Z$, and a direct computation gives $2^kA=Z^{\eta_k}2^{1-k/(2^k-1)}\le2Z^{\eta_k}$.
    Therefore
    \begin{equation*}
        \Lambda\le\tfrac522^kA+\tfrac32Z\le5Z^{\eta_k}+\tfrac32Z\le7Z^{\eta_k}\le112R^{\eta_k},
    \end{equation*}
    using $1<\eta_k\le2$.
    Now \eqref{eq:RadixTime} gives $J\le448R^{\beta_k}+32R^2+1\le481R^{\beta_k}$.
    Finally, $q_i-1\le\Pi_i/P_{i-1}\le q_i^*\le\sqrt{2A}\le2Z^{\gamma_k}$, and \eqref{eq:RadixPool} gives the bound on $m$.
\end{proof}

\subsection{Realization and the block bound}
\label{sec:Realization}

To use the game, we track a fixed pool of coordinates and realize each request by the next arrival.
In the signed model, we track coordinates directly.
In the binary model, all entries are nonnegative, so we track differences of disjoint pairs; such a difference can reach $2(H-1)$ while both heights stay below $H$, which is why the binary model needs a cube of twice the radius.
The binary realization uses the actual support size $s$ and only the requested touch count $r\le s$; the unused positions are filled outside the pool.
This is why both models lead to the same exponents.
We use the model parameter
\begin{equation*}
    \chi\colonequals1\text{ in the signed model},\qquad\chi\colonequals2\text{ in the binary model}.
\end{equation*}

\begin{lemma}[Exact realization]
    \label{lem:Realization}
    Let $H\ge2$, $m\ge1$ and $1\le r\le s$ be integers, put $R\colonequals\chi(H-1)$ and $r_0\colonequals\min\{r,m\}$, and assume $n-\chi m\ge s-1$.
    Fix $m$ distinct coordinates $c_1,\ldots,c_m$ in the signed model, or $m$ disjoint pairs $(c_1^+,c_1^-),\ldots,(c_m^+,c_m^-)$ in the binary model, and define the tracked state $Y\in\Z^m$ by $Y_i\colonequals S_{c_i}$, respectively $Y_i\colonequals S_{c_i^+}-S_{c_i^-}$.
    For every $v\in\Vc_{m,r}$ there is a set $E(v)$ of possible arrivals with the following properties.
    \begin{enumerate}[label=(\roman*)]
        \item\label{item:RealizationProbability} $\P(V\in E(v))=p_{\abs{\supp v}}$ for an arrival $V$ with law $\mu_s$, where, for $1\le j\le r_0$,
        \begin{equation}
            \label{eq:RealizationProbability}
            p_j\colonequals
            \begin{cases}
                2^{1-j}\binom{n-m}{s-j}\Big/\binom ns, & \chi=1,\\[1ex]
                \binom{n-2m}{s-j}\Big/\binom ns, & \chi=2.
            \end{cases}
        \end{equation}
        \item\label{item:RealizationMove} If the arrival lies in $E(v)$, then, whatever sign the algorithm chooses, the tracked state moves to $Y+v$ or to $Y-v$.
        \item\label{item:RealizationCube} If all heights are smaller than $H$, then $Y\in Q_0$.
    \end{enumerate}
    Moreover, $p_*\colonequals\min_{1\le j\le r_0}p_j=\min\{p_1,p_{r_0}\}>0$.
    If $n\ge2s\chi m$, then
    \begin{equation}
        \label{eq:RealizationLower}
        p_*\ge\Big(\frac s{2en}\Big)^r=e^{-\lambda},\qquad\lambda\colonequals r\log\frac{2en}s.
    \end{equation}
\end{lemma}

\begin{proof}
    In the signed model, let $E(v)$ be the set of arrivals whose support meets $\{c_1,\ldots,c_m\}$ exactly in the coordinates $c_i$ with $v_i\ne0$, and whose marks on these coordinates equal $(v_i)_{v_i\ne0}$ or $(-v_i)_{v_i\ne0}$.
    Marks outside the pool are unrestricted.
    This gives the first formula in \eqref{eq:RealizationProbability}, and an arrival in $E(v)$ changes $Y$ by $\pm xv$ for the chosen sign $x$.

    In the binary model, a requested mark $v_i=+1$ is realized by including $c_i^+$, and $v_i=-1$ by including $c_i^-$.
    Let $E(v)$ be the set of arrivals whose support meets the $2m$ pool coordinates exactly in these $\abs{\supp v}$ endpoints.
    The remaining $s-\abs{\supp v}$ positions are unrestricted outside the pool coordinates, which gives the second formula, and the update is exactly $Y'=Y+xv$.
    If all heights are smaller than $H$, then $\abs{Y_i}\le\chi(H-1)=R$ in both models, which is \ref{item:RealizationCube}.

    The ratios
    \begin{equation*}
        \frac{p_{j+1}}{p_j}=\frac{\chi(s-j)}{2(n-\chi m-s+j+1)}
    \end{equation*}
    decrease in $j$.
    Hence $(p_j)$ is log-concave, and its minimum over an interval is attained at an endpoint.
    The assumption $n-\chi m\ge s-1$ makes all $p_j$ positive.

    For \eqref{eq:RealizationLower}, the requested coordinates are all included with probability $(s)_j/(n)_j\ge(s/(en))^j$, because the geometric mean of the $j$ largest factors of $s!$ is at least $(s!)^{1/s}\ge s/e$.
    Conditionally on their inclusion, the remaining $s-j$ coordinates of the support form a uniform subset of the other $n-j$ coordinates, and they meet one of the other $\chi m-j$ pool coordinates with probability at most $(s-j)(\chi m-j)/(n-j)\le s\chi m/n\le1/2$.
    In the signed model, the antipodal mark patterns contribute the factor $2^{1-j}$, so $p_j\ge(s/(2en))^j$.
    In the binary model, $p_j\ge\frac12(s/(en))^j\ge(s/(2en))^j$ since $j\ge1$.
    As $j\le r$ and $s/(2en)<1$, this proves \eqref{eq:RealizationLower}.
    The margin $n\ge2s\chi m$ also implies $n-\chi m\ge s-1$.
\end{proof}

\begin{proposition}[Block bound at every horizon]
    \label{prop:BlockBound}
    Assume the hypotheses of \cref{lem:Realization}, and suppose that the safety sets of $\Vc_{m,r}$ with cube radius $R=\chi(H-1)$ satisfy $Q_J=\emptyset$.
    Then, for every integer $T\ge1$ and every online algorithm,
    \begin{equation}
        \label{eq:BlockBound}
        \P(D_T<H)\le\big(1-p_*^J\big)^{\floor{T/J}}\le\exp\Big\{-\floor{\frac TJ}p_*^J\Big\}.
    \end{equation}
    If $n\ge2s\chi m$, one may replace $p_*^J$ by $e^{-\lambda J}$, with $\lambda$ from \eqref{eq:RealizationLower}.
\end{proposition}

\begin{proof}
    Let $F$ be the forcing map of \cref{lem:SafetyForcing}, and write $Y(u)$ for the tracked state at time $u$.
    Divide the time axis into consecutive blocks of $J$ arrivals and consider block $b$, which consists of the arrivals $(b-1)J+1,\ldots,bJ$.
    Let $G_b$ be the event that, for every time $u$ of the block such that all heights are smaller than $H$ at time $u-1$, the arrival $V_u$ lies in $E(F(Y(u-1)))$.
    The request $F(Y(u-1))$ is $\Fc_{u-1}$-measurable, and $V_u$ is independent of $\Fc_{u-1}$, so each such requirement holds with conditional probability at least $p_*$; at the other times of the block, nothing is required.
    Hence $\P(G_b\mid\Fc_{(b-1)J})\ge p_*^J$.

    Suppose that all heights are smaller than $H$ at the beginning of the block.
    On $G_b$, as long as no height reaches $H$, the tracked state stays in $Q_0$ and moves, by \cref{lem:Realization}\ref{item:RealizationMove}, to one of the two successors of the requested direction.
    By \cref{lem:SafetyForcing}, it leaves $Q_0$ within the block, and \cref{lem:Realization}\ref{item:RealizationCube} shows that some height reaches $H$.
    Therefore the conditional probability that all heights stay below $H$ until the end of block $b$ is at most $1-p_*^J$, uniformly over all histories and all choices of the algorithm.
    Iterating over the $\floor{T/J}$ complete blocks proves \eqref{eq:BlockBound}.
\end{proof}

The proof neither resets the state at block boundaries nor uses any independence between the blocks beyond the conditional bound for each block.

\subsection{Proof of the lower bound}
\label{sec:ProofUniversalLower}

We now insert the optimized radices.
For $1\le k\le s-1$ and $\chi\in\{1,2\}$ put
\begin{equation}
    \label{eq:HeightConstants}
    C_\chi\colonequals481\chi^3,\qquad A_{k,\chi}\colonequals2+8\chi k,\qquad\lambda_k\colonequals(k+1)\log\frac{2en}s.
\end{equation}

\begin{lemma}[Admissible heights]
    \label{lem:AdmissibleHeights}
    Let $1\le k\le s-1$, and let $H\ge2$ be an integer with
    \begin{equation}
        \label{eq:AdmissibleHeights}
        2^k\le4\chi(H-1),\qquad H^{\gamma_k}\le\frac n{2s\chi A_{k,\chi}}.
    \end{equation}
    Then, for $J\colonequals\floor{C_\chi H^{\beta_k}}$, every integer $T\ge1$ and every online algorithm,
    \begin{equation*}
        \P(D_T<H)\le\exp\Big\{-\floor{\frac TJ}e^{-\lambda_kJ}\Big\}.
    \end{equation*}
\end{lemma}

\begin{proof}
    Apply \cref{prop:OptimizedRadices} with $R\colonequals\chi(H-1)$, which is possible by the first condition.
    Since $R\le\chi H$, $(4\chi)^{\gamma_k}\le4\chi$ and $\beta_k\le3$, the resulting game has $m\le A_{k,\chi}H^{\gamma_k}$ coordinates and touch count $k+1\le s$, and its forcing time \eqref{eq:RadixTime} is an integer at most $481\chi^3H^{\beta_k}$, hence at most $J$; since the safety sets are nested, $Q_J=\emptyset$.
    The second condition gives $n\ge2s\chi m$, and \cref{prop:BlockBound} applies with $\lambda=\lambda_k$.
\end{proof}

\begin{proof}[Proof of \cref{thm:Prefix}\ref{item:PrefixLower}]
    \emph{The horizon term.}
    Fix $s\ge2$ and $c>0$, let $n\le T\le e^{cn}$, and put $k\colonequals s-1$.
    Then $1/\beta_{s-1}=\alpha_s$ and $\gamma_{s-1}\alpha_s=2^{s-2}/(2^s-1)<1$, and $\log n\le\lambda_{s-1}\le2s\log n$ for large $n$.
    Put
    \begin{equation*}
        y\colonequals\Big(\frac{\log T}{2C_\chi\lambda_{s-1}}\Big)^{\alpha_s}\ge\kappa_s\Big(\frac{\log T}{\log n}\Big)^{\alpha_s},\qquad\kappa_s\colonequals(4sC_2)^{-\alpha_s},
    \end{equation*}
    and $H\colonequals\floor y$; the inequality holds in both models, since $C_\chi\le C_2$.
    Suppose first that $H\ge2$ and $2^{s-1}\le4\chi(H-1)$.
    Since $H\le(cn)^{\alpha_s}$ and $\gamma_{s-1}\alpha_s<1$, the second condition in \eqref{eq:AdmissibleHeights} holds for large $n$, uniformly in $T$.
    For $J\colonequals\floor{C_\chi H^{\beta_{s-1}}}$ we have $2\lambda_{s-1}J\le\log T<T$, so $e^{-\lambda_{s-1}J}\ge T^{-1/2}$ and $\floor{T/J}\ge T/(2J)\ge\lambda_{s-1}T/\log T$.
    \Cref{lem:AdmissibleHeights} and $T\ge n$ give
    \begin{equation*}
        \P(D_T<H)\le\exp\Big\{-\frac{\lambda_{s-1}\sqrt T}{\log T}\Big\}\le\exp\Big\{-\frac{\sqrt n}{\log n}\Big\},
    \end{equation*}
    and $H\ge y/2\ge\frac{\kappa_s}2(\log T/\log n)^{\alpha_s}$.
    Otherwise $y<2+2^{s-1}$, so $(\log T/\log n)^{\alpha_s}<(2+2^{s-1})/\kappa_s$, and the deterministic bound $D_T\ge1$ suffices.
    In both cases, outside an event of probability at most $\exp(-\sqrt n/\log n)$ for large $n$,
    \begin{equation*}
        D_T\ge\frac{\kappa_s}{2+2^{s-1}}\Big(\frac{\log T}{\log n}\Big)^{\alpha_s}.
    \end{equation*}

    \emph{The doubly logarithmic term.}
    For fixed $s$ and large $n$ we have $s\le(\log\log n)^2$, and \cref{prop:Baseline}\ref{item:BaselineLogLog} gives $D_T\ge\norm{S_T}_\infty\ge c_{\mathrm{bl}}\log\log n$ with probability at least $1-C_{\mathrm{bl}}n^{-0.8}$, where $c_{\mathrm{bl}}$ and $C_{\mathrm{bl}}$ denote the constants $c$ and $C$ of \cref{prop:Baseline}.
    On the intersection of both events, $\max\{u,v\}\ge(u+v)/2$ proves \eqref{eq:MainLower} with $c'\colonequals\frac12\min\{c_{\mathrm{bl}},\kappa_s/(2+2^{s-1})\}$ and failure probability $\e_n\colonequals C_{\mathrm{bl}}n^{-0.8}+\exp(-\sqrt n/\log n)$ for large $n$, and $\e_n\colonequals1$ otherwise.
\end{proof}

The exponents $\alpha_s=1/\beta_{s-1}$ are $1/3,3/7,7/15,15/31,\ldots$ for $s=2,3,4,5,\ldots$; they increase to $1/2$, and they are exactly the limit of \cref{lem:WidthCertificate} with static coefficients (\cref{prop:StaticBarrier}).
For $s=2$, the cubic forcing time is reminiscent of the deterministic lower bound for online carpooling \cite[Theorem~2.3]{Ajtai98}, which rests on the chip game of Anderson, Lov\'asz, Shor, Spencer, Tardos and Winograd \cite{Anderson89}: for $h$ up to about $n/4$, at most $h^3$ requests force unfairness $h$.
The constants of the horizon term are explicit, and the restriction $T\le e^{cn}$ serves only to guarantee the second condition in \eqref{eq:AdmissibleHeights}.

\paragraph{Complements.}
We record four complements: pairs beyond exponential horizons, growing sparsity, random edge orientations, and terminal lower bounds at small failure probabilities.

\begin{remark}[Pairs beyond exponential horizons]
    \label{rem:PairsBeyondExponential}
    For $s=2$ we have $\gamma_1=1$, so the second condition in \eqref{eq:AdmissibleHeights} holds for all $H\le n/(4\chi A_{1,\chi})$, independently of $T$.
    The proof of \cref{thm:Prefix}\ref{item:PrefixLower}, with $H$ capped at this value, therefore gives $D_T\ge c\min\{(\log T/\log n)^{1/3},n\}$ with probability $1-o(1)$, uniformly for all $T\ge n$.
    This matches the deterministic bound $D_T\le2n$ of quadratic Greedy (\cref{lem:DoubleCover}) once $\log T\ge Cn^3\log n$; in particular, the order $n$ of the limit in \cref{cor:BeyondExponential} is reached at such horizons.
\end{remark}

\begin{remark}[Growing sparsity]
    \label{rem:GrowingSparsity}
    The construction also gives lower bounds when $s$ grows with $n$, provided the number of touches and the size of the pool are kept under control.
    Put $U\colonequals\log T/\log(4en/s)$, $k\colonequals\ceil{\log_2\log U}$ and $H\colonequals\floor{10^{-3}\sqrt{U/(k+1)}}$.
    If $U$ exceeds a sufficiently large absolute constant, $k\le s-1$ and $n\ge8s(16k+2)H^{\gamma_k}$, then $H^{1/(2^k-1)}\le e$, so the forcing time of \cref{lem:AdmissibleHeights} is at most $3848eH^2$, and the proof above gives $\P(D_T\ge H)\ge1-\exp\{-\sqrt T/\log T\}$ in both models.
    Thus the prefix discrepancy is at least of order $\sqrt{\log T/(\log(4en/s)\log\log U)}$.
    For instance, $T=\ceil{e^n}$ and $s=\ceil{2\log n}$ give the order $\sqrt{n/(\log n\log\log n)}$, whereas the generic bound of \cref{rem:GenericBenchmarks} is $O(\sqrt{n\log n})$.
    The convergence $\alpha_s\to\frac12$ alone does not justify inserting a growing $s$ into \cref{thm:Prefix}\ref{item:PrefixLower}, which is stated for fixed $s$, since the cost of the $k+1$ touches per request and the size of the pool both enter.
\end{remark}

\begin{remark}[Random edge orientations]
    \label{rem:EdgeOrientations}
    The method applies whenever each request of a forcing strategy is realized by the next arrival with a probability bounded below uniformly over the past, for instance for arrivals $e_i-e_j$ with a uniformly random pair $\{i,j\}$, as in stochastic online carpooling on the complete graph \cite{Ajtai98} (see \cite{Gupta20,BansalCarpool26} for general graphs).
    There, with $R\colonequals H-1$ and $M\colonequals4R$, track a distinguished coordinate with coefficient $0$ and $2M$ further coordinates $[\ell]$ with coefficients $\ell=\pm1,\ldots,\pm M$; the directions $e_1-e_{[\ell]}$ and $e_1$ have the $2M+1$ distinct roots $-M,\ldots,M$, so by the last assertion of \cref{lem:WidthCertificate} the safety sets of this alphabet are empty after $O(H^3)$ requests.
    Each request is realized with probability at least $\binom n2^{-1}$ (the request $e_1$ by a pair of the distinguished coordinate and a coordinate outside the pool), so the proof of \cref{prop:BlockBound} gives prefix discrepancy $\Omega((\log T/\log n)^{1/3})$ for every online orientation algorithm, with probability $1-o(1)$, uniformly for $n\le T\le e^{cn}$.
    This makes the dependence on $n$ explicit in the lower bound $\Omega((\log T)^{1/3})$ for random edges that Bansal, Prabhu, Singla and Sundaram state with reference to \cite{Ajtai98} and derive in this way \cite[Section~1]{BansalCarpool26}.
\end{remark}

\begin{remark}[Terminal times at small failure probabilities]
    \label{rem:TerminalSmallConfidence}
    The forcing argument also shows that the terminal discrepancy is large with some probability, but the probability that it guarantees is small, consistently with \cref{thm:Terminal}: of order $n^{-O(1)}$ for bounded heights, and smaller than every fixed power of $1/n$ for heights that grow with $n$.
    Assume the hypotheses of \cref{prop:BlockBound} with $s\le n/2$ and $n\ge2s\chi m$, and let $T\ge J$.
    Consider the last $J$ arrivals up to time $T$.
    As long as all heights are smaller than $H$, require the arrival at time $u$ to lie in $E(F(Y(u-1)))$, as in the proof of \cref{prop:BlockBound}.
    From the first time at which some height is at least $H$, which may already be the beginning of the block, require every later arrival of the block not to select a fixed coordinate $i_*$ of height at least $H$, chosen at that time by a fixed criterion, for instance the smallest such index.
    Each requirement holds with conditional probability at least $\min\{p_*,1-s/n\}=p_*$, since $p_*\le p_1\le s/n\le1-s/n$.
    On the intersection of all requirements, a height at least $H$ is reached within the block, by \cref{lem:SafetyForcing,lem:Realization}, and the coordinate $i_*$ is not selected afterwards, so its height stays at least $H$.
    Hence every online algorithm satisfies $\P\{\norm{S_T}_\infty\ge H\}\ge p_*^J\ge e^{-\lambda J}$.

    For fixed $s\ge2$, take $k=s-1$ and the height
    \begin{equation*}
        H\colonequals\floor{\Big(\frac{\log(1/\delta)}{4sC_\chi\log n}\Big)^{\alpha_s}}.
    \end{equation*}
    The proof of \cref{lem:AdmissibleHeights} provides $Q_J=\emptyset$ and $n\ge2s\chi m$, and $\lambda_{s-1}J\le\frac12\log(1/\delta)$ and $J\le\log(1/\delta)/(4s\log n)\le n\le T$ when $n$ is large.
    This gives constants $c_s,C_s>0$ such that, for all large $n$, every $T\ge n$ and every $\delta$ with $e^{-n}\le\delta\le n^{-C_s}$, every online algorithm satisfies
    \begin{equation*}
        \P\bc{\norm{S_T}_\infty\ge c_s\Big(\frac{\log(1/\delta)}{\log n}\Big)^{\alpha_s}}\ge\sqrt\delta>\delta.
    \end{equation*}
    Thus the terminal radius at failure probability $\delta$, the smallest $r$ for which some online algorithm satisfies $\P\{\norm{S_T}_\infty>r\}\le\delta$, must grow with $\log(1/\delta)/\log n$.
    By \cref{thm:Terminal}, it does not grow with the horizon at each fixed polynomial level $\delta=n^{-k}$; for smaller $\delta$, \cref{thm:Terminal} gives no information (see \cref{q:TerminalConfidence}).
\end{remark}

\section{Upper bounds for pairs}
\label{sec:Pairs}

Throughout this section $s=2$.
The support of each arrival is a uniformly random pair $\{i,j\}$ of distinct coordinates, and its two entries are independent signs (signed model) or both equal to one (binary model).
We analyze \emph{quadratic Greedy},
\begin{equation*}
    x_t\in\argmin_{x\in\{\pm1\}}\norm{S_{t-1}+xV_t}_2^2,\qquad\text{that is,}\qquad x_t=-\sgn\inp{S_{t-1}}{V_t},
\end{equation*}
with an arbitrary fixed rule at ties; in this section we call it simply Greedy.
This is a single deterministic rule; it uses no parameters and knows neither $T$ nor $\delta$.
We write $N_b(t)\colonequals N_b(S_t)$ and use the excess energy $F_b(t)\colonequals\sum_{i=1}^n(h_i(t)-b)_+^2$.
\Cref{sec:ChipGames} reduces large heights to an accumulation of excess energy, \cref{sec:CubicExcursions} shows that this accumulation is unlikely while few coordinates are high, \cref{sec:GeometricMoments,sec:GlobalQuadratic} control the number of high coordinates over the whole horizon, and \cref{sec:ProofPairs} combines these estimates into the proof of \cref{thm:Prefix}\ref{item:PrefixPairs}; \cref{sec:GreedyTerminal} shows that Greedy also attains the optimal order of the terminal discrepancy at every prescribed time $T\ge n$.

\paragraph{Outcomes of one arrival.}
Let $a\ge c$ be the heights of the two selected coordinates before the arrival.
If $a>c$, Greedy moves the higher one inward.
More precisely, the new pair of heights, as an unordered pair, is as follows.
\begin{enumerate}[label=(O\arabic*)]
    \item\label{item:OutcomePositive} If $c\ge1$, the outcome is $(a-1,c-1)$ or $(a-1,c+1)$.
    In the signed model each occurs with conditional probability $1/2$, given the past and the support.
    In the binary model the first occurs if $S_i$ and $S_j$ have equal signs and the second otherwise.
    \item\label{item:OutcomeZero} If $c=0<a$, the outcome is $(a-1,1)$; if $a=c=0$, it is $(1,1)$.
\end{enumerate}
When $a=c\ge1$, the outcome $(a-1,a+1)$ arises from a tie $\inp{S_{t-1}}{V_t}=0$; apart from the outcome $(1,1)$ for $a=c=0$, this is the only way in which the larger selected height can increase.
In particular, every state is $(b,0)$-isolating for Greedy, in the sense of \cref{sec:Tools}, for every $b\ge1$; this is the only property of the rule that the layer iteration in \cref{sec:ProofPairs} uses.
The clearance estimate of \cref{sec:Tools} and its iteration are stated for rules that depend only on the current state and the arrival.
They apply verbatim to Greedy even if its rule at ties depends on the past, with conditioning on $\Fc_t$ in place of the current state, since an arrival in which exactly one selected coordinate has height at least $b\ge1$ never produces a tie.

\subsection{Chip games and a cubic energy witness}
\label{sec:ChipGames}

The only deterministic inputs from the literature are two estimates of Bansal, Prabhu, Singla and Sundaram for Greedy, viewed as a game of chips on the line: their cube-root bound and their spreadness estimate.
Chips sit at real positions.
A \emph{Greedy move} with weight $w\in[0,1]$ selects two chips, moves the one at the smaller position (either one at equal positions) to the right by $w$, and moves the other to the left by $w$.
The move is \emph{expanding} if the chip moving to the right ends strictly to the right of the previous position of the other chip.

\begin{literaturetheorem}[{\cite[Lemmas~3.1 and 3.3]{BansalCarpool26}}]
    \label{thm:ChipGame}
    Consider $N$ chips that start at nonpositive positions and are moved by Greedy moves.
    \begin{enumerate}[label=(\roman*)]
        \item\label{item:ChipBound} If at most $p$ expanding moves have occurred, every chip is at a position at most $\min\{4p^{1/3},N\}$.
        \item\label{item:ChipSpread} If the largest position at some time is $D>0$, then for every integer $1\le r\le\min\{\floor D,N\}$ there is an earlier or equal time at which the sum of the $r$ largest positions is at least $r(D-r+1)$.
    \end{enumerate}
\end{literaturetheorem}

\begin{lemma}[Double cover and cubic witness]
    \label{lem:DoubleCover}
    Consider an arbitrary sequence of arrivals with exactly two nonzero entries in $\{\pm1\}$, and let the signs be chosen by quadratic Greedy.
    \begin{enumerate}[label=(\roman*)]
        \item\label{item:DoubleCoverBound} $D_T\le\min\{2n,4(2T)^{1/3}\}$ for every $T$.
        \item\label{item:CubicWitness} Let $b\ge1$ and $u\ge4$ be integers.
        If some height reaches $b+u$ for the first time at time $t_0$, then $F_b(t')\ge u^3/32$ for some $t'\le t_0$.
    \end{enumerate}
\end{lemma}

\begin{proof}
    Represent the state by its \emph{double cover}, the $2n$ chips $(i,\sigma)$, $\sigma\in\{\pm1\}$, where chip $(i,\sigma)$ sits at $\sigma S_i$.
    Let the arrival be $\alpha e_i+\gamma e_j$ with $\alpha,\gamma\in\{\pm1\}$, and let $x$ be the chosen sign.
    The update is realized by two chip moves: chip $(i,\alpha)$ gains $x$ while chip $(j,-\gamma)$ loses $x$, and then chip $(j,\gamma)$ gains $x$ while chip $(i,-\alpha)$ loses $x$.
    In both moves the difference between the two positions before the move is $\alpha S_i+\gamma S_j=\inp{S_{t-1}}{V_t}$.
    Greedy chooses $x$ with $x\inp{S_{t-1}}{V_t}\le0$, so in both moves the chip at the smaller position moves to the right; these are Greedy moves of weight one.
    After both moves the chips are again at $(S,-S)$.
    We index the chip game by \emph{chip times}, one after each chip move; the chip time after both moves of arrival $t$ is called a \emph{complete time} and corresponds to the state $S_t$.
    Positions are integers, so a move is expanding only if the two chips are at the same position.

    \ref{item:DoubleCoverBound}
    There are $2n$ chips and at most $2T$ expanding moves, and the largest chip position at a complete time is $\norm{S_t}_\infty$.
    \Cref{thm:ChipGame}\ref{item:ChipBound} gives the claim.

    \ref{item:CubicWitness}
    Shift all chip positions by $-b$; the shifted chips start at $-b\le0$, and the game is translation invariant.
    At the end of arrival $t_0$, the largest shifted position is $u$, and $u\le2n$ by part~\ref{item:DoubleCoverBound}.
    Apply \cref{thm:ChipGame}\ref{item:ChipSpread} with $D=u$ and $r\colonequals\floor{u/2}$.
    At some chip time not after the end of arrival $t_0$, the $r$ largest shifted positions have sum at least $r(u-r+1)$, and by convexity, as in the proof of \cite[Lemma~3.1]{BansalCarpool26}, the shifted positive energy $\Ec\colonequals\sum_{\mathrm{chips}}(\text{shifted position})_+^2$ is at least $r(u-r+1)^2\ge\frac{3u}8\big(\frac u2\big)^2\ge u^3/32$ there, since $u\ge4$.
    At complete times $\Ec=F_b$: of the two chips of coordinate $i$, one sits at $h_i$ and contributes $(h_i-b)_+^2$, and the other sits at $-h_i<b$ and contributes nothing.

    It remains to compare an intermediate chip time, between the two moves of an arrival, with the adjacent complete times.
    Let $\Delta_1$ and $\Delta_2$ be the changes of $\Ec$ caused by the first and the second move.
    The second move displaces the mirror images of the chips displaced by the first move, in the opposite directions.
    If $\Delta_1>0$, the chip moving right in the first move starts at a shifted position at least $0$, that is, at an unshifted position at least $b\ge1$; its mirror sits at a position at most $-1$ and moves left, contributing nothing, while the other mirror moves right and cannot decrease $\Ec$.
    Hence $\Delta_1>0$ implies $\Delta_2\ge0$, so the intermediate value of $\Ec$ is at most the value before the arrival if $\Delta_1\le0$, and at most the value after it if $\Delta_1>0$.
    In either case there is a complete time $t'\le t_0$ with $F_b(t')\ge u^3/32$.
\end{proof}

By \cref{lem:DoubleCover}\ref{item:CubicWitness}, no height can reach $b+u$ unless the excess energy above $b$ has first accumulated to order $u^3$.
The remaining subsections show that this accumulation is unlikely as long as few coordinates exceed $b$, and that this population condition holds throughout the horizon at a level $b$ that is small enough for the claimed radius: $b=O(\log\log n+u)$ if $L\le\sqrt n$, and $b=O(\log n+u)$ otherwise.

\subsection{Cubic excursions}
\label{sec:CubicExcursions}

We now show that the energy accumulation required by \cref{lem:DoubleCover}\ref{item:CubicWitness} is unlikely while at most $q$ coordinates have height at least $b$.

\begin{lemma}[Cubic excursions]
    \label{lem:CubicExcursions}
    Let $b\ge1$ and $u\ge8$ be integers, let $1\le q\le n/2$, and suppose that $z\colonequals\sqrt{nu}/(64q)\ge2$.
    In both models,
    \begin{equation}
        \label{eq:CubicExcursions}
        \P\Big\{D_T\ge b+u,\ \max_{t\le T}N_b(t)\le q\Big\}\le T\exp\Big\{-\frac{u^3\log z}{128}\Big\}.
    \end{equation}
\end{lemma}

\begin{proof}
    Let $\Ac_t$ be the event that $N_b(t)\le q$ and all heights at time $t$ are smaller than $b+u$, and consider a pre-state in $\Ac_t$ with $F_b>u^3/64$.
    Every excess $(h_i-b)_+$ is then smaller than $u$, so $\mathsf m\colonequals N_{b+1}>F_b/u^2>u/64$.
    By \ref{item:OutcomePositive}, \ref{item:OutcomeZero} and the remark on ties after them, $F_b$ can increase only on a tie between two equal heights at least $b$, and then by at most two; such a pair consists of two coordinates of height at least $b$, so this happens with probability at most $q(q-1)/[n(n-1)]$.
    A coordinate with positive excess paired with a coordinate of height smaller than $b$ moves inward and decreases $F_b$ by at least one; this happens with probability at least $2\mathsf m(n-q)/[n(n-1)]$.
    Consequently
    \begin{equation*}
        \E\big[z^{F_b(t+1)-F_b(t)}\bigm|\Fc_t\big]\le1+(z^2-1)\frac{q(q-1)}{n(n-1)}-(1-z^{-1})\frac{2\mathsf m(n-q)}{n(n-1)}\le1,
    \end{equation*}
    because the last inequality is equivalent to $z(z+1)q(q-1)\le2\mathsf m(n-q)$, whose left side is at most $\frac32z^2q^2=3nu/8192$ and whose right side exceeds $nu/64$.
    \Cref{lem:Excursions} with $\Phi=F_b$, $K_\Phi=2$, $E_-=u^3/64$ and $E_+=u^3/32$ shows that, except with probability $Tz^{2-u^3/64}\le T\exp\{-u^3\log z/128\}$, there is no time $t'\le T$ such that $F_b(t')\ge u^3/32$ and $\Ac_t$ holds for all $t<t'$; here we used $u\ge8$.
    On the event in \eqref{eq:CubicExcursions}, let $t_0\le T$ be the first time at which a height reaches $b+u$.
    By \cref{lem:DoubleCover}\ref{item:CubicWitness}, $F_b(t')\ge u^3/32$ for some $t'\le t_0$, and $\Ac_t$ holds for all $t<t'$.
    Hence the event in \eqref{eq:CubicExcursions} is contained in this exceptional event.
\end{proof}

\subsection{Moments of the heights}
\label{sec:GeometricMoments}

In the binary model we use twice the following elementary inequality: for a positive semidefinite matrix $(K_{ij})$ and signs $\sigma_i\in\{\pm1\}$,
\begin{equation}
    \label{eq:PSD}
    \sum_{i<j}\sigma_i\sigma_jK_{ij}\ge-\frac12\sum_iK_{ii},
\end{equation}
since $\sum_{i,j}\sigma_i\sigma_jK_{ij}\ge0$.

\begin{lemma}[Geometric moments]
    \label{lem:GeometricMoments}
    Let $n\ge10$ and put $Y_i(t)\colonequals2^{h_i(t)}$.
    In both models,
    \begin{equation}
        \label{eq:GeometricMoments}
        \begin{aligned}
            \E e_r(Y(t))&\le32^r\binom nr&&(0\le r\le n,\ t\ge0),\\
            \P\{N_b(t)\ge r\}&\le\Big(\frac{32en}r\Big)^r2^{-br}&&(b\ge0,\ 1\le r\le n).
        \end{aligned}
    \end{equation}
\end{lemma}

\begin{proof}
    We track the generating polynomial $\prod_i(1+\xi Y_i)$ in a formal variable $\xi$.
    Consider first the signed model, and let the selected heights be $a\ge c$ with $\mathsf A\colonequals2^a$ and $\mathsf B\colonequals2^c$.
    If $c\ge1$, the outcomes \ref{item:OutcomePositive} replace $(\mathsf A,\mathsf B)$ by $(\mathsf A/2,\mathsf B/2)$ or $(\mathsf A/2,2\mathsf B)$ with equal probabilities, so the expected change of $(1+\xi\mathsf A)(1+\xi\mathsf B)$ is $\xi(-\mathsf A/2+\mathsf B/4)-\frac38\xi^2\mathsf A\mathsf B$.
    For $c=0<a$ it is $\xi(1-\mathsf A/2)$, and for $a=c=0$ it is $2\xi+3\xi^2$.
    Since $\mathsf B\le\mathsf A$, in all three cases the expected change is coefficientwise at most
    \begin{equation*}
        -\frac18\big\{\xi(\mathsf A+\mathsf B)+2\xi^2\mathsf A\mathsf B\big\}+2\xi\big\{(1+\xi\mathsf A)+(1+\xi\mathsf B)\big\}.
    \end{equation*}
    Multiply by the product over the unselected coordinates and average over the uniform pair.
    In the first term, a monomial of degree $r$ is charged once for each of its variables that is selected, which happens on average $2r/n$ times.
    In the second term, the coefficient of $\xi^r$ is $2\{e_{r-1}(Y_{-j})+e_{r-1}(Y_{-i})\}$, whose average is $\frac4n\sum_ie_{r-1}(Y_{-i})$.
    By \eqref{eq:ElementaryIdentities}, the signed model satisfies
    \begin{equation*}
        \E\big[e_r(Y(t+1))-e_r(Y(t))\bigm|\Fc_t\big]\le-\frac r{4n}e_r(Y(t))+\frac{4(n-r+1)}ne_{r-1}(Y(t)).
    \end{equation*}

    In the binary model the outcome for a pair of positive heights is deterministic.
    Let $P\colonequals\{i:h_i>0\}$ and $\sigma_i\colonequals\sgn S_i$ for $i\in P$.
    A direct comparison with the signed average shows that, on a selected pair $\{i,j\}\subseteq P$, the binary change of $(1+\xi Y_i)(1+\xi Y_j)$ equals the signed average plus
    \begin{equation*}
        -\sigma_i\sigma_j\Big\{\tfrac34\xi\min(Y_i,Y_j)+\tfrac38\xi^2Y_iY_j\Big\},
    \end{equation*}
    and that there is no correction if a selected height is zero.
    After multiplication by the other factors and summation over pairs, the total correction to the coefficient of $\xi^r$ is $-\binom n2^{-1}\{\frac34U_r+\frac38W_r\}$, where
    \begin{equation*}
        U_r\colonequals\sum_{\{i,j\}\subseteq P}\sigma_i\sigma_j\min(Y_i,Y_j)\,e_{r-1}(Y_{-\{i,j\}}),\qquad W_r\colonequals\sum_{\{i,j\}\subseteq P}\sigma_i\sigma_jY_iY_j\,e_{r-2}(Y_{-\{i,j\}}).
    \end{equation*}
    Both are at least $-\frac r2e_r(Y)$.
    For $U_r$, expand $e_{r-1}(Y_{-\{i,j\}})$ over sets $I$ of size $r-1$.
    For fixed $I$, the kernel $\min(Y_i,Y_j)=\int_0^\infty\mathbf1\{y\le Y_i\}\mathbf1\{y\le Y_j\}\,dy$ is positive semidefinite on $P\setminus I$, so by \eqref{eq:PSD} its signed off-diagonal sum over pairs is at least $-\frac12\sum_{i\in P\setminus I}Y_i$.
    Weighting by $\prod_{k\in I}Y_k$ and summing over $I$ gives at least $-\frac r2e_r(Y)$.
    For $W_r$, expand $e_{r-2}(Y_{-\{i,j\}})$ over sets $I'$ of size $r-2$ and group the terms according to the $r$-set $J\colonequals\{i,j\}\cup I'$; then use $\sum_{\{i,j\}\subseteq J\cap P}\sigma_i\sigma_j=\frac12\{(\sum_{i\in J\cap P}\sigma_i)^2-\abs{J\cap P}\}\ge-r/2$.
    Hence the binary correction is at most $\frac9{16}re_r/\binom n2\le\frac r{8n}e_r$ for $n\ge10$, and both models satisfy
    \begin{equation*}
        \E\big[e_r(Y(t+1))\bigm|\Fc_t\big]\le\Big(1-\frac r{8n}\Big)e_r(Y(t))+\frac{4(n-r+1)}ne_{r-1}(Y(t)).
    \end{equation*}
    \Cref{lem:TriangularRecursion} with $\mathsf u=1/(8n)$, $\mathsf v=4/n$, $\mathsf M=32$, $r_*=n$ and $\varphi(h)=2^h$ proves \eqref{eq:GeometricMoments}.
\end{proof}

The second bound in \eqref{eq:GeometricMoments} shows that, at each fixed time, at most $q$ coordinates exceed the level $\log_2(32en/q)+O(L/q)$, except with probability $e^{-L}$.
\Cref{lem:LayerIteration} then reduces the population to $O(L)$ within $O(\log\log n)$ further levels.

\subsection{A global quadratic moment}
\label{sec:GlobalQuadratic}

For $L>\sqrt n$, a single exponential moment controls the populations above a level $b_*\asymp\log n$ uniformly in time; in this range $\log n$ is at most of order $(L/\log n)^{1/3}$, so no layer iteration is needed.

\begin{lemma}[Global quadratic moment]
    \label{lem:GlobalQuadratic}
    Let $n\ge65$.
    Define slopes $d_1\colonequals1/(64n)$ and $d_{h+1}\colonequals d_h+\min\{d_h,1/16\}$, put $f(0)\colonequals0$ and $f(h)\colonequals\sum_{j=1}^hd_j$, and let $b_*\colonequals1+\ceil{\log_2(4n)}$ and $K_\star\colonequals3e^{32}$.
    In both models,
    \begin{equation}
        \label{eq:GlobalQuadratic}
        \sup_{t\ge0}\E\exp\Big\{\sum_{i=1}^nf(h_i(t))\Big\}\le K_\star,\qquad f(h)\ge\frac{(h-b_*)_+^2}{32}.
    \end{equation}
    Consequently, for all integers $r,v\ge1$,
    \begin{equation}
        \label{eq:GlobalQuadraticPopulation}
        \P\Big\{\max_{t\le T}N_{b_*+v}(t)\ge r\Big\}\le K_\star Te^{-rv^2/32}.
    \end{equation}
\end{lemma}

\begin{proof}
    The slopes double until they exceed $1/16$ and then increase by $1/16$ per step.
    If $d_h<1$, geometric summation before this crossing and arithmetic summation afterwards give $f(h)\le2d_h+16d_h^2\le18d_h$.
    Since $d_{b_*}\ge4n\,d_1=1/16$, the slopes beyond $b_*$ increase by $1/16$ per step, so $f(b_*+v)\ge\frac1{16}\sum_{j=1}^vj\ge v^2/32$.

    Write $\Psi\colonequals\sum_if(h_i)$, $Z\colonequals e^\Psi$, and $G\colonequals\sum_{i:h_i\ge2}\min\{d_{h_i},1\}$.
    Consider the signed model and a pair of positive selected heights $a\ge c$.
    By \ref{item:OutcomePositive}, the conditional expectation of the ratio $Z'/Z$ is
    \begin{equation*}
        \mathsf R(a,c)\colonequals\tfrac12e^{-d_a-d_c}+\tfrac12e^{-d_a+d_{c+1}}.
    \end{equation*}
    If $a>c$, then $d_a\ge d_{c+1}$ and $\mathsf R\le(1+e^{-d_a})/2\le1-\frac14\min\{d_a,1\}$.
    If $a=c=h$, put $d\colonequals d_h$; then $\mathsf R=(e^{-2d}+e^{\min\{d,1/16\}})/2$.
    For $d\le1/16$, Taylor bounds give $\mathsf R\le1-d/2+3d^2/2\le1-d/4$.
    For $1/16\le d\le1$, the function $1-(e^{-2d}+e^{1/16})/2-d/4$ is concave and positive at both endpoints, and for $d\ge1$ the deficit $1-\mathsf R$ is at least $1/4$.
    Hence $\mathsf R(a,c)\le1-\frac14\min\{d_a,1\}$ for every pair of positive heights.
    For one zero the ratio is $e^{-d_a+d_1}$, which equals one for $a=1$ and is at most $1-\frac14\min\{d_a,1\}$ for $a\ge2$.
    Two zeros give the ratio $e^{2d_1}\le1+4d_1$.
    Averaging over the uniform pair, note that for a pair whose larger height $a$ is at least two, the deficit $\frac14\min\{d_a,1\}$ is at least $\frac18$ times the sum of $\min\{d_{h_k},1\}$ over its coordinates $k$ with $h_k\ge2$, since the slopes are nondecreasing; each coordinate of height at least two lies in $n-1$ pairs, so
    \begin{equation}
        \label{eq:GlobalSignedDrift}
        \E\big[Z_{t+1}/Z_t\bigm|\Fc_t\big]\le1+\frac1{16n}-\frac G{4n}.
    \end{equation}

    In the binary model, a pair of positive heights $a\ge c$ has the deterministic ratio $e^{-d_a-d_c}$ for equal signs and $e^{-d_a+d_{c+1}}$ for opposite signs.
    The difference from the signed average is $-\sigma_i\sigma_j\mathsf D(a,c)/2$, where
    \begin{equation*}
        \mathsf D(a,c)\colonequals e^{-d_a}\big(e^{d_{c+1}}-e^{-d_c}\big)=\mathsf g(a)\mathsf g(c)\min\{\mathsf k(a),\mathsf k(c)\},\qquad\mathsf g(h)\colonequals e^{-d_h},\quad\mathsf k(h)\colonequals e^{d_h+d_{h+1}}-1.
    \end{equation*}
    Since $\mathsf k$ is increasing and nonnegative, $\mathsf D$ is a positive semidefinite kernel, and by \eqref{eq:PSD} its signed off-diagonal sum is at least minus half its diagonal sum.
    Moreover $\mathsf D(h,h)=e^{\min\{d_h,1/16\}}-e^{-2d_h}\le4\min\{d_h,1\}$, and the coordinates of height one contribute at most $4nd_1=1/16$ to the diagonal sum.
    Hence the binary correction to \eqref{eq:GlobalSignedDrift} is at most $(G+1/64)/\binom n2$.
    For $n\ge65$, both models therefore satisfy
    \begin{equation*}
        \E\big[Z_{t+1}/Z_t\bigm|\Fc_t\big]\le1+\frac{129}{2048\,n}-\frac{7G}{32n}.
    \end{equation*}
    If $\Psi\ge32$, then $G\ge1$.
    Indeed, if some slope $d_{h_i}$ is at least one, this is immediate; otherwise $f(h)\le18d_h$ for all heights, so $32\le\Psi\le nd_1+18G=1/64+18G$.
    The ratio is therefore at most $1-1/(8n)$ when $\Psi\ge32$, and at most $1+1/(8n)$ in general.
    Hence, at every state,
    \begin{equation*}
        \E\big[Z_{t+1}\bigm|\Fc_t\big]\le\Big(1-\frac1{16n}\Big)Z_t+\frac{3e^{32}}{16n},
    \end{equation*}
    and iteration from $Z_0=1$ gives \eqref{eq:GlobalQuadratic}.
    Finally, $N_{b_*+v}(t)\ge r$ implies $\Psi\ge rv^2/32$, and Markov's inequality with a union bound over $1\le t\le T$ gives \eqref{eq:GlobalQuadraticPopulation}.
\end{proof}

\subsection{Proof of Theorem~\ref{thm:Prefix}\ref{item:PrefixPairs}}
\label{sec:ProofPairs}

Let $n\ge e^{256}$, $T\ge n$ and $0<\delta<1$, and suppose that $L=\log(16n(T+1)/\delta)\le n^2$; note that $L\ge2\log n$.
We consider two ranges of $L$.
For $L\le\sqrt n$, the population at a level $b=O(\log\log n+u)$ is controlled by \cref{lem:GeometricMoments} and the layer iteration.
For $L>\sqrt n$, the global quadratic moment of \cref{lem:GlobalQuadratic} controls the population above a level of order $\log n$, which is at most of order $(L/\log n)^{1/3}$ in this range.

\paragraph{The range $L\le\sqrt n$.}
\emph{Parameters.}
Put
\begin{equation*}
    \nu\colonequals\log(n/L),\qquad u\colonequals\ceil{64(L/\nu)^{1/3}},\qquad q\colonequals u^2,\qquad z\colonequals\sqrt{nu}/(64q)=\sqrt{n/u^3}/64.
\end{equation*}
Then $\nu\ge\frac12\log n\ge128$ and $L/\nu\ge1$, so $64(L/\nu)^{1/3}\le u\le65(L/\nu)^{1/3}$.
Moreover $q/n\le65^2\nu^{-2/3}n^{-2/3}<1/512$, and
\begin{equation*}
    \log z\ge\frac\nu2+\frac12\log\nu-\frac32\log65-\log64\ge\frac\nu4.
\end{equation*}

\emph{Populations.}
Put $q_0\colonequals\floor{n/(16A_{\mathrm{cl}})}$, $R\colonequals\ceil L$ and
\begin{equation*}
    b_0\colonequals\ceil{\log_2(32en/q_0)+L/(q_0\log2)}\le40.
\end{equation*}
By \eqref{eq:GeometricMoments} and a union bound over $0\le t\le T$, the event $\Vc\colonequals\{N_{b_0}(t)\le q_0\text{ for all }t\le T\}$ fails with probability at most $(T+1)e^{-L}$.
Apply \cref{lem:LayerIteration} with $g=0$, $R$ and $q_0$, which satisfy $2R<q_0\le n/(8A_{\mathrm{cl}}s)$ with $s=2$, with $\Gc_t\colonequals\{N_{b_0}(t)\le q_0\}$, and with $t_0=\cdots=t_{J+1}=0$ and $M=0$; every state is $(b,0)$-isolating.
With $b_j=b_0+2j$, all conditions $N_{b_j}(t)\le q_j$, $1\le j\le J$, $t\le T$, hold on $\Vc$ outside an event of probability at most $J(T+1)e^{-c_*R}\le J(T+1)e^{-L}$, and $q_J\le2R\le3L$.

\emph{The final level.}
If $q_J\le q$, put $d\colonequals0$; otherwise put
\begin{equation*}
    d\colonequals\ceil{\frac{\log(8eq_J/q)+L/q}{\log\vartheta_J}},
\end{equation*}
and note that \eqref{eq:LayerFinal} with $r=q$ and a union bound over $t\le T$ show that $N_{b_J+d}(t)<q$ for all $t\le T$, outside an event of probability at most $(T+1)e^{-L}$.
Put $b\colonequals b_J+d$ and $\Cc_{\mathrm{fin}}\colonequals\{N_b(t)\le q\text{ for all }t\le T\}$.
Then $\P(\Cc_{\mathrm{fin}}^c)\le(J+2)(T+1)e^{-L}\le\delta/8$, since $(T+1)e^{-L}=\delta/(16n)$ and $J+2\le n$.
To bound $d$, note that $\log\vartheta_J=\frac12\log(n/(2C_{\mathrm{cl}}q_J))\ge\frac12(\nu-\log(6C_{\mathrm{cl}}))\ge\nu/4$.
If $d\ge1$, then $x\colonequals L/q>1/3$, since $q<q_J\le3L$, and the numerator in the definition of $d$ is at most $\log(24ex)+x\le15x$.
Since $L/(u^2\nu)\le(L/\nu)^{1/3}/4096\le u/2^{18}$, we obtain
\begin{equation*}
    d\le1+\frac{60L}{u^2\nu}\le1+\frac u{4096},\qquad b\le41+2J+\frac u{4096}.
\end{equation*}

\emph{The excursion.}
Finally, \cref{lem:CubicExcursions} applies to $b,u,q,z$, since $u\ge8$, $q\le n/2$ and $z\ge2$, and gives
\begin{equation*}
    \P\big(\{D_T\ge b+u\}\cap\Cc_{\mathrm{fin}}\big)\le T\exp\{-u^3\nu/512\}\le Te^{-512L}<\delta/8.
\end{equation*}
Hence $D_T<b+u\le41+2J+2u$ with probability at least $1-\delta/4$.
Since $\nu\ge\frac12\log n$ and $J=O(\log\log n)$, this radius is $O(\log\log n+(L/\log n)^{1/3})$.

\paragraph{The range $L>\sqrt n$.}
Write $\ell\colonequals\log n$ and $\Lambda\colonequals\log(8K_\star(T+1)/\delta)$, with $K_\star$ from \cref{lem:GlobalQuadratic}.
Then $\Lambda=L-\log(2n/K_\star)<L\le n^2$, and $\Lambda\ge\sqrt n-\ell\ge\ell^4$ because $n\ge e^{256}$.
Put
\begin{equation*}
    w\colonequals(\Lambda/\ell)^{1/3},\qquad u\colonequals\ceil{128w},\qquad q\colonequals\ceil{64\Lambda/u^2},\qquad b\colonequals b_*+u,\qquad z\colonequals\sqrt{nu}/(64q).
\end{equation*}
Then $w\ge\ell$, $128w\le u\le129w$, and $64\Lambda/u^2\le q\le65\Lambda/u^2$, because $\Lambda/u^2\ge\ell^2/129^2\ge1$.
The bound $\Lambda\le n^2$ gives $q/n\le(65/128^2)n^{-1/3}\ell^{2/3}<1/2$ and $u<n$.
Moreover,
\begin{equation*}
    \log z\ge\frac\ell2-\frac16\log\Lambda-\frac56\log\ell+\frac52\log128-\log4160\ge\frac\ell6-\frac56\log\ell+\frac52\log128-\log4160\ge\frac\ell{12},
\end{equation*}
where the last inequality holds at $\ell=256$ and the difference of its two sides is increasing in $\ell$.
\Cref{lem:GlobalQuadratic}, applied with $u$ in place of the offset $v$ in \eqref{eq:GlobalQuadraticPopulation} and with $r=q+1$, and \cref{lem:CubicExcursions} give
\begin{equation*}
    \P\Big\{\max_{t\le T}N_b(t)>q\Big\}\le K_\star Te^{-2\Lambda},\qquad\P\Big\{D_T\ge b+u,\ \max_{t\le T}N_b(t)\le q\Big\}\le Te^{-4\Lambda}.
\end{equation*}
Their sum is smaller than $\delta$.
Since $b_*\le2\ell\le2w$ and $u\le129w$, the radius $b+u$ is at most $260w\le260(L/\log n)^{1/3}$.

Every admissible $(T,\delta)$ falls into one of the two ranges, and the same rule is used throughout.
This proves \cref{thm:Prefix}\ref{item:PrefixPairs}.
\qed

Quadratic Greedy also satisfies a bound outside the range of \cref{thm:Prefix}\ref{item:PrefixPairs}.

\begin{proposition}[All horizons and failure probabilities]
    \label{prop:PairsAllParameters}
    Let $n\ge65$, $T\ge1$ and $0<\delta<1$, let $b_*$ and $K_\star$ be as in \cref{lem:GlobalQuadratic}, and put $u\colonequals\ceil{(1024\log(K_\star T/\delta))^{1/3}}$.
    In both models, the prefix discrepancy $D_T$ of quadratic Greedy satisfies $\P\{D_T\ge b_*+u\}\le\delta$.
\end{proposition}

\begin{proof}
    Since $K_\star\ge e^{32}$, we have $u\ge32$, so \cref{lem:DoubleCover}\ref{item:CubicWitness} applies with $b=b_*$.
    If $D_T\ge b_*+u$, it gives a time $1\le t'\le T$ with $F_{b_*}(t')\ge u^3/32$, and then $\sum_if(h_i(t'))\ge u^3/1024\ge\log(K_\star T/\delta)$ by \eqref{eq:GlobalQuadratic}.
    By Markov's inequality and a union bound over $t'$, this has probability at most $TK_\star\cdot\delta/(K_\star T)=\delta$.
\end{proof}

\section{Upper bounds for fixed sparsity}
\label{sec:FixedSparsity}

In this section we prove \cref{thm:Prefix}\ref{item:PrefixFixed}.
Fix an integer $s\ge3$.
We use one family of rules and one proof for both models.
\Cref{sec:FreshOrientations} reduces both models to estimates averaged over the support, \cref{sec:StoppedMoments} controls the number of high coordinates over the whole horizon, \cref{sec:Profile,sec:Layers} bound the excursions of a potential, by a tilted supermartingale for $L\ge\sqrt n$ and by a squared hinge for $L<\sqrt n$, and \cref{sec:ProofFixedSparsity} combines these estimates into the proof of \cref{thm:Prefix}\ref{item:PrefixFixed}.
Throughout, we use the constants
\begin{equation}
    \label{eq:FixedConstants}
    \rho\colonequals\frac1{10^4s^2},\qquad\kappa\colonequals s\rho,\qquad K\colonequals s\kappa=10^{-4};
\end{equation}
here $\rho$ is the growth rate of the core weights, $\kappa$ the slope of the tail weights, and $K$ bounds the one-step increment of the potential of \cref{sec:Profile}.

\paragraph{The rule.}
Fix an integer $B\ge1$, the height at which the weights change from exponential to linear growth; we call it the \emph{knee}.
It is chosen deterministically from $n$, $s$ and $L$ in \cref{sec:ProofFixedSparsity}.
Define weights
\begin{equation}
    \label{eq:Weights}
    a_0\colonequals0,\qquad a_h\colonequals e^{\rho(h-B)}\quad(1\le h\le B),\qquad a_{B+j}\colonequals1+\kappa j\quad(j\ge0),
\end{equation}
which consist of an exponential \emph{core} below the knee and a linear \emph{tail} above it; both definitions give $a_B=1$.
The core is \emph{flat} in the sense that it grows slowly: consecutive weights differ by the factor $e^\rho=1+O(s^{-2})$.
For an arrival $V$ with support $A$, the rule chooses
\begin{equation}
    \label{eq:WeightRule}
    x\colonequals-\sgn\Big(\sum_{i\in A}V(i)\sgn(S_i)\,a_{h_i}\Big),
\end{equation}
and $x\colonequals+1$ if the sum vanishes, where $V(i)=1$ in the binary model and $V(i)$ is the observed mark in the signed model.
We call the sum in \eqref{eq:WeightRule} the \emph{projection} of the arrival.
The rule is deterministic and its parameters never change during a run.
The linear tail makes the potential of \cref{sec:Profile} grow quadratically above the knee.

\subsection{Fresh orientations}
\label{sec:FreshOrientations}

All one-step estimates below are averaged over the uniform support, for an arbitrary fixed orientation of the marks; in the binary model, drift bounds cannot hold support by support (\cref{rem:BinaryTransfer}).
For a selected coordinate $i$ with $h_i>0$, put $\sigma_i\colonequals V(i)\sgn(S_i)$.
Then the height of a selected coordinate changes by $x\sigma_i$ if $h_i>0$ and by $+1$ if $h_i=0$.

\begin{lemma}[Fresh orientations]
    \label{lem:FreshOrientations}
    Fix the past up to time $t$, and let $F$ be a function of the next arrival.
    For $\xi\in\{\pm1\}^n$, called an \emph{orientation}, let $F_\xi(A)$ be the value of $F$ for the arrival with support $A$ and marks $\xi_i$, $i\in A$.
    If $\E_AF_\xi(A)\le c$ for every orientation $\xi$, where $A$ is a uniform $s$-subset of $[n]$, then $\E[F\mid\Fc_t]\le c$ in both models.
\end{lemma}

\begin{proof}
    In the binary model all marks are equal to one, so $\E[F\mid\Fc_t]=\E_AF_\xi(A)$ with $\xi=(1,\ldots,1)$.
    In the signed model, generate the arrival by drawing a vector $\xi\in\{\pm1\}^n$ of Rademacher marks, independent of the past and of $A$, and revealing only its restriction to $A$; this has the required law, so $\E[F\mid\Fc_t]=\E_\xi\E_AF_\xi(A)$.
\end{proof}

\begin{remark}[Binary inputs]
    \label{rem:BinaryTransfer}
    In the signed model, the projection is anti-concentrated already for each fixed support, by \eqref{eq:SignedAC} applied to the marks.
    In the binary model, a fixed support can make every symmetric potential that is strictly increasing in each height grow for both signs, even at reachable states.
    For instance, let $s=3$.
    Starting from $S_0=0$, the rule \eqref{eq:WeightRule} turns the supports $\{1,3,4\}$ and $\{2,3,4\}$ into the state $e_1-e_2$.
    For the next support $\{1,2,5\}$ the projection vanishes, and the signs $x=+1$ and $x=-1$ turn the heights $(1,1,0)$ of the selected coordinates into $(2,0,1)$ and $(0,2,1)$, respectively.
    In both cases the multiset of heights changes from $\{1,1,0\}$ to $\{2,1,0\}$, so the potential $\Psi$ of \cref{sec:Profile} and the moments $E_r$ of \cref{lem:StoppedMoments} increase for both signs.
    The same distinction appears in the work of Altschuler and Tikhomirov: their sampling estimate averages over the support and holds for arbitrary signs, since for binary inputs the signs of the running discrepancies are not independent, whereas random marks as in \cite{Bansal20a} would allow a central limit argument \cite[Section~3.2 and Lemma~3.4]{Altschuler26Binary}.
\end{remark}

\subsection{Stopped geometric moments}
\label{sec:StoppedMoments}

We use the constants
\begin{equation}
    \label{eq:MomentConstants}
    c_s\colonequals\frac{\sqrt s}{144},\qquad\e_s\colonequals\frac{c_s}{128s^3},\qquad M_s\colonequals\frac{8s}{c_s}=1152\sqrt s,
\end{equation}
where $c_s$ is the anti-concentration constant of \eqref{eq:AveragedService}.
The next lemma controls the number of coordinates above a level $b\le B$ over the whole horizon.
The difficulty, compared with \cref{lem:GeometricMoments}, is that the rule \eqref{eq:WeightRule} uses the weights $a_h$, which differ from the exponential weights in the tail, and that the drift bound must hold simultaneously for monomials of every degree up to the maximal degree $r_*$ in the lemma.

\begin{lemma}[Stopped geometric moments]
    \label{lem:StoppedMoments}
    Assume $n\ge2s$, $1\le b\le B$ and $1\le q\le r_*\colonequals\floor{\e_sn}$.
    For the rule \eqref{eq:WeightRule} started at $S_0=0$, let $\tau\colonequals\inf\{t\ge0:N_b(S_t)>q\}$.
    In both models, for every integer $T\ge1$,
    \begin{equation}
        \label{eq:StoppedPopulation}
        \P(\tau\le T)\le T\Big(\frac{M_sen}qe^{-\rho b}\Big)^q.
    \end{equation}
\end{lemma}

\begin{proof}
    Put $Y_i\colonequals e^{\rho h_i}$ and $E_r(S)\colonequals e_r(Y_1,\ldots,Y_n)$.
    We prove a one-step drift bound for $E_r$ at a pre-state with $\tau>t$, for every fixed orientation as in \cref{lem:FreshOrientations}.
    Fix $1\le r\le r_*$, abbreviate $E\colonequals E_r(S)$, and let $\chi_i$ and $v_i$ be the weights \eqref{eq:NewtonWeights} of $Y$.

    \emph{Step 1: service of the moment weights.}
    Apply \cref{lem:NewtonService} with $\mathsf k(y)\colonequals\mathbf1\{y>1\}$ and $\e_0\colonequals0$; thus $\pi_i=\chi_i$ and $\hat w_i=v_i$ at positive heights, and $\pi_i=\hat w_i=0$ at height zero.
    For a support $A$, let $f_Y$ be a sign with $f_Y\sum_{i\in A}\sigma_i\hat w_i\ge0$.
    By \eqref{eq:AveragedService} with $\mathsf C=[n]$,
    \begin{equation}
        \label{eq:StoppedService}
        \E_A\Big[f_Y\sum_{i\in A}\sigma_i\pi_i\Big]\ge\Big[\frac{c_s}n-\frac{8s(s-1)r}{n(n-1)}\Big]\sum_{h_i>0}\chi_i.
    \end{equation}

    \emph{Step 2: comparison with the actual decision.}
    Let $g_i\colonequals E\chi_i=Y_ie_{r-1}(Y_{-i})$ for $h_i>0$ and $g_i\colonequals0$ otherwise, and let $G_+\colonequals\sum_ig_i$.
    By \cref{lem:NewtonService}\ref{item:NewtonBasic}, the $g_i$ increase and the ratios $g_i/Y_i=e_{r-1}(Y_{-i})$ decrease with the height.
    Let $f\colonequals-x$ be the actual sign.
    The weights $v_i$ are proportional to $Y_i$, and $Y_i$ and $a_{h_i}$ are proportional on the core $1\le h_i\le B$.
    Call a coordinate with $h_i>B$ a \emph{tail coordinate}, and let $\mathsf m$ be the number of tail coordinates.
    If no tail coordinate is selected, then $\hat w_i$ and $a_{h_i}$ are proportional on $A$, so $f_Y=f$, provided that $f_Y$ is chosen equal to $f$ at ties.

    Suppose exactly one tail coordinate $i_0$ is selected.
    If $g_{i_0}\ge\sum_{i\in A\setminus\{i_0\}}g_i$, then the monotonicity of $Y_i/g_i$ at positive heights gives $Y_{i_0}\ge\sum_{i\in A\setminus\{i_0\},\,h_i>0}Y_i$, so $i_0$ dominates the exponential weights, which vanish at height zero.
    It also dominates the actual weights: if $a_{h_{i_0}}>s-1$, this is clear, since core weights are at most one; otherwise, with $j\colonequals h_{i_0}-B$, we have $1+\kappa j\le s-1$ and
    \begin{equation*}
        \log(1+\kappa j)=\int_0^j\frac\kappa{1+\kappa y}\,dy\ge\frac{\kappa j}{s-1}>\rho j,
    \end{equation*}
    so $a_{h_{i_0}}>e^{-\rho B}Y_{i_0}\ge\sum_{i\in A\setminus\{i_0\}}a_{h_i}$, since $a_0=0$.
    Then the actual decision has the sign $\sigma_{i_0}$, and its service $f\sum_A\sigma_ig_i$ equals $\lvert\sum_A\sigma_ig_i\rvert$.
    If instead $g_{i_0}<\sum_{i\in A\setminus\{i_0\}}g_i$, a sign change costs at most $2\lvert\sum_A\sigma_ig_i\rvert\le4\sum_{i\in A\setminus\{i_0\}}g_i$.
    If two or more tail coordinates are selected, we use the crude bound $2\sum_Ag_i$.
    Thus a loss occurs only if $g_{i_0}<\sum_{i\in A\setminus\{i_0\}}g_i$ or if at least two tail coordinates are selected.
    In both cases the loss is at most $4\sum g_i$, where the sum runs over those $i\in A$ for which $A$ contains a tail coordinate other than $i$.
    For each $i$, this happens with probability at most $s(s-1)\mathsf m/[n(n-1)]$.
    Multiplying \eqref{eq:StoppedService} by $E$, we obtain
    \begin{equation*}
        \E_A\Big[f\sum_{i\in A}\sigma_ig_i\Big]\ge\Big[\frac{c_s}n-\frac{8s(s-1)r+4s(s-1)\mathsf m}{n(n-1)}\Big]G_+.
    \end{equation*}
    Before $\tau$, every tail coordinate has height at least $b$, so $\mathsf m\le q\le\e_sn$; also $r\le\e_sn$.
    The error terms are therefore at most $24\e_ss^2G_+/n\le c_sG_+/(8n)$, and
    \begin{equation}
        \label{eq:ActualService}
        \E_A\Big[f\sum_{i\in A}\sigma_ig_i\Big]\ge\frac{c_s}{2n}G_+.
    \end{equation}

    \emph{Step 3: expansion.}
    Let $G_0\colonequals\sum_{h_i=0}Y_ie_{r-1}(Y_{-i})$.
    Then $G_++G_0=rE$ and $G_0\le(n-r+1)E_{r-1}(S)$ by \eqref{eq:ElementaryIdentities}.
    For a monomial $Y_I$ of degree $r$, the logarithmic change is $\rho D_I$, where $D_I$ is the sum over $I\cap A$ of the height increments; these are $-f\sigma_i$ at positive heights and $+1$ at height zero.
    Thus $D_I^2\le s\abs{I\cap A}$, and $e^{\rho D}\le1+\rho D+\frac12\rho^2D^2e^{s\rho}$ gives, after summation over $I$ and averaging over $A$,
    \begin{equation*}
        \E\big[E_r(S_{t+1})-E\bigm|\Fc_t,\xi\big]\le-\frac{\rho c_s}{2n}G_++\frac{s\rho}nG_0+\frac{s^2\rho^2e^{s\rho}}{2n}(G_++G_0)\le-\frac{\rho c_sr}{4n}E+\frac{2s\rho(n-r+1)}nE_{r-1}(S).
    \end{equation*}
    Here we wrote $G_+=rE-G_0$ in the first term and $G_++G_0=rE$ in the last term, and we used $e^{s\rho}\le2$, $\rho\le c_s/(4s^2)$, $s+c_s/2\le2s$ and $G_0\le(n-r+1)E_{r-1}(S)$.
    By \cref{lem:FreshOrientations}, the bound holds without conditioning on $\xi$.

    \emph{Step 4: induction.}
    \Cref{lem:TriangularRecursion} applies with $X_r(t)\colonequals E_r(S_t)$, $\Ac_t\colonequals\{\tau>t\}$, $\mathsf u\colonequals\rho c_s/(4n)$, $\mathsf v\colonequals2s\rho/n$ and $\mathsf M=8s/c_s=M_s$, since $\mathsf ur_*\le1$ and $E_r(S_0)=\binom nr$.
    Since $\{\tau=t\}\subseteq\{N_b(S_t)\ge q\}\cap\Ac_{t-1}$ for $t\ge1$, \eqref{eq:TriangularPopulation} with $\varphi(h)=e^{\rho h}$ and a union bound over $1\le t\le T$ prove \eqref{eq:StoppedPopulation}.
\end{proof}

\subsection{The potential of the rule and a tilted supermartingale}
\label{sec:Profile}

The rule \eqref{eq:WeightRule} is the exact gradient rule for a potential whose centered differences are the weights $a_h$.
Put $\alpha\colonequals\tanh(\rho/2)$ and define the profile $\psi:\Z_{\ge0}\to[0,\infty)$ by
\begin{equation}
    \label{eq:Profile}
    \begin{aligned}
        \psi(h)&\colonequals\frac{e^{\rho(h-B)}-e^{-\rho B}}{\sinh\rho}&&(0\le h\le B),\\
        \psi(B+j)&\colonequals\psi(B)+j+\frac\kappa2j^2+\Big(\alpha-\frac\kappa2\Big)\mathbf1\{j\text{ odd}\}&&(j\ge0).
    \end{aligned}
\end{equation}
A direct computation shows that $\frac12(\psi(h+1)-\psi(h-1))=a_h$ for every $h\ge1$.
The \emph{curvature} $d_h\colonequals\frac12(\psi(h+1)+\psi(h-1))-\psi(h)$ equals $\alpha a_h$ for $1\le h\le B$ and alternates between $\alpha$ and $\kappa-\alpha$ above $B$.
We also put $d_0\colonequals\psi(1)=(1+\alpha)e^{-\rho B}$, the increment of $\psi$ at height zero, which takes the place of the curvature there.
Since $\alpha\le\rho/2\le\kappa$,
\begin{equation}
    \label{eq:ProfileBounds}
    0\le d_h\le\min\{\kappa a_h,\kappa\}\quad(h\ge1),\qquad\psi(B+j)\ge\frac\kappa2j^2\quad(j\ge1).
\end{equation}
Let $\Psi(S)\colonequals\sum_i\psi(h_i)$.
Because $\psi(h\pm1)-\psi(h)=\pm a_h+d_h$ for $h\ge1$ and $a_0=0$, the rule \eqref{eq:WeightRule} produces the exact increment
\begin{equation}
    \label{eq:ProfileIncrement}
    \Psi(S')-\Psi(S)=\sum_{i\in A}d_{h_i}-\Big|\sum_{i\in A}\sigma_ia_{h_i}\Big|\le K\qquad\text{if }d_0\le\kappa,
\end{equation}
where $\sigma_i$ is arbitrary at height zero.
This is the split of the increment into a curvature term and a linear term used by Bansal and Spencer \cite[Section~2.1]{Bansal20a} and by Altschuler and Tikhomirov \cite[Section~3.2]{Altschuler26Binary}.
Since the increment is at most the small constant $K$, the exponential tilt $e^{\eta\Psi}$ with a large $\eta$ is a supermartingale as long as few coordinates carry non-negligible weight; in the next lemma, the error terms $\Gamma$ and $\zeta$ account for pairs of such coordinates and for selected coordinates at height zero.

\begin{lemma}[Tilted drift]
    \label{lem:TiltedDrift}
    Assume $n\ge2s$ and $d_0\le\kappa$.
    For $\eta>0$ put
    \begin{equation*}
        \mathsf v_i\colonequals\min\{\eta a_{h_i},1\},\qquad \mathsf V\colonequals\sum_i\mathsf v_i,\qquad\Gamma\colonequals\frac{s(s-1)}{n(n-1)}\sum_{i<j}\mathsf v_i\mathsf v_j\le\frac{s(s-1)\mathsf V^2}{2n(n-1)},
    \end{equation*}
    and let $c_\star\colonequals1/(1152s)$, $C_\star\colonequals1024s^2$ and $\zeta\colonequals s\eta d_0e^{K\eta}$.
    In both models,
    \begin{equation}
        \label{eq:TiltedDrift}
        \E\big[e^{\eta(\Psi(S_{t+1})-\Psi(S_t))}\bigm|\Fc_t\big]\le1-\frac{sc_\star}n\mathsf V+C_\star e^{K\eta}\Gamma+\zeta.
    \end{equation}
    If, in addition, $\Psi(S_t)\ge1$, $\eta\ge1$, $\eta\sinh\rho\ge1$,
    \begin{equation}
        \label{eq:TiltedConditions}
        \mathsf V\le\frac{c_\star(n-1)}{2(s-1)C_\star e^{K\eta}}\qquad\text{and}\qquad\zeta\le\frac{sc_\star}{4n},
    \end{equation}
    then the right side of \eqref{eq:TiltedDrift} is at most one.
\end{lemma}

\begin{proof}
    Fix the orientation.
    For a support $A$ put $\mathsf b_i\colonequals\eta a_{h_i}$, $Z\colonequals\sum_{i\in A}\sigma_i\mathsf b_i$ and $\mathsf H\colonequals\eta\sum_{i\in A,h_i>0}d_{h_i}$.
    By \eqref{eq:ProfileBounds}, $\mathsf H\le\kappa\sum_A\mathsf b_i$ and $\mathsf H-\abs Z\le K\eta$.
    We claim that, pointwise,
    \begin{equation}
        \label{eq:TiltedPointwise}
        e^{\mathsf H-\abs Z}\le e^{-\abs Z/2}+2\kappa\sum_{i\in A}\mathsf v_i+C_\star e^{K\eta}\sum_{\{i,j\}\subseteq A}\mathsf v_i\mathsf v_j.
    \end{equation}
    Write $\mathsf b_{\max}\colonequals\max_A\mathsf b_i$ and $\mathsf v_{\max}\colonequals\min\{\mathsf b_{\max},1\}$.
    If $\mathsf b_{\max}\le1/(16s)$, then $\mathsf H\le\kappa/16$, so $e^{\mathsf H-\abs Z}\le e^{-\abs Z}(1+2\mathsf H)\le e^{-\abs Z/2}+2\kappa\sum_A\mathsf v_i$.
    If $\mathsf b_{\max}>1/(16s)$ and the other selected values of $\mathsf b$ sum to at most $\mathsf b_{\max}/4$, then $\abs Z\ge\frac34\mathsf b_{\max}$ and $\mathsf H\le\frac54\kappa\mathsf b_{\max}\le\abs Z/2$.
    In the remaining case the clipped values $\mathsf v_i$ of the other selected coordinates sum to more than $\mathsf v_{\max}/4$, so the pair sum exceeds $\mathsf v_{\max}^2/4>1/(1024s^2)$, and the last term alone bounds $e^{\mathsf H-\abs Z}\le e^{K\eta}$.

    Apply \eqref{eq:ExponentialAC} to the coefficients $\sigma_i\mathsf b_i/2$ and use $\min\{\norm{\mathsf b_A}_2,1\}\ge\frac1s\sum_A\mathsf v_i$:
    \begin{equation*}
        \E_A\big[1-e^{-\abs Z/2}\big]\ge\frac1{576}\,\E_A\min\{\norm{\mathsf b_A}_2,1\}\ge\frac1{576s}\,\E_A\sum_{i\in A}\mathsf v_i.
    \end{equation*}
    Since $1/(576s)-2\kappa\ge c_\star$ and $\E_A\sum_A\mathsf v_i=s\mathsf V/n$, averaging \eqref{eq:TiltedPointwise} gives \eqref{eq:TiltedDrift} without the term $\zeta$, that is, ignoring the increment $d_0$ of selected coordinates at height zero.
    Each selected zero coordinate adds $d_0$ to the increment of $\Psi$, which raises the exponent by at most $s\eta d_0$; since both exponents are at most $K\eta$ by \eqref{eq:ProfileIncrement}, the mean value theorem bounds the resulting increase of the exponential by $\zeta$.
    \Cref{lem:FreshOrientations} removes the conditioning on the orientation.

    Under the additional assumptions, $C_\star e^{K\eta}\Gamma\le\frac{sc_\star}{4n}\mathsf V$.
    Moreover $\mathsf V\ge1$: otherwise every $\mathsf v_i<1$, so $a_{h_i}<1/\eta\le1$ for all $i$, hence all heights are smaller than $B$, and $\Psi\le\sum_ia_{h_i}/\sinh\rho\le \mathsf V/(\eta\sinh\rho)<1$.
    Hence $\zeta\le\frac{sc_\star}{4n}\mathsf V$, and the right side of \eqref{eq:TiltedDrift} is at most $1-\frac{sc_\star}{2n}\mathsf V\le1$.
\end{proof}

\subsection{Layers and squared-hinge excursions}
\label{sec:Layers}

For $L<\sqrt n$, we place the knee by the layer iteration of \cref{lem:LayerIteration}; this requires that a lone high coordinate dominates the flat core, which holds with a gap of order $s^2\log s$.

\begin{lemma}[Isolation with a gap]
    \label{lem:GapIsolation}
    Put
    \begin{equation}
        \label{eq:LayerGap}
        g\colonequals\ceil{\frac{\log(2s)}\rho}.
    \end{equation}
    If $b+g\le B$, then every state is $(b,g)$-isolating for the rule \eqref{eq:WeightRule}, in the sense of \cref{sec:Tools}.
\end{lemma}

\begin{proof}
    Suppose that exactly one selected coordinate has height at least $b$, and that its height is at least $b+g$.
    The other selected coordinates have heights smaller than $b\le B$, so their weights sum to at most $(s-1)e^{\rho(b-1-B)}$, which is smaller than $2s\,e^{\rho(b-B)}$ and, since $e^{\rho(b-B)}\le e^{-\rho g}\le1/(2s)$, smaller than $\frac12$.
    The weight of the high coordinate is at least $e^{\rho(b+g-B)}\ge2s\,e^{\rho(b-B)}$ if it lies in the core, and at least one otherwise.
    In either case it dominates the sum in \eqref{eq:WeightRule}, so the rule moves it inward.
\end{proof}

Hence \cref{lem:LocalClearance,lem:LayerIteration} apply with gap $g$ at all levels $b$ with $b+g\le B$.
Above the knee, a lone high coordinate dominates through its linear weight, so the squared hinge in the next lemma can increase only when two high coordinates meet; a priority rule plays the same role for the rows above the threshold in the algorithm of Altschuler and Tikhomirov \cite[Lemma~3.9]{Altschuler26Binary}.

\begin{lemma}[Squared-hinge excursions]
    \label{lem:HingeExcursions}
    Define
    \begin{equation}
        \label{eq:HingeConstants}
        \Delta_s\colonequals\ceil{s/\kappa},\qquad K_s\colonequals\ceil{\frac{2s(1+\kappa\Delta_s)}\kappa+s},\qquad\ell_B\colonequals B+\Delta_s.
    \end{equation}
    Let $1\le q\le n/(4s)$ and $u\ge1$ be integers with $u^2\ge4K_s$, and suppose that $z\colonequals(n/(4sq^2))^{1/K_s}\ge2$.
    Then, in both models,
    \begin{equation}
        \label{eq:HingeExcursions}
        \P\big\{D_T\ge\ell_B+u,\ N_B(S_t)\le q\text{ for all }t\le T\big\}\le T\exp\Big\{-\frac{u^2\log z}4\Big\}.
    \end{equation}
\end{lemma}

\begin{proof}
    Put $\Phi(S)\colonequals\sum_i(h_i-\ell_B)_+^2$ and, for a support $A$, $Z\colonequals\sum_{i\in A}\sigma_i(h_i-\ell_B)_+$.
    For $h\ge\ell_B$ we have $a_h=\kappa(h-\ell_B)+1+\kappa\Delta_s$, and for $h<\ell_B$ we have $a_h\le1+\kappa\Delta_s$.
    Hence the projection is $\kappa Z+\mathsf R$ with $\abs{\mathsf R}\le s(1+\kappa\Delta_s)$, and the decision satisfies $xZ\le s(1+\kappa\Delta_s)/\kappa$.
    The exact finite differences of the squared hinge, including a possible crossing at height $\ell_B$, give $\Phi(S')-\Phi(S)\le2xZ+s\le K_s$.

    A coordinate of height at least $\ell_B$ has weight at least $1+\kappa\Delta_s\ge s+1$.
    If it is the only selected coordinate of height at least $B$, the other selected weights sum to less than $s-1$, so it moves inward.
    Thus $\Phi$ can increase only if at least two selected coordinates have height at least $B$, which, at a state with $N_B\le q$, has probability at most $s^2q^2/n^2$.
    At a pre-state with $N_B\le q$ and $\Phi>u^2/2$, fix a coordinate $i$ of positive excess.
    With probability at least $s/(2n)$ the arrival selects $i$ and no other coordinate of height at least $B$, and then $\Phi$ decreases by at least one.
    Therefore, at every such pre-state,
    \begin{equation*}
        \E\big[z^{\Phi(S')-\Phi(S)}\bigm|\Fc_t\big]\le1+(z^{K_s}-1)\frac{s^2q^2}{n^2}-(1-z^{-1})\frac s{2n}\le1;
    \end{equation*}
    here $z^{K_s}s^2q^2/n^2=s/(4n)$ and $1-z^{-1}\ge\frac12$, since $z\ge2$.
    \Cref{lem:Excursions} with $E_-=u^2/2$, $E_+=u^2$, $K_\Phi=K_s$ and $\Ac_t\colonequals\{N_B(S_t)\le q\}$ bounds the probability that $\Phi$ reaches $u^2$ at some time $t\le T$ while $N_B(S_{t'})\le q$ for all $t'<t$ by $Tz^{K_s-u^2/2}\le Tz^{-u^2/4}$, since $u^2\ge4K_s$.
    On the event in \eqref{eq:HingeExcursions}, a height $\ell_B+u$ produces $\Phi\ge u^2$ on its own, which proves the claim.
\end{proof}

\subsection{Proof of Theorem~\ref{thm:Prefix}\ref{item:PrefixFixed}}
\label{sec:ProofFixedSparsity}

Let $T\ge n$ and $0<\delta<1$, and suppose that $L=\log(16n(T+1)/\delta)\le C_0n$.
We choose $n_0(s,C_0)$ so large that every statement below qualified by ``for large $n$'' holds for all $n\ge n_0$, uniformly in the respective range of $L$, and assume $n\ge n_0$.
All parameters below, including the knee $B$, are computed from $n$, $s$ and $L$ before the first arrival.
We consider two ranges of $L$.
For $L\ge\sqrt n$, the knee $B\asymp\sqrt{sL/\log n}$ is so high that $\rho B\gg\log n$; then the weights below a level $b$ well below the knee are negligible, and the tilted supermartingale of \cref{sec:Profile} bounds the excursions directly.
For $L<\sqrt n$, the knee is placed by a layer iteration at height $O_s(\log\log n+\sqrt{L/\log n})$, and the excursions above it are bounded by the squared hinge of \cref{sec:Layers}.

\paragraph{The range $\sqrt n\le L\le C_0n$.}
\emph{Parameters.}
Choose
\begin{equation}
    \label{eq:LongRangeParameters}
    B\colonequals\ceil{16\sqrt{\frac{sL}{\log n}}},\qquad\eta\colonequals\frac{\log n}{4K},\qquad q\colonequals\ceil{n^{2/3}},\qquad b\colonequals B-\ceil{\frac{3\log n}\rho},
\end{equation}
and let $\tau\colonequals\inf\{t:N_b(S_t)>q\}$.
For large $n$, we have $B/\log n\ge16\sqrt s\,n^{1/4}/(\log n)^{3/2}\to\infty$, $b\ge B/2\ge1$, $q\le\e_sn$, $d_0\le2e^{-\rho B}\le\kappa$, $\eta\ge1$, and $\eta\sinh\rho\ge\eta\rho=\log n/(4s^2)\ge1$.
Before $\tau$, every height below $b$ has $a_h\le e^{-\rho\ceil{3\log n/\rho}}\le n^{-3}$, so $\mathsf V\le q+\eta n^{-2}$.
Since $e^{K\eta}=n^{1/4}$, the right side of the first condition in \eqref{eq:TiltedConditions} is of order $n^{3/4}$, with a positive constant depending on $s$; hence that condition holds before $\tau$.
Moreover $\zeta\le s\eta\cdot2e^{-\rho B}n^{1/4}\le sc_\star/(4n)$, because $\rho B/\log n\to\infty$.
Thus \eqref{eq:TiltedConditions} holds at every pre-state before $\tau$ with $\Psi\ge1$.

\emph{Populations.}
\Cref{lem:StoppedMoments} controls the population:
\begin{equation*}
    \P(\tau\le T)\le T\exp\big\{-q\big[\rho b-\log(M_sen/q)\big]\big\}\le(T+1)e^{-4L}.
\end{equation*}
Indeed, $\log(M_sen/q)\le\rho B/4$ for large $n$, so the exponent is at least $\rho Bq/4$, and $Bq/L\ge16\sqrt s\,n^{1/6}/\sqrt{C_0\log n}\to\infty$.

\emph{The excursion.}
A height $2B$ implies $\Psi\ge E_\star\colonequals\kappa B^2/2$ by \eqref{eq:ProfileBounds}.
By \eqref{eq:ProfileIncrement} and \cref{lem:TiltedDrift}, \cref{lem:Excursions} applies with $\Phi=\Psi$, $z=e^\eta$, $K_\Phi=K$, $E_-=1$, $E_+=E_\star$ and $\Ac_t\colonequals\{\tau>t\}$.
For large $n$,
\begin{equation*}
    \eta(E_\star-1-K)\ge\frac{B^2\log n}{16s}\ge16L,
\end{equation*}
and therefore
\begin{equation*}
    \P(D_T\ge2B)\le\P(\tau\le T)+Te^{-\eta(E_\star-1-K)}\le(T+1)e^{-4L}+Te^{-16L}<\delta.
\end{equation*}
In this range $2B=O(\sqrt{sL/\log n})$, which proves the theorem in this range.

\paragraph{The range $L<\sqrt n$.}
\emph{Parameters.}
We keep the same family of rules and choose the knee through a layer iteration.
With $g$ from \cref{lem:GapIsolation} and $\Delta_s$, $K_s$ from \eqref{eq:HingeConstants}, put
\begin{equation}
    \label{eq:ShortRangeParameters}
    \begin{gathered}
        \nu\colonequals\log(n/L),\qquad\lambda_s\colonequals\ceil{16\sqrt{K_s}},\qquad u\colonequals\ceil{\lambda_s\sqrt{L/\nu}},\qquad q\colonequals u,\qquad R\colonequals\ceil L,\\
        D_s\colonequals\ceil{\max\{8A_{\mathrm{cl}}s,\,2/\e_s\}},\qquad q_0\colonequals\floor{n/D_s},\qquad b_0\colonequals\ceil{\frac{\log(M_sen/q_0)+L/q_0}\rho}.
    \end{gathered}
\end{equation}
Let $q_j$, $\vartheta_j$ and $J$ be as in \cref{lem:LayerIteration} for $g$, $R$ and $q_0$, put $b_j\colonequals b_0+j(g+2)$, and let $d\colonequals0$ if $q_J\le q$ and otherwise
\begin{equation}
    \label{eq:ShortRangeJump}
    d\colonequals\ceil{\frac{\log(8eq_J/q)+L/q}{\log\vartheta_J}}.
\end{equation}
The knee is $B\colonequals b_J+g+d$.
It is determined from $n$, $s$ and $L$ before the run, and every level $b_j$, $j\le J$, satisfies $b_j+g\le B$.

Since $T\ge n$ and $\delta<1$, we have $L\ge\log(16n(n+1))$.
For large $n$, uniformly for $L<\sqrt n$, we have $\nu\ge\frac12\log n$, $L/\nu\ge1$, $q\le L\le R$ and $2R<q_0$.
Here $q\le L$ follows from $q/L\le\lambda_s/\sqrt{L\nu}+1/L\to0$.
Moreover $b_0=O_s(1)$, $q_0\le\e_sn/2$ and $q_0\le n/(8A_{\mathrm{cl}}s)$.

\emph{Populations.}
By \cref{lem:StoppedMoments} at level $b_0$ and cutoff $q_0$, the event $\Vc\colonequals\{N_{b_0}(S_t)\le q_0\text{ for all }t\le T\}$ fails with probability at most $Te^{-L}$.
By \cref{lem:GapIsolation}, \cref{lem:LayerIteration} applies with $\Gc_t\colonequals\{N_{b_0}(S_t)\le q_0\}$, $t_0=\cdots=t_{J+1}=0$ and $M=0$.
Its failure probability \eqref{eq:LayerFailure} is at most $J(T+1)e^{-c_*R}\le J(T+1)e^{-L}$, and if $d\ge1$, \eqref{eq:LayerFinal} with $r=q$ and a union bound over $t\le T$ add at most $(T+1)e^{-L}$.
Hence the event $\Cc_{\mathrm{fin}}\colonequals\{N_B(S_t)\le q\text{ for all }t\le T\}$, which for $d=0$ follows from $N_B\le N_{b_J}\le q_J\le q$, satisfies $\P(\Cc_{\mathrm{fin}}^c)\le(J+2)(T+1)e^{-L}\le\delta/8$.

\emph{The knee.}
We have $q_J\le2R\le3L$ and, for large $n$, $\log\vartheta_J=\frac12\log(n/(C_{\mathrm{cl}}sq_J))\ge\nu/4$.
If $d\ge1$, the numerator in \eqref{eq:ShortRangeJump} is at most $\log(24eL/q)+L/q\le6L/q$, since $L/q\ge1$; hence
\begin{equation*}
    d\le1+\frac{24L}{q\nu}\le1+\frac{24}{\lambda_s}\sqrt{\frac L\nu}\le1+\frac{24}{\lambda_s^2}u,\qquad B=O_s\big(\log\log n+u\big).
\end{equation*}

\emph{The excursion.}
Let $z\colonequals(n/(4sq^2))^{1/K_s}$.
Since $u\le(\lambda_s+1)\sqrt{L/\nu}$,
\begin{equation*}
    \log z\ge\frac{\nu+\log\nu-\log(4s)-2\log(\lambda_s+1)}{K_s}\ge\frac\nu{2K_s}
\end{equation*}
for large $n$; in particular $z\ge2$.
Moreover $q\le n/(4s)$ and $u^2\ge\lambda_s^2\ge4K_s$.
The exponent in \eqref{eq:HingeExcursions} is at least $u^2\nu/(8K_s)\ge\lambda_s^2L/(8K_s)\ge32L$.
Therefore
\begin{equation*}
    \P(D_T\ge B+\Delta_s+u)\le\frac\delta8+Te^{-32L}<\delta.
\end{equation*}
Since $\nu\ge\frac12\log n$, the radius $B+\Delta_s+u$ is $O_s(\log\log n+\sqrt{L/\log n})$.

\paragraph{Dependence on $s$.}
The quantities $g$, $b_0$ and $\Delta_s$ are $O(s^2\log(es))$, $J=O(\log\log n)$, $d\le1+u$ and $u=O(s^{3/2}\sqrt{L/\log n})$.
Hence the radius in the second range is at most $Cs^2\log(es)\,(\log\log n+\sqrt{L/\log n})$ for an absolute constant $C$.
Since $2B=O(\sqrt{sL/\log n})$ in the first range, one can take $A_s=O(s^2\log(es))$.
This completes the proof of \cref{thm:Prefix}\ref{item:PrefixFixed}.
\qed

\section{Terminal discrepancy at every prescribed time}
\label{sec:Terminal}

In this section we prove \cref{thm:Terminal}; the matching lower bound at every terminal time $T\ge n$ is \cref{prop:Baseline}.
The bounds concern each prescribed time $T$ separately; they do not control the prefix discrepancy, which for fixed $s$ grows with the horizon once $\log T\gg\log n\,(\log\log n)^{1/\alpha_s}$, by \cref{thm:Prefix}\ref{item:PrefixLower}.

Throughout this section $s\ge2$; upper limits on $s$ are stated in each result.
We write
\begin{equation*}
    \ell\colonequals\log(en),\qquad K_0\colonequals4,
\end{equation*}
and we use the notation $p=s/n$ and $N_b(S)=\abs{\{i:\abs{S_i}\ge b\}}$ of \cref{sec:Tools}.
The letters $c,C$ denote absolute positive constants, which may change between occurrences; they do not depend on $n$, $s$, the horizon or the failure probability.
The notation of \cref{sec:FixedSparsity} is not used here.

The rule of \cref{thm:Terminal} is one of two deterministic rules, selected before the first arrival according to $n$ and $s$ only: it is the hybrid rule of \cref{def:HybridRule} with the parameters $b_0,q_0$ of \cref{prop:CoshTerminal} if $s\le\log\ell$ and $n\ge2^{40}s^{3/2}$, and the gauge rule of \cref{def:GaugeRule} otherwise.
The condition $n\ge2^{40}s^{3/2}$ means that the integer $q_0$ of \cref{prop:CoshTerminal} is positive, and it implies the standing assumptions of \cref{sec:CoshHybrid}; it holds for all $s\le\log\ell$ once $n$ exceeds an absolute constant.
Both rules use no information about the horizon or the failure probability, and neither is ever restarted.
The proof distinguishes four ranges of $s$, the last of which extends to $s\le n/2$ in the binary model and to $s\le n$ in the signed model; in each range, for every fixed $k\ge1$, all $n\ge n_k$ and every $T$, the terminal discrepancy at time $T$ exceeds $C_k$ times the radius in the following table with probability at most $n^{-k}$.
\begin{center}
    \renewcommand{\arraystretch}{1.15}
    \begin{tabular}{@{}llll@{}}
        \toprule
        \textbf{Range of $s$} & \textbf{Rule and method} & \textbf{Radius} & \textbf{Result}\\
        \midrule
        $s\le\log\ell$ & hybrid rule, hyperbolic moments & $\sqrt s\log(s+1)+\log\ell$ & \cref{prop:CoshTerminal}\\
        $\log\ell<s<\ell^{2/3}$ & gauge rule, bootstrap & $\sqrt s+\log\ell$ & \cref{prop:Bootstrap}\\
        $\ell^{2/3}\le s\le\ell^2$ & gauge rule, moderate exponent & $\sqrt s+\log\ell+\ell/s$ & \cref{prop:CoarseTerminal}\\
        $s>\ell^2$ & gauge rule, absolute exponent & $\sqrt s$ & \cref{prop:CoarseTerminal}\\
        \bottomrule
    \end{tabular}
\end{center}
\Cref{sec:Gauge,sec:GaugeEstimates,sec:Conjugation,sec:GaugeService,sec:GaugeFoster,sec:ModerateExponents} construct the gauge rule and prove Foster inequalities \cite{Foster53,Hajek82} for exponential moments of its gauge at every state, first with an absolute and then with a moderate exponent.
\Cref{sec:Composition,sec:Bootstrap} use these moments in a deterministic window before $T$: \cref{sec:Composition} combines them with the layer iteration of \cref{sec:Tools}, and \cref{sec:Bootstrap} adds a bootstrap with a large exponent inside a controlled height region.
\Cref{sec:CoshHybrid} treats the hybrid rule with the triangular recursion and the weights of a moment from \cref{sec:Tools}, \cref{sec:ProofTerminal} combines the four ranges, and \cref{sec:GreedyTerminal} shows that, for $s=2$, quadratic Greedy has the same terminal guarantee.

The four ranges are dictated by the failure probability.
For $s>\ell^2$, the radius $\sqrt s$ exceeds $\ell$, and the absolute exponent already gives failure probability $n^{-k}$.
For $s\le\ell^2$, it gives the sharp radius only with failure probability $(\log n)^{-k}$ (see the end of \cref{sec:Composition}); in the three remaining ranges, the moderate exponent, the bootstrap and the hybrid rule improve this to $n^{-k}$, which the union bound behind \cref{cor:PolynomialPrefix} requires.

\subsection{The translated gauge and the rule}
\label{sec:Gauge}

Fix a sufficiently large absolute constant $a$ and then a sufficiently large absolute constant $D_0\ge2^{40}$; both may be increased finitely many times below.
The profile is a continuous variant of the potential of Altschuler and Tikhomirov \cite[Section~3.1]{Altschuler26Binary}, which already consists of an inverse-square core and an exponential continuation.
Put
\begin{equation}
    \label{eq:GaugeProfile}
    \begin{gathered}
        \rho\colonequals a(\sqrt s+2),\qquad h_c\colonequals\rho-a,\qquad\lambda\colonequals\frac{4h_c}{\rho^2-h_c^2},\\
        f(u)\colonequals
        \begin{cases}
            \big(\rho^2/(\rho^2-u^2)\big)^2, & 0\le u\le h_c,\\
            f(h_c)\,e^{\lambda(u-h_c)}, & u\ge h_c.
        \end{cases}
    \end{gathered}
\end{equation}
On the core $[0,h_c]$, $f'/f=4u/(\rho^2-u^2)$, which equals $\lambda$ at the junction $h_c$; hence $f$ is increasing, convex and continuously differentiable, with $f(0)=1$ and $f'(0)=0$.
At the junction, an upper bound on $f''$ refers to the larger one-sided value.
Their potential uses the same $\rho$, but it is defined on the integers, its two parts agree in value at $\floor{\rho-a}+1$, and it is capped at their threshold for exceptional coordinates; here the two parts are joined with matching derivatives, no cap is imposed, and the profile is applied to the excess over the gauge defined below.
With $\sigma\colonequals\sqrt s+2$ we have $\rho=a\sigma$, $h_c=a(\sigma-1)$, $\lambda=4(\sigma-1)/(a(2\sigma-1))$ and $f(h_c)=(\sigma^2/(2\sigma-1))^2$; the last expression is increasing in $\sigma\ge1$ and equals $4$ at $\sigma=2+\sqrt2$.
Consequently, since $s\ge2$ gives $\sigma\ge2+\sqrt2$,
\begin{equation}
    \label{eq:GaugeProfileConstants}
    a^{-1}\le\lambda\le2a^{-1},\qquad s/4\le f(h_c)\le6s,\qquad f(h_c)\ge4.
\end{equation}

For every state $S\in\R^n$, the \emph{gauge} $B=B(S)$ is defined by
\begin{equation}
    \label{eq:GaugeDefinition}
    \sum_{i=1}^nf(u_i)=K_0n,\qquad u_i\colonequals(\abs{S_i}-B)_+,\qquad g_i\colonequals f'(u_i),\qquad\Gamma\colonequals\sum_ig_i.
\end{equation}
The left side of the first equation, as a function of $B$, is the \emph{profile sum} at level $B$; it decreases continuously from infinity to $n$, and it strictly decreases wherever it exceeds $n$; hence the root is unique, and $\Gamma>0$.
The root is real and need not be nonnegative.
A level defined by such a scalar normalization equation also determines the gradient weights of the regularized maximum in our earlier work \cite[Lemma~1]{Wengiel26}; here the level itself serves as the potential.
Put $h_0\colonequals f^{-1}(K_0)$.
Then $B(S)\ge-h_0$ at every state, since otherwise every coordinate would contribute more than $K_0$, and $h_0\le h_c<\rho$ by \eqref{eq:GaugeProfileConstants}.
We call $u_i$ the \emph{excess} of coordinate $i$.
The rule uses the fixed parameters
\begin{equation}
    \label{eq:GaugeThreshold}
    Y_*\colonequals2K_0D_0s\ell^8,\qquad h_*\colonequals f^{-1}(Y_*),\qquad\Ec(S)\colonequals\{i:f(u_i)\ge Y_*\}=\{i:\abs{S_i}\ge B(S)+h_*\}.
\end{equation}
The coordinates in $\Ec(S)$ are called \emph{exceptional}.
Since $D_0$ is large, $Y_*\ge C_YK_0s$ for a sufficiently large absolute constant $C_Y$, and $h_*\ge h_c+2$.
A coordinate with $S_i=0$ has $f(u_i)\le f((-B)_+)\le K_0<Y_*$ and is never exceptional.
Each exceptional coordinate contributes at least $Y_*$ to the profile sum, so there are at most $K_0n/Y_*$ of them; by \cref{lem:GaugeGeometry} below, $h_*=O(\sqrt s+\log\ell)$, and no height exceeds the gauge by more than $O(\sqrt s+\log n)$.

\begin{definition}[The gauge rule]
    \label{def:GaugeRule}
    Let $V$ be an arrival with support $A$, and write $\e_i\colonequals V(i)$ for $i\in A$, so that $\e_i=1$ in the binary model.
    If $A\cap\Ec(S)\ne\emptyset$, let $j$ be the element of $A\cap\Ec(S)$ with the largest height, ties broken by the smallest index, and choose $x\colonequals-\e_j\sgn(S_j)$; this moves $j$ inward.
    Otherwise choose
    \begin{equation*}
        x\colonequals-\sgn\Big(\sum_{i\in A}\e_i\sgn(S_i)\,g_i\Big),
    \end{equation*}
    with $\sgn(0)=0$ in the sum and $x\colonequals+1$ if the sum vanishes.
\end{definition}

The first case is the priority of Altschuler and Tikhomirov for exceptional coordinates \cite[Algorithm~3.2]{Altschuler26Binary}; in their algorithm a coordinate stays exceptional once its discrepancy has reached the threshold, whereas $\Ec(S)$ depends only on the current state.
The rule is defined at every state, depends only on $n$ and $s$, and knows neither the horizon nor the failure probability.
All selected coordinates move by $x\e_i$ with the same $x$, which we call the \emph{common sign} of the update.
For a fixed arrival we put
\begin{equation*}
    Z\colonequals\sum_{i\in A}\e_i\sgn(S_i)\,g_i,
\end{equation*}
so that $x=-\sgn Z$ whenever $A$ contains no exceptional coordinate.

\subsection{Deterministic estimates}
\label{sec:GaugeEstimates}

This subsection collects estimates for the profile and the gauge that hold at every state, for every arrival and every common sign; only \eqref{eq:GaugeInjection} averages over the support.
At a state $S$ with excesses $u_i$, we put
\begin{equation*}
    \nu_i\colonequals\sup\{f''(v):v\ge0,\ \abs{v-u_i}\le2\}.
\end{equation*}
The curvature bounds in the next lemma are continuous counterparts of the finite-difference estimates of Altschuler and Tikhomirov \cite[Lemma~3.3]{Altschuler26Binary}, with the same split at $\rho/2$.

\begin{lemma}[Profile estimates]
    \label{lem:ProfileEstimates}
    Uniformly for $s\ge2$,
    \begin{equation}
        \label{eq:ProfileCurvature}
        \begin{aligned}
            f''(u)&\le\frac C{a^2s}&&(u\le\rho/2),\\
            f''(u)&\le\frac Caf'(u),\qquad f''(u)\le Cf'(u)\sqrt{\frac{f(u)}{a^2s}}&&(u>\rho/2),\\
            f''(u)&\le\frac Caf'(u)+\frac C{a^2s},\qquad\frac{c(f(u)-1)}{a\sqrt s}\le f'(u)\le\lambda f(u)&&(u\ge0).
        \end{aligned}
    \end{equation}
    In particular $f(u)\le e^{\lambda u}$ for $u\ge0$.
    If $a$ is large enough, then at every state
    \begin{equation}
        \label{eq:GaugeNormalization}
        \frac{cn}{a\sqrt s}\le\Gamma\le\lambda K_0n,\qquad\nu_i\le Cf''(u_i)\le\frac Cag_i+\frac C{a^2s},\qquad\sum_i\nu_i\le\frac Ca\Gamma,
    \end{equation}
    and for all nonnegative $\tilde u_i$ with $\abs{\tilde u_i-u_i}\le2$,
    \begin{equation}
        \label{eq:GaugeStability}
        \sum_if'(\tilde u_i)\ge\frac\Gamma2,\qquad f'(u_i+2)\le C\Big(g_i+\frac1{a^2s}\Big).
    \end{equation}
\end{lemma}

\begin{proof}
    On the core,
    \begin{equation*}
        \frac{f'(u)}{f(u)}=\frac{4u}{\rho^2-u^2},\qquad\frac{f''(u)}{f(u)}=\frac{4\rho^2+20u^2}{(\rho^2-u^2)^2}.
    \end{equation*}
    For $u\le\rho/2$ these give $f\le16/9$ and $f''\le C/\rho^2\le C/(a^2s)$.
    For $\rho/2<u\le h_c$ we have $\rho-u\ge a$, and $f''/f'=(\rho^2+5u^2)/(u(\rho^2-u^2))\le12/(\rho-u)\le12/a$; moreover $\sqrt f=\rho^2/(\rho^2-u^2)\ge\rho/(2(\rho-u))\ge a\sqrt s/(2(\rho-u))$, so that $f''/f'\le24\sqrt f/(a\sqrt s)$.
    On the exponential continuation, $f'=\lambda f$ and $f''=\lambda f'$, and \eqref{eq:GaugeProfileConstants} gives $\lambda\le2/a$ and $\lambda\le4\sqrt{f(h_c)}/(a\sqrt s)$.
    This proves the curvature bounds.
    Below $\rho/2$, $f(u)-1\le Cu^2/\rho^2$ and $f'(u)\ge4u/\rho^2$ with $u\le2a\sqrt s$; above $\rho/2$, $f'/f\ge\min\{2/\rho,\lambda\}\ge c/(a\sqrt s)$.
    This proves the lower bound on $f'$, and the upper bound holds because $f'/f$ increases on the core up to its terminal value $\lambda$.
    Integrating $(\log f)'\le\lambda$ from $f(0)=1$ gives $f(u)\le e^{\lambda u}$.

    A displacement of at most two changes the factors in these formulas by bounded factors: on the core the distance to the pole $\rho$ stays at least $a-2\ge a/2$, near zero $f''\asymp\rho^{-2}$, and the two one-sided curvatures at the junction are both of order $f(h_c)/a^2$.
    This proves $\nu_i\le Cf''(u_i)$.
    With $\psi\colonequals f'+1/(as)$, the global curvature bound reads $\psi'\le(C/a)\psi$; integrating over an interval of length two gives
    \begin{equation*}
        f'((u-2)_+)\ge e^{-2C/a}f'(u)-\frac C{a^2s},\qquad f'(u+2)\le C\Big(f'(u)+\frac1{a^2s}\Big),
    \end{equation*}
    where for $u<2$ we also used $f'\le2\sup_{[0,2]}f''\le C/(a^2s)$ on $[0,2]$.
    Summing the lower bound on $f'$ and using \eqref{eq:GaugeDefinition} gives $\Gamma\ge c(K_0-1)n/(a\sqrt s)$, while $\Gamma\le\lambda\sum_if(u_i)=\lambda K_0n$.
    The additive loss $Cn/(a^2s)$ in the display is therefore at most $C\Gamma/(a\sqrt s)$, which proves \eqref{eq:GaugeStability} once $a$ is large, since $f'$ is nondecreasing.
    Summing the curvature bound and using the same lower bound on $\Gamma$ proves the last assertion of \eqref{eq:GaugeNormalization}.
\end{proof}

\begin{lemma}[Gauge geometry]
    \label{lem:GaugeGeometry}
    The gauge is convex and $1$-Lipschitz with respect to $\norm\cdot_\infty$, and $B(S)\le\norm S_\infty$.
    For every state, with an absolute constant $C_{\mathrm{env}}$,
    \begin{equation}
        \label{eq:GaugeEnvelope}
        \norm S_\infty\le B(S)+C_{\mathrm{env}}\big(\sqrt s+\log(en/s)\big),\qquad h_*\le C\big(\sqrt s+\log(Y_*/s)\big)\le C\big(\sqrt s+\log\ell\big).
    \end{equation}
    If $B(S)\le\beta$ and $b\ge\ceil{\beta+h_*}$ is an integer, then $N_b(S)\le K_0n/Y_*$, and whenever exactly one selected coordinate of an arrival has height at least $b$, the gauge rule moves it inward, in either model; that is, $S$ is $(b,0)$-isolating in the sense of \cref{sec:Tools}.
\end{lemma}

\begin{proof}
    The epigraph of $B$ is $\{(S,b):\sum_if((\abs{S_i}-b)_+)\le K_0n\}$, a convex set.
    If $\norm{S'-S}_\infty\le d$, substituting $B(S)+d$ into the defining equation for $S'$ shows $B(S')\le B(S)+d$; exchanging the roles gives the Lipschitz bound.
    At the level $\norm S_\infty$ all excesses vanish and the profile sum is $n<K_0n$, so $B(S)\le\norm S_\infty$.
    Since $\max_i\abs{S_i}>B$, the normalization gives $f(\max_i\abs{S_i}-B)\le K_0n$; inverting $f$ with its exponential continuation and \eqref{eq:GaugeProfileConstants} gives both bounds in \eqref{eq:GaugeEnvelope}.
    Every coordinate counted by $N_b(S)$ is exceptional, so $N_b(S)\le K_0n/Y_*$.
    Finally, the unique selected coordinate with height at least $b$ is exceptional and strictly higher than every other selected exceptional coordinate; the rule therefore moves it inward.
\end{proof}

\begin{lemma}[One-step comparison and remainder]
    \label{lem:GaugeRemainder}
    Let $S$ be a state, and let $S'=S+xV$ for an arrival $V$ with support $A$ and arbitrary marks, and an arbitrary common sign $x$.
    Then $\abs{B(S')-B(S)}\le1$ and
    \begin{equation}
        \label{eq:GaugeComparison}
        \big(B(S')-B(S)\big)_+\le\frac C\Gamma\sum_{i\in A}\Big(g_i+\frac1{a^2s}\Big).
    \end{equation}
    In particular, if the support is a uniform $s$-subset and the sign may depend on the arrival, then, at every state,
    \begin{equation}
        \label{eq:GaugeInjection}
        \E\big[(B(S')-B(S))_+\bigm|S\big]\le Cp.
    \end{equation}
    If $B(S)>1$, then, with $Z$ as in \cref{sec:Gauge} and $Q_A\colonequals\Gamma^{-1}\sum_{i\in A}\nu_i$,
    \begin{equation}
        \label{eq:GaugeRemainder}
        B(S')-B(S)=\frac{xZ}\Gamma+R_A,\qquad0\le R_A\le CQ_A+\frac Ca\cdot\frac{\abs Z}\Gamma.
    \end{equation}
\end{lemma}

\begin{proof}
    The bound $\abs{B(S')-B(S)}\le1$ is the Lipschitz property, since $\norm{S'-S}_\infty\le1$.
    For \eqref{eq:GaugeComparison} we compare roots, without differentiating at coordinates with $S_i=0$.
    Write $B\colonequals B(S)$ and $B'\colonequals B(S')$, and suppose that $B'>B$.
    At level $B$, the excesses of $S'$ are at most $u_i+1$ for $i\in A$ and equal to $u_i$ otherwise, so the profile sum of $S'$ at level $B$ exceeds $K_0n$ by at most
    \begin{equation*}
        \sum_{i\in A}\big(f(u_i+1)-f(u_i)\big)\le\sum_{i\in A}f'(u_i+1)\le C\sum_{i\in A}\Big(g_i+\frac1{a^2s}\Big),
    \end{equation*}
    by convexity and \eqref{eq:GaugeStability}.
    For each level $b\in[B,B']$, the excesses $(\abs{S_i'}-b)_+$ differ from $u_i$ by at most two, since $B'\le B+1$; hence the derivative of the profile sum of $S'$ with respect to $b$ is at most $-\Gamma/2$ by \eqref{eq:GaugeStability}.
    Since this profile sum equals $K_0n$ at level $B'$, integration over $[B,B']$ proves \eqref{eq:GaugeComparison}.
    The argument permits arbitrary marks and signs, negative roots, and zero coordinates.
    For \eqref{eq:GaugeInjection}, average \eqref{eq:GaugeComparison} over the support: since each coordinate lies in $A$ with probability $p$, the right side has expectation $C(p+1/(a^2\Gamma))$, and $1/(a^2\Gamma)\le\sqrt s/(can)\le Cp$ by \eqref{eq:GaugeNormalization}.

    For \eqref{eq:GaugeRemainder}, follow the path $S(t)\colonequals S+txV$, $0\le t\le1$, and put $B_t\colonequals B(S(t))$ and $u_i(t)\colonequals(\abs{S_i(t)}-B_t)_+$.
    Suppose $B_0>1$; then $B_t>0$ by the Lipschitz property.
    If $u_i(t)>0$, then $\abs{S_i(t)}>B_t>0$, so $S_i(t)\ne0$ and $\abs{S_i(t)}$ moves with constant speed $v_i(t)\colonequals x\e_i\sgn(S_i(t))\mathbf1\{i\in A\}$ near $t$; where the excess vanishes, $f'(0)=0$ makes $f(u_i(t))$ continuously differentiable.
    Since $\abs{S_i}>B_0>1$ whenever $u_i>0$, the speed $v_i(t)$ can differ from $v_i(0)$ only if $u_i=0$; a selected coordinate with $S_i=0$, for instance, has $v_i(t)=1$ for $t>0$.
    Implicit differentiation of the normalization gives, for almost every $t$,
    \begin{equation}
        \label{eq:GaugeDerivatives}
        B_t'=\frac{\sum_if'(u_i(t))v_i(t)}{\Gamma_t},\qquad B_t''=\frac{\sum_{i:u_i(t)>0}f''(u_i(t))(v_i(t)-B_t')^2}{\Gamma_t},\qquad\Gamma_t\colonequals\sum_if'(u_i(t)).
    \end{equation}
    The first derivative is Lipschitz on the compact path, since $f'$ is locally Lipschitz and $\Gamma_t$ stays positive, so these identities integrate across the profile junctions.
    In particular $\abs{B_t'}\le1$ and $B_t''\ge0$.
    Excesses change by at most two, so $\Gamma_t\ge\Gamma/2$ and $f''(u_i(t))\le\nu_i$ by \cref{lem:ProfileEstimates}.
    Splitting the numerator of $B_t''$ into selected and unselected coordinates, and using $\sum_i\nu_i\le(C/a)\Gamma$, gives
    \begin{equation*}
        B_t''\le8Q_A+\frac Ca(B_t')^2.
    \end{equation*}
    Since $(B_t')^2\le\abs{B_t'}\le\abs{B_0'}+\int_0^tB_v''\,dv$, integration and absorption of the last term for large $a$ give $\int_0^1B_t''\,dt\le CQ_A+(C/a)\abs{B_0'}$.
    Now $B_0'=xZ/\Gamma$, since $g_i=0$ whenever $u_i=0$, and $R_A=\int_0^1(1-t)B_t''\,dt$, which proves \eqref{eq:GaugeRemainder}.
\end{proof}

\subsection{Reduction to binary inputs}
\label{sec:Conjugation}

The service estimates of \cref{sec:GaugeService} concern binary inputs, and the signed model is reduced to the binary one by a conjugation with the fresh marks.
We consider rules for binary inputs that depend on the state only through the magnitudes $\abs{S_i}$ and through the signs $\sgn(S_i)$ of the selected coordinates.
The \emph{signed version} of such a rule replaces $\sgn(S_i)$ by $\e_i\sgn(S_i)$ for the selected coordinates; both rules of \cref{thm:Terminal} have this form, and in the signed model they act as the signed versions of their binary forms.
Unlike \cref{lem:FreshOrientations}, which keeps the state and requires an estimate for every orientation of the marks, the following lemma conjugates the state and thereby reduces every orientation to the binary one; this requires the binary estimates at \emph{every} state, because the conjugated state need not be reachable.

\begin{lemma}[Fresh-mark conjugation]
    \label{lem:MarkConjugation}
    Let $2s\le n$, consider a rule for binary inputs as above, and let $\Sc\subseteq\R^n$ be invariant under sign changes of coordinates.
    Let $\Phi(S,S')$ depend only on the magnitudes of $S$ and $S'$, let $\Psi$ be a function on $\Sc$, and suppose that the binary rule satisfies
    \begin{equation*}
        \E\big[\Phi(W,W')\bigm|W\big]\le\Psi(W)\qquad\text{at every state }W\in\Sc,
    \end{equation*}
    where $W'$ is the successor of $W$.
    Then its signed version satisfies, at every state $S\in\Sc$,
    \begin{equation*}
        \E\big[\Phi(S,S')\bigm|S\big]\le\E_\e\Psi(\diag(\e)S),
    \end{equation*}
    where $\e$ is uniformly distributed on $\{\pm1\}^n$.
    In particular, if $\Psi$ depends only on the magnitudes, then the signed version satisfies the same inequality as the binary rule.
\end{lemma}

\begin{proof}
    Condition on the past and generate the entire fresh Rademacher vector $\e\in\{\pm1\}^n$, independently of the support $A$; the arrival is $V=\sum_{i\in A}\e_ie_i$.
    After fixing $\e$, put $W\colonequals\diag(\e)S\in\Sc$.
    The support is still uniform.
    The binary rule at $W$ chooses exactly the common sign of the signed version at $S$, since the magnitudes of $W$ and $S$ agree and $\sgn(W_i)=\e_i\sgn(S_i)$.
    Moreover $\diag(\e)(S+xV)=W+x\mathbf1_A$, so the magnitudes after the signed step at $S$ coincide with those after the binary step at $W$.
    Hence the conditional expectation of $\Phi(S,S')$ given $\e$ equals $\E[\Phi(W,W')\mid W]\le\Psi(\diag(\e)S)$, and averaging over $\e$ proves the claim.
    The argument is repeated separately at each step; unselected marks serve only as a conditioning device and are not used by the rule.
\end{proof}

\subsection{Service and curvature in the binary model}
\label{sec:GaugeService}

Besides \cref{lem:ClippedAntiConcentration}, we use the following head--tail form of the sampling inequality of Altschuler and Tikhomirov, which controls the curvature remainder.

\begin{literaturetheorem}[{\cite[Lemma~3.4]{Altschuler26Binary}}]
    \label{thm:HeadTail}
    Let $U$ be a set of size $N$, let $A$ be a uniform $s$-subset of $U$ with $2\le s\le N/2$, and let $(c_i)_{i\in U}$ be real numbers with decreasing rearrangement $c_1^*\ge\cdots\ge c_N^*$ of $(\abs{c_i})$.
    Put $m\colonequals\floor{N/s}$.
    Then, for an absolute constant $c_D>0$,
    \begin{equation*}
        \E\Big|\sum_{i\in A}c_i\Big|\ge c_D\Big(\frac sN\sum_{j\le m}c_j^*+\Big(\frac sN\sum_{j>m}(c_j^*)^2\Big)^{1/2}\Big).
    \end{equation*}
\end{literaturetheorem}

In this subsection we consider binary inputs with $2\le s\le n/2$ and a state with $B(S)>1$.
Write $U\colonequals[n]\setminus\Ec$, $N\colonequals\abs U$, and
\begin{equation}
    \label{eq:ServiceQuantities}
    \begin{gathered}
        \Gamma_U\colonequals\sum_{i\in U}g_i,\qquad\Gamma_\Ec\colonequals\sum_{i\in\Ec}g_i,\qquad P_0\colonequals\P(A\subseteq U),\\
        m_U\colonequals\E\big[\abs Z\bigm|A\subseteq U\big],\qquad\xi\colonequals\frac{P_0m_U+p\,\Gamma_\Ec}\Gamma.
    \end{gathered}
\end{equation}
Up to constant factors, $\xi\Gamma$ is the expected service $\E[-xZ]$ (\cref{lem:GaugeService}): the term $P_0m_U$ comes from supports avoiding $\Ec$, on which the rule follows the gradient, and $p\Gamma_\Ec$ from exceptional coordinates, which move inward when they are selected alone.
As noted in \cref{sec:Gauge}, $\abs\Ec\le K_0n/Y_*\le n/(C_Ys)$, so $P_0\ge1-p\abs\Ec\ge23/24$.
If $\Ec\ne\emptyset$, then $Y_*\le K_0n$ forces $s\le n/C_Y$, hence $N\ge n/2$ and $2s\le N$; if $\Ec=\emptyset$, then $N=n\ge2s$.
Conditionally on $A\subseteq U$, the support is a uniform $s$-subset of $U$.

\begin{lemma}[Service and curvature]
    \label{lem:GaugeService}
    For the gauge rule in the binary model, at every state with $B(S)>1$,
    \begin{equation}
        \label{eq:GaugeServiceBounds}
        \frac{c\sqrt s}n\le\xi\le\frac sn,\qquad\E[-xZ]\ge\frac{\xi\Gamma}2,\qquad\E\abs Z\le2\xi\Gamma.
    \end{equation}
    Furthermore,
    \begin{equation}
        \label{eq:GaugeCurvatureBounds}
        \E\big[Q_A\bigm|A\subseteq U\big]\le\frac Ca\cdot\frac{m_U}\Gamma+\frac C{a^2\Gamma},\qquad\E Q_A\le\frac Ca\xi.
    \end{equation}
\end{lemma}

\begin{proof}
    At $B>1$, a coordinate with $S_i=0$ has $g_i=0$, so \eqref{eq:UnclippedAC}, applied to the uniform $s$-subset of $U$ with $c_i=\sgn(S_i)g_i$, gives $m_U\ge\frac1{144}\E[\norm{c_A}_2\mid A\subseteq U]\ge\sqrt s\,\Gamma_U/(144N)$.
    This gives the lower bound on $\xi$, and the upper bound follows from $P_0m_U\le\E\sum_{i\in A\cap U}g_i=p\Gamma_U$.

    On $\{A\subseteq U\}$ the rule is the gradient rule, so $-xZ=\abs Z$.
    Conditionally on selecting an exceptional coordinate $i$, the probability of selecting a second one is at most $(s-1)(\abs\Ec-1)/(n-1)\le s\abs\Ec/n\le1/24$.
    A uniquely selected exceptional coordinate moves inward and contributes $g_i$ to $-xZ$; otherwise its contribution is at least $-g_i$.
    Hence the exceptional coordinates contribute at least $p(1-2s\abs\Ec/n)\Gamma_\Ec$ to $\E[-xZ]$.
    The possible loss from unexceptional coordinates on supports meeting $\Ec$ is at most
    \begin{equation*}
        p\frac{(s-1)\abs\Ec}{n-1}\Gamma_U\le\frac{CK_0s^2\abs\Ec}{an}\le\frac{CK_0s}{Y_*}\,p\,\Gamma_\Ec\le\frac p4\Gamma_\Ec,
    \end{equation*}
    because $\Gamma_U\le\lambda K_0n\le2K_0n/a$ and $\Gamma_\Ec\ge\lambda\abs\Ec Y_*\ge\abs\Ec Y_*/a$; if $\Ec=\emptyset$ the loss is zero.
    This proves the service lower bound, and the same split, with absolute values on supports meeting $\Ec$, proves $\E\abs Z\le2\xi\Gamma$.

    For the curvature we follow the proof of \cite[Lemma~3.5]{Altschuler26Binary}.
    Put $\Phi_U\colonequals\sum_{i\in U}f(u_i)\le K_0n$, and sort the unexceptional $g_i$ decreasingly; since $f'$ is increasing, the coordinates with $u_i>\rho/2$ come first.
    The coordinates with $u_i\le\rho/2$ contribute at most $C/a^2$ to $(s/N)\sum_U\nu_i$.
    Among the other coordinates, the $m=\floor{N/s}$ largest derivatives contribute at most $(C/a)(s/N)\sum_{j\le m}g_j^*$.
    By \eqref{eq:ProfileCurvature} and the Cauchy--Schwarz inequality, the remaining ones satisfy
    \begin{equation*}
        \frac sN\sum_{j>m,\,u_j>\rho/2}\nu_j\le C\Big(\frac sN\sum_{j>m}(g_j^*)^2\Big)^{1/2}\Big(\frac{\Phi_U}{a^2N}\Big)^{1/2}\le\frac Ca\Big(\frac sN\sum_{j>m}(g_j^*)^2\Big)^{1/2}.
    \end{equation*}
    \Cref{thm:HeadTail}, applied on $U$ to $c_i=\sgn(S_i)g_i$, therefore proves the conditional assertion.
    For the unconditional bound, the unexceptional part is $p\sum_U\nu_i\le(s/N)\sum_U\nu_i$, which is bounded as above, now with $m_U=P_0m_U/P_0\le\frac{24}{23}\xi\Gamma$.
    On exceptional coordinates the two-unit neighborhood of $u_i$ lies in the exponential part, so $\nu_i\le Cg_i/a$, and their part is at most $(C/a)p\Gamma_\Ec\le(C/a)\xi\Gamma$.
    Dividing by $\Gamma$ gives $\E Q_A\le C\xi/a+C/(a^2\Gamma)$, and finally $1/(a^2\Gamma)\le C\sqrt s/(an)\le C\xi/a$.
\end{proof}

\subsection{Drift and a Foster inequality at every state}
\label{sec:GaugeFoster}

We now combine \cref{lem:GaugeRemainder,lem:GaugeService} into a drift inequality for the gauge in both models, and then into a Foster inequality for an exponential moment of the gauge.

\begin{lemma}[Gauge drift]
    \label{lem:GaugeDrift}
    For the gauge rule, in the binary model with $2\le s\le n/2$ and in the signed model with $2\le s\le n$, every state $S$ with $B(S)>1$ admits a number $\bar\xi=\bar\xi(S)$ such that $\Delta B\colonequals B(S')-B(S)$ satisfies
    \begin{equation}
        \label{eq:GaugeDrift}
        \frac{c\sqrt s}n\le\bar\xi\le\frac sn,\qquad\E\big[\Delta B\bigm|S\big]\le-\frac{\bar\xi}4,\qquad\E\big[\abs{\Delta B}\bigm|S\big]\le3\bar\xi.
    \end{equation}
\end{lemma}

\begin{proof}
    \emph{Binary inputs.}
    Take $\bar\xi\colonequals\xi$ from \eqref{eq:ServiceQuantities}.
    By \cref{lem:GaugeRemainder,lem:GaugeService},
    \begin{equation*}
        \E\Delta B=\frac{\E[xZ]}\Gamma+\E R_A\le-\frac\xi2+C\,\E Q_A+\frac Ca\cdot\frac{\E\abs Z}\Gamma\le-\frac\xi2+\frac Ca\xi,
    \end{equation*}
    and similarly $\E\abs{\Delta B}\le\E\abs Z/\Gamma+\E R_A\le2\xi+C\xi/a$.
    After increasing $a$ if necessary, this proves \eqref{eq:GaugeDrift}: the remainder costs at most $C\xi/a$, which the service absorbs once $a$ is large.

    \emph{Signed inputs, $s\le n/2$.}
    The gauge rule has the form required in \cref{lem:MarkConjugation}, since its exceptional set and its order of priority depend only on the magnitudes.
    Apply that lemma on the sign-invariant set $\Sc\colonequals\{S:B(S)>1\}$, with $\Phi=B(S')-B(S)$ and with $\Phi=\abs{B(S')-B(S)}$, and put $\bar\xi(S)\colonequals\E_\e\,\xi(\diag(\e)S)$; the binary bounds apply at every conjugate state because $B(\diag(\e)S)=B(S)>1$.

    \emph{Signed inputs, $s>n/2$.}
    Then $Y_*>K_0n$, so no state has exceptional coordinates and the rule is the gradient rule; thus $\E[xZ]=-\E\abs Z$.
    Put $m_\pm\colonequals\E\abs Z$ and $\bar\xi\colonequals m_\pm/\Gamma$.
    Conditionally on $A$, the signed sum $Z$ has second moment $\norm{g\mathbf1_A}_2^2$ and fourth moment at most $3\norm{g\mathbf1_A}_2^4$, so H\"older's inequality gives $\E_\e\abs Z\ge\norm{g\mathbf1_A}_2/\sqrt3$.
    Consequently,
    \begin{equation*}
        m_\pm\ge\frac p{\sqrt3}\norm g_2,\qquad m_\pm\ge\frac{\sqrt s}{\sqrt3\,n}\Gamma,\qquad m_\pm\le p\Gamma;
    \end{equation*}
    the first inequality uses $\E_A\norm{g\mathbf1_A}_2\ge\norm{\E_A(g\mathbf1_A)}_2=p\norm g_2$, and the second bounds the selected $\ell_2$-norm by the selected $\ell_1$-norm divided by $\sqrt s$.
    Directly from \eqref{eq:ProfileCurvature}, the Cauchy--Schwarz inequality and $\sqrt{n/s}<\sqrt2$,
    \begin{equation*}
        p\sum_i\nu_i\le\frac C{a^2}+\frac{Cp}{a\sqrt s}\sum_{u_i>\rho/2}g_i\sqrt{f(u_i)}\le\frac C{a^2}+\frac Cap\sqrt{n/s}\,\norm g_2\le\frac C{a^2}+\frac Cam_\pm.
    \end{equation*}
    After division by $\Gamma$, the first term is absorbed by $\bar\xi\ge\sqrt s/(\sqrt3n)$, which gives $\E Q_A\le C\bar\xi/a$.
    Since \cref{lem:GaugeRemainder} holds for arbitrary selected marks and signs, $\E\Delta B\le-\bar\xi+C\bar\xi/a$ and $\E\abs{\Delta B}\le\bar\xi+C\bar\xi/a$, which proves \eqref{eq:GaugeDrift} for large $a$.
\end{proof}

\begin{proposition}[Foster inequality for the gauge]
    \label{prop:GaugeFoster}
    There are absolute constants $\theta_0,c,C>0$ such that the gauge rule satisfies
    \begin{equation}
        \label{eq:GaugeFoster}
        \E\big[e^{\theta_0B(S')_+}\bigm|S\big]\le\Big(1-\frac{c\sqrt s}n\Big)e^{\theta_0B(S)_+}+\frac{Cs}n
    \end{equation}
    at every state $S\in\R^n$, for binary inputs with $2\le s\le n/2$ and for signed inputs with $2\le s\le n$.
\end{proposition}

\begin{proof}
    Let $B(S)>1$, and put $\theta_0\colonequals1/24$.
    Since $\abs{\Delta B}\le1$, \cref{lem:GaugeDrift} gives
    \begin{equation*}
        \E e^{\theta_0\Delta B}\le1+\theta_0\E\Delta B+\tfrac12\theta_0^2e^{\theta_0}\E(\Delta B)^2\le1-\frac{\theta_0\bar\xi}4+\frac32\theta_0^2e^{\theta_0}\bar\xi\le1-\frac{\theta_0\bar\xi}8\le1-\frac{c\sqrt s}n.
    \end{equation*}
    Here $B(S')>0$, so this is the multiplicative drift of $e^{\theta_0B_+}$.
    If $B(S)\le1$, then $B(S')\le2$, the function $b\mapsto e^{\theta_0b_+}$ is $\theta_0e^{2\theta_0}$-Lipschitz on $(-\infty,2]$, and \eqref{eq:GaugeInjection} gives an increase of at most $Cs/n$.
    To write this in the form \eqref{eq:GaugeFoster}, we add $(c\sqrt s/n)e^{\theta_0B(S)_+}\le c\sqrt se^{\theta_0}/n\le Cs/n$ to the right side.
\end{proof}

\begin{corollary}[Global moments and deterministic windows]
    \label{cor:GaugeMoments}
    Starting from $S_0=0$, in the ranges of \cref{prop:GaugeFoster},
    \begin{equation}
        \label{eq:GaugeGlobalMoment}
        \sup_{t\ge0}\E e^{\theta_0B(S_t)_+}\le C\sqrt s.
    \end{equation}
    For all deterministic integers $t_0,w\ge0$ and every $\beta\ge0$,
    \begin{equation}
        \label{eq:GaugeWindow}
        \P\Big\{\max_{t_0\le t\le t_0+w}B(S_t)>\beta\Big\}\le C\big(\sqrt s+sw/n\big)e^{-\theta_0\beta}.
    \end{equation}
\end{corollary}

\begin{proof}
    Since $B(0)=-h_0<0$, the initial exponential moment is one, and iteration of \eqref{eq:GaugeFoster} gives \eqref{eq:GaugeGlobalMoment}.
    Dropping the negative drift, $e^{\theta_0B(S_t)_+}+(Cs/n)(t_0+w-t)$, $t_0\le t\le t_0+w$, is a nonnegative supermartingale.
    Stopping it at the first time with $B>\beta$, or at $t_0+w$, and using bounded optional stopping together with \eqref{eq:GaugeGlobalMoment} at time $t_0$ proves \eqref{eq:GaugeWindow}.
\end{proof}

No event controlling the entire elapsed history is imposed: the window bound only uses the moment at its deterministic starting time, as in \cite[Lemma~27]{Wengiel26}.

\subsection{Moderate exponents}
\label{sec:ModerateExponents}

We now strengthen the exponent in \cref{prop:GaugeFoster} without changing the rule.
Throughout this subsection,
\begin{equation}
    \label{eq:ModerateExponent}
    2\le s\le\ell^2,\qquad\theta\colonequals c_0\min\{s,\log(n/s)\},
\end{equation}
where $c_0>0$ is a sufficiently small absolute constant, and $n\ge n_0$ for an absolute $n_0$; all statements hold uniformly in this range.
The exponent $\theta$ is an analysis parameter only.

\begin{proposition}[Moderate-exponent Foster inequality]
    \label{prop:ModerateFoster}
    There are absolute constants $c,C>0$ such that, in both models, at every state with $B(S)>1$,
    \begin{equation}
        \label{eq:ModerateFosterHigh}
        \E\big[e^{\theta(B(S')-B(S))}\bigm|S\big]\le1-\frac{c\min\{\theta\sqrt s,s\}}n.
    \end{equation}
    Consequently, with $\mathsf V_\theta(S)\colonequals e^{\theta B(S)_+}$ and $d_\theta\colonequals c\min\{\theta\sqrt s,s\}/n$,
    \begin{equation}
        \label{eq:ModerateFoster}
        \E\big[\mathsf V_\theta(S')\bigm|S\big]\le(1-d_\theta)\mathsf V_\theta(S)+C\theta e^{2\theta}p\qquad\text{at every state.}
    \end{equation}
\end{proposition}

\begin{proof}
    We first consider binary inputs and a state with $B>1$, and we keep the notation of \cref{sec:GaugeService}.
    Put $\hat g_i\colonequals g_i/\Gamma$, so that $\sum_i\hat g_i=1$.
    Since $g_i\le\lambda f(u_i)$, with equality on $\Ec$, and $\Gamma\le\lambda K_0n$,
    \begin{equation}
        \label{eq:ModerateSetting}
        \abs\Ec\le\frac{K_0n}{Y_*},\quad N\ge\frac n2,\quad2s\le N,\quad P_0\ge1-p\abs\Ec\ge\frac12,\quad\hat g_i\ge\hat g_*\colonequals\frac{Y_*}{K_0n}=2D_0p\ell^8\quad(i\in\Ec).
    \end{equation}
    By \eqref{eq:GaugeRemainder} and \eqref{eq:GaugeNormalization}, after increasing $a$,
    \begin{equation}
        \label{eq:ModerateRemainder}
        \Delta B=\frac{xZ}\Gamma+R_A,\qquad0\le R_A\le\frac1{32}\sum_{i\in A}\hat g_i+r_\Gamma,\qquad r_\Gamma\colonequals\frac C{a^2\Gamma}.
    \end{equation}

    \emph{Supports meeting $\Ec$.}
    Order the exceptional coordinates by decreasing height, with the index tie order of the rule.
    Let $E_i$ be the event that $i$ is the exceptional coordinate of highest priority in $A$; these events partition $\{A\cap\Ec\ne\emptyset\}$.
    On $E_i$ the rule moves $i$ inward, and every other selected coordinate has $\hat g_j\le\hat g_i$; hence
    \begin{equation}
        \label{eq:ModerateHot}
        \Delta B\le-\frac{31}{32}\hat g_i+\frac{33}{32}\sum_{j\in A\setminus\{i\}}\hat g_j+r_\Gamma.
    \end{equation}
    Conditionally on $E_i$, the remaining $s-1$ selected coordinates form a uniform subset of the population obtained by removing $i$ and all exceptional coordinates of higher priority; its inclusion probability $p'$ is at most $2p$.
    Put $\theta'\colonequals(33/32)\theta$.
    Expanding $\prod_j[1+(e^{\theta'\hat g_j}-1)\mathbf1\{j\in A\}]$ and using $(s-1)_r/(N')_r\le(p')^r$ term by term, where $N'$ is the size of this population, we obtain
    \begin{equation}
        \label{eq:ModerateMGF}
        \E\Big[\exp\Big(\theta'\sum_{j\in A\setminus\{i\}}\hat g_j\Big)\Bigm|E_i\Big]\le\exp\Big(p'\sum_j\big(e^{\theta'\hat g_j}-1\big)\Big)\le\exp\Big(\frac{2p}{\hat g_i}\big(e^{\theta'\hat g_i}-1\big)\Big),
    \end{equation}
    where the second inequality uses $e^{\theta'z}-1\le z(e^{\theta'\hat g_i}-1)/\hat g_i$ for $0\le z\le\hat g_i$ and $\sum_j\hat g_j\le1$.
    No independence of inclusion indicators is used.
    Uniformly for $\alpha\in[\hat g_*,1]$, we have
    \begin{equation}
        \label{eq:ModerateUniform}
        \frac{2p\big(e^{\theta'\alpha}-1\big)}{\theta\alpha^2}\le\frac14.
    \end{equation}
    Indeed, the function $(e^z-1)/z^2$ first decreases and then increases on $z>0$, since its derivative has the sign of $e^z(z-2)+2$, which has exactly one positive zero; so the maximum in \eqref{eq:ModerateUniform} is attained at an endpoint.
    At $\hat g_*$, we have $\theta\hat g_*=O(s\ell^9/n)=o(1)$, and the left side is at most $4p/\hat g_*=O(\ell^{-8})$.
    At $1$, it is at most $2p^{1-(33/32)c_0}/\theta=o(1)$, provided $c_0<16/33$; here $\theta\ge2c_0$ and $p\le\ell^2/n$ for large $n$.
    Moreover, $r_\Gamma/\hat g_i=C/(a^2g_i)\le C/(a^2\lambda Y_*)\le1/32$ for exceptional $i$.
    Combining \eqref{eq:ModerateHot}--\eqref{eq:ModerateUniform} gives
    \begin{equation}
        \label{eq:ModerateHotMGF}
        \E\big[e^{\theta\Delta B}\bigm|E_i\big]\le e^{-\theta\hat g_i/2},\qquad\P(E_i)\ge\frac p2,
    \end{equation}
    where the probability bound holds because, conditionally on selecting $i$, a coordinate of higher priority is selected with probability at most $2p\abs\Ec=o(1)$.
    Supports with several exceptional coordinates are included in the conditional moment generating function; none have been discarded.

    \emph{Supports avoiding $\Ec$.}
    On $\{A\subseteq U\}$ the rule is the gradient rule.
    Put $\xi_U\colonequals m_U/\Gamma$ and $\Phi_U\colonequals\sum_{i\in U}f(u_i)$.
    \Cref{lem:GaugeService} and \eqref{eq:GaugeRemainder} give $\E[R_A\mid A\subseteq U]\le(C/a)\xi_U+C/(a^2\Gamma)$, and after increasing $a$,
    \begin{equation}
        \label{eq:ModerateCold}
        \E\big[\Delta B\bigm|A\subseteq U\big]\le-\frac34\xi_U+\frac C{a^2\Gamma},\qquad\E\big[\abs{\Delta B}\bigm|A\subseteq U\big]\le2\xi_U+\frac C{a^2\Gamma}.
    \end{equation}
    Since $g_i\le\lambda Y_*$ on $U$, \eqref{eq:ModerateRemainder} gives $\abs{\Delta B}\le\delta_c\colonequals CsY_*\sqrt s/n=O(\ell^{13}/n)$ on every support avoiding $\Ec$, and in particular $\theta\delta_c=o(1)$.
    The inequality $e^{\theta z}\le1+\theta z+\frac12\theta^2z^2e^{\theta\abs z}$, with $z^2\le\delta_c\abs z$, and \eqref{eq:ModerateCold} imply
    \begin{equation}
        \label{eq:ModerateColdMGF}
        \E\big[e^{\theta\Delta B}\bigm|A\subseteq U\big]\le1-\frac12\theta\xi_U+\frac{C\theta}{a^2\Gamma}.
    \end{equation}

    \emph{Absorbing the curvature term.}
    The positive term $C\theta/(a^2\Gamma)$ in \eqref{eq:ModerateColdMGF} comes from the curvature remainder and grows with $\theta$; we now show that the deficit dominates it.
    Write $\eta_U\colonequals\Gamma_U/\Gamma$ and $\eta_\Ec\colonequals1-\eta_U$.
    If $\Phi_U\ge2n$, then $\sum_{i\in U}(f(u_i)-1)\ge n$, and the lower bound on $f'$ in \eqref{eq:ProfileCurvature} gives $\Gamma_U\ge cn/(a\sqrt s)$.
    The service bound on $U$ gives $m_U\ge\sqrt s\,\Gamma_U/(144N)\ge c/a$, so $C/(a^2\Gamma)\le\xi_U/4$ after increasing $a$.
    Combining \eqref{eq:ModerateHotMGF}, \eqref{eq:ModerateColdMGF} and $P_0\ge1/2$, the total deficit is at least
    \begin{equation*}
        c\Big[\frac{\theta\sqrt s}n\eta_U+\frac sn\min\{\theta\eta_\Ec,1\}\Big];
    \end{equation*}
    here we used $1-e^{-z/2}\ge c\min\{z,1\}$, $\sum_{i\in\Ec}\min\{\theta\hat g_i,1\}\ge\min\{\theta\eta_\Ec,1\}$ and $\xi_U\ge\sqrt s\,\eta_U/(144n)$.
    If $\eta_U\ge1/2$, the first term suffices; otherwise the second is at least $c\min\{\theta\sqrt s,s\}/n$.
    If $\Phi_U<2n$, the exceptional profile mass exceeds $(K_0-2)n=2n$, so
    \begin{equation*}
        \Gamma\ge\lambda(K_0-2)n\ge2n/a,\qquad\eta_\Ec\ge(K_0-2)/K_0=1/2.
    \end{equation*}
    The exceptional deficit is then at least $cp\min\{\theta,1\}$, whereas the positive term is at most $C\theta/(an)$.
    For $\theta\le1$ it is absorbed by choosing $a$ large; for $\theta\ge1$ we use $\theta\le c_0s$ and choose $c_0$ small.
    The remaining deficit again dominates $c\min\{\theta\sqrt s,s\}/n$, which proves \eqref{eq:ModerateFosterHigh} for binary inputs.

    \emph{Low states and signed inputs.}
    For $B>1$, the Lipschitz property gives $B(S')>0$, so \eqref{eq:ModerateFosterHigh} applies directly to $\mathsf V_\theta$.
    For $B\le1$, we have $B(S')\le2$, and \eqref{eq:GaugeInjection} gives $\E(\Delta B)_+\le Cp$.
    The function $b\mapsto e^{\theta b_+}$ is $\theta e^{2\theta}$-Lipschitz on $(-\infty,2]$, whence $\E[\mathsf V_\theta(S')-\mathsf V_\theta(S)\mid S]\le C\theta e^{2\theta}p$.
    To write this in the form \eqref{eq:ModerateFoster}, we add $d_\theta\mathsf V_\theta(S)\le d_\theta e^\theta\le C\theta e^{2\theta}p$ to the right side.
    This proves \eqref{eq:ModerateFoster} for binary inputs at every state.
    Both \eqref{eq:ModerateFosterHigh} and \eqref{eq:ModerateFoster} have right sides that depend only on magnitudes, and $2s\le n$ in the range \eqref{eq:ModerateExponent}; \cref{lem:MarkConjugation} transfers them to signed inputs.
\end{proof}

\begin{corollary}[Moderate moments and deterministic windows]
    \label{cor:ModerateMoments}
    Starting from $S_0=0$, in either model and in the range \eqref{eq:ModerateExponent},
    \begin{equation}
        \label{eq:ModerateGlobal}
        \sup_{t\ge0}\E\mathsf V_\theta(S_t)\le Ce^{2\theta}\max\{\sqrt s,\theta\}.
    \end{equation}
    For all deterministic integers $t_0,w\ge0$ and every $\beta\ge0$,
    \begin{equation}
        \label{eq:ModerateWindow}
        \P\Big\{\max_{t_0\le t\le t_0+w}B(S_t)>\beta\Big\}\le Ce^{2\theta}\big[\max\{\sqrt s,\theta\}+\theta pw\big]e^{-\theta\beta}.
    \end{equation}
\end{corollary}

\begin{proof}
    At zero, $B<0$ and $\mathsf V_\theta=1$.
    Iterating \eqref{eq:ModerateFoster} proves \eqref{eq:ModerateGlobal}, since $\theta s/\min\{\theta\sqrt s,s\}=\max\{\sqrt s,\theta\}$.
    On the window, $\mathsf V_\theta(S_t)+C\theta e^{2\theta}p\,(t_0+w-t)$ is a nonnegative supermartingale by \eqref{eq:ModerateFoster}; stopping it at the first time with $B>\beta$ or at the end of the window and using \eqref{eq:ModerateGlobal} at time $t_0$ proves \eqref{eq:ModerateWindow}.
\end{proof}

The restriction $\theta=O(s)$ comes from this all-state certificate and is not a discrepancy lower bound; \cref{sec:Bootstrap} uses a much larger exponent, but only inside a region of bounded heights.

\subsection{Composition at a prescribed time}
\label{sec:Composition}

We now combine bounds that hold at every fixed time with the layer iteration of \cref{lem:LayerIteration} in a deterministic window before the prescribed time $T$, as in our analysis of a rare-activation law \cite[Appendix~B.3--B.4]{Wengiel26}.
The following proposition applies to every rule that chooses the sign as a function of the current state and the arrival, as in \cref{sec:Tools}; its good events require a bounded population, bounded heights and the isolating property.
The three terms on the right side of \eqref{eq:Composition} correspond to three sources of failure: a good event fails somewhere in the final window, the population fails to drop at one of the levels, or a coordinate still exceeds the last level $b_J$ by at least $u$ at time $T$.
The length of the window depends on the height bound $\bar H$, but not on $T$.

\begin{proposition}[Composition]
    \label{prop:Composition}
    Consider a rule as in \cref{sec:Tools}, in either model.
    Fix $k\ge1$, put $R\colonequals\ceil{100(k+10)\ell}$, and let $b_0\ge1$, $\bar H\ge0$ and $q_0$ be integers with $2R<q_0\le n/(8A_{\mathrm{cl}}s)$.
    Let $J$, $q_j$ and $\vartheta_j$ be as in \cref{lem:LayerIteration}, and put
    \begin{equation}
        \label{eq:CompositionParameters}
        b_J\colonequals b_0+2J,\qquad M\colonequals n^{1+\bar H/2},\qquad w\colonequals\ceil{\frac{8n}s\log(1+M)},\qquad\bar w\colonequals(J+1)w.
    \end{equation}
    Let $\Gc_t\in\Fc_t$ be events on which $N_{b_0}(S_t)\le q_0$, all heights are at most $\bar H$, and $S_t$ is $(b,0)$-isolating for every integer $b\ge b_0$.
    Then, for every deterministic integer $T\ge0$ and every integer $u\ge1$,
    \begin{equation}
        \label{eq:Composition}
        \P\big(\norm{S_T}_\infty\ge b_J+u\big)\le\P\big(\Gc_t\text{ fails for some }t\in[(T-\bar w)_+,T]\big)+J(\bar w+1)e^{-c_*R}+\frac{8eq_J}{\vartheta_J^u}.
    \end{equation}
\end{proposition}

\begin{proof}
    \Cref{lem:LayerIteration} applies with $g=0$, so that its levels are $b_j=b_0+2j$.
    On $\Gc_t$, since all heights are at most $\bar H$ and $\vartheta_j\le\sqrt n$, we have $\sum_i\vartheta_j^{(\abs{S_i}-b_j)_+}\le n\vartheta_j^{\bar H}\le M$ for every $j$.
    If $T\ge\bar w$, take $t_j\colonequals T-(J+1-j)w$ for $0\le j\le J+1$; consecutive times differ by $w\ge(8n/s)\log(1+M)$.
    If $T<\bar w$, take $t_0=\cdots=t_{J+1}\colonequals0$ and replace $M$ by zero; this is admissible because all heights vanish at time zero.
    In both cases $t_0=(T-\bar w)_+$, $t_{J+1}\le T$ and $T-t_0\le\bar w$, and the event $\Vc$ of \cref{lem:LayerIteration} is the complement of the first event in \eqref{eq:Composition}.
    On $\Vc$, either a population condition $N_{b_j}(S_t)\le q_j$ with $1\le j\le J$ and $t\in[t_j,T]$ fails, which has probability at most $J(\bar w+1)e^{-c_*R}$ by \eqref{eq:LayerFailure}, or $\Lc_T$ holds; in the latter case, \eqref{eq:LayerFinal} with $t=T$, $r=1$ and $d=u$ bounds the probability that some height at time $T$ is at least $b_J+u$.
\end{proof}

For the gauge rule, a bound on the gauge implies all three requirements.

\begin{corollary}[Composition for the gauge rule]
    \label{cor:GaugeComposition}
    Fix $k\ge1$ and $C_B\ge1$, suppose $2\le s\le\ell^2$, and put $q_0\colonequals\floor{n/(D_0s\ell^8)}$.
    For a deterministic level $1\le B_0\le C_B\ell$, put
    \begin{equation*}
        b_0\colonequals\ceil{B_0+h_*},\qquad\bar H\colonequals\ceil{B_0+C_{\mathrm{env}}(\sqrt s+\ell)},
    \end{equation*}
    and let $R$, $J$, $q_j$, $\vartheta_j$, $b_J$ and $\bar w$ be as in \cref{prop:Composition}.
    Then, for all $n\ge n_0(k,C_B)$, we have $J=O(\log\ell)$, $\log\vartheta_J\ge(\log n)/4$ and $\bar w\le n^2$, and the gauge rule satisfies, in either model, for every deterministic integer $T\ge0$ and every integer $u\ge1$,
    \begin{equation}
        \label{eq:GaugeComposition}
        \P\big(\norm{S_T}_\infty\ge b_J+u\big)\le\P\Big(\max_{(T-\bar w)_+\le t\le T}B(S_t)>B_0\Big)+J(\bar w+1)e^{-c_*R}+\frac{8eq_J}{\vartheta_J^u}.
    \end{equation}
\end{corollary}

\begin{proof}
    Uniformly for $s\le\ell^2$ and large $n$, we have $K_0n/Y_*\le q_0\le n/(8A_{\mathrm{cl}}s)$, since $D_0\ell^8\ge8A_{\mathrm{cl}}$, and $q_0\ge n/(2D_0\ell^{10})>2R$.
    By \cref{lem:GaugeGeometry}, on $\Gc_t\colonequals\{B(S_t)\le B_0\}$ the state is $(b,0)$-isolating for every integer $b\ge b_0$ and satisfies $N_{b_0}(S_t)\le K_0n/Y_*\le q_0$, and all heights are at most $\bar H$ by \eqref{eq:GaugeEnvelope}, since $\log(en/s)\le\ell$.
    Hence \cref{prop:Composition} gives \eqref{eq:GaugeComposition}.
    Moreover, $J\le\log_2\log_2n+3$ by \cref{lem:LayerIteration}, and $q_J\le2R$ gives $\vartheta_J^2=n/(C_{\mathrm{cl}}sq_J)\ge\sqrt n$.
    Since $\log(1+n^{1+\bar H/2})\le(2+\bar H)\ell$ and $\bar H=O_{C_B}(\ell)$, we have $\bar w=O_{C_B}((n/s)\ell^2\log\ell)\le n^2$.
\end{proof}

\begin{proposition}[Coarse terminal bound]
    \label{prop:CoarseTerminal}
    For every fixed $k\ge1$ there are constants $C_k$ and $n_k$ such that the gauge rule satisfies, for all $n\ge n_k$ and all integers $T\ge0$, in the binary model with $2\le s\le n/2$ and in the signed model with $2\le s\le n$,
    \begin{equation}
        \label{eq:CoarsePoly}
        \P\bc{\norm{S_T}_\infty>C_k\Big(\sqrt s+\log\ell+\frac\ell s\Big)}\le n^{-k}.
    \end{equation}
\end{proposition}

\begin{proof}
    If $s\ge\ell^2$, put $B_0\colonequals C_k'\sqrt s$.
    By \cref{cor:GaugeMoments} and Markov's inequality, $\P\{B(S_T)>B_0\}\le C\sqrt se^{-\theta_0C_k'\ell}\le n^{-k}$ for $C_k'$ large enough, and on the complement \eqref{eq:GaugeEnvelope} with $\ell\le\sqrt s$ gives $\norm{S_T}_\infty\le C_k\sqrt s$.
    This range includes the signed range above $n/2$.

    Let $s\le\ell^2$, let $\theta$ be as in \eqref{eq:ModerateExponent}, and put $B_0\colonequals4+(k+20)\log n/\theta$.
    Uniformly in this range, $\log(n/s)\ge\ell/2$ for large $n$, so $\theta\ge c_0\min\{s,\ell/2\}$ and $B_0=O_k(1+\ell/s)\le C_k''\ell$ for a constant $C_k''$.
    We apply \cref{cor:GaugeComposition} with $C_B=C_k''$.
    By \eqref{eq:ModerateWindow}, the gauge term in \eqref{eq:GaugeComposition} is at most
    \begin{equation*}
        Ce^{2\theta}\big\{\max\{\sqrt s,\theta\}+\theta p\bar w\big\}e^{-\theta B_0}\le n^{-k-10},
    \end{equation*}
    since $p\bar w=O_k(\ell^2\log\ell)$, $\theta=O(\ell)$ and $e^{2\theta-\theta B_0}=e^{-2\theta}n^{-k-20}$.
    The population term is at most $n^{-k-10}$, since $\bar w\le n^2$, $J\le n$ and $c_*R\ge R\ge100(k+10)\log n$, as $c_*=\log(32/e)>1$.
    At the last level choose
    \begin{equation}
        \label{eq:FinalExcess}
        u_*\colonequals\ceil{\frac{(k+20)\log n+\log(8eq_J)}{\log\vartheta_J}};
    \end{equation}
    then $u_*=O_k(1)$ and $8eq_J/\vartheta_J^{u_*}\le n^{-k-20}$.
    Since $b_J+u_*\le B_0+h_*+1+2J+u_*\le C_k(\sqrt s+\log\ell+\ell/s)$ by \eqref{eq:GaugeEnvelope} and $J=O(\log\ell)$, this proves \eqref{eq:CoarsePoly}.
    The signed model is covered because \cref{prop:ModerateFoster,lem:LayerIteration} hold in both models.
\end{proof}

Since $\ell/s\le\sqrt s$ for $s\ge\ell^{2/3}$, the bound \eqref{eq:CoarsePoly} has the sharp radius $O_k(\sqrt s+\log\log n)$ with polynomially small failure probability in this range.
With the exponent $\theta_0$ of \cref{prop:GaugeFoster} in place of $\theta$, and with $B_0$ of order $\log\ell$, the same argument gives the sharp radius for all $s\le\ell^2$, but only with failure probability $(\log n)^{-k}$; for $\log\ell<s<\ell^{2/3}$, \cref{sec:Bootstrap} improves this to $n^{-k}$.

\subsection{Polynomially small failure probability at the sharp radius: a bootstrap}
\label{sec:Bootstrap}

In this subsection we remove the term $\ell/s$ from \eqref{eq:CoarsePoly} in the range
\begin{equation}
    \label{eq:BootstrapRange}
    \log\ell<s<\ell^{2/3}.
\end{equation}
We use the exponential moment of the gauge with the large exponent $\hat\theta\colonequals n^{1/4}$, but only at states whose heights are at most $H=O(\ell/\log\ell)$, where the gauge moves by at most $n^{-3/4}$ per step; we control the exit from this height region by the local moment of degree one, and the coarse bound \eqref{eq:CoarsePoly} supplies the starting state.

\begin{lemma}[A large exponent in a height region]
    \label{lem:HighDrift}
    Let $2\le s\le n/2$, in either model, let $H\ge0$, and put
    \begin{equation}
        \label{eq:SmallJumps}
        \delta_H\colonequals\frac{Cs^{3/2}}n\,e^{\lambda(H+1)+2\sqrt s+4},\qquad\hat{\mathsf V}(S)\colonequals e^{\hat\theta B(S)_+},\qquad\hat d\colonequals\frac{c\hat\theta\sqrt s}n,\qquad\hat\beta\colonequals C\hat\theta pe^{\hat\theta(1+\delta_H)},
    \end{equation}
    with suitable absolute constants $c,C>0$.
    At every state $S$ with $\norm S_\infty\le H$, every arrival and either common sign give $\abs{B(S')-B(S)}\le\delta_H$.
    If $\delta_H\le n^{-3/4}$ and $n\ge n_0$, then the gauge rule satisfies, at every such state,
    \begin{equation}
        \label{eq:HighDrift}
        \E\big[\hat{\mathsf V}(S')\bigm|S\big]\le(1-\hat d)\hat{\mathsf V}(S)+\hat\beta,
    \end{equation}
    and $0<\hat d\le1/2$ and $\hat\beta/\hat d\le C\sqrt se^{\hat\theta(1+\delta_H)}$.
\end{lemma}

\begin{proof}
    Since $h_0<\rho$ and $\lambda\le2/a$, we have $\lambda h_0\le2\sqrt s+4$.
    At a state with heights at most $H'$, the bound $B\ge-h_0$ gives $u_i\le H'+h_0$, and \cref{lem:ProfileEstimates} gives $g_i\le\lambda f(u_i)\le\lambda e^{\lambda(H'+h_0)}$.
    By \eqref{eq:GaugeComparison}, $\Gamma\ge cn/(a\sqrt s)$ and $a\lambda\le2$, every update of such a state increases the gauge by at most
    \begin{equation*}
        \frac{Ca\sqrt s}n\Big(s\lambda e^{\lambda H'+2\sqrt s+4}+\frac1{a^2}\Big)\le\frac{Cs^{3/2}}ne^{\lambda H'+2\sqrt s+4}.
    \end{equation*}
    Apply this to the update $S\to S'$ with $H'=H$, and to the reverse update $S'\to S=S'+(-x)V$, which has the same support and marks and the opposite common sign, with $H'=H+1$.
    This proves the bound on $\abs{B(S')-B(S)}$; it holds for every sign, at coordinates with $S_i=0$ and at negative gauges.

    Let $B(S)>1$.
    Then $B(S')>0$, $(\Delta B)^2\le\delta_H\abs{\Delta B}$ and $\hat\theta\delta_H\le n^{-1/2}$, so the inequality $e^z\le1+z+\frac12z^2e^{\abs z}$ and \cref{lem:GaugeDrift} give
    \begin{equation*}
        \E\big[e^{\hat\theta\Delta B}\bigm|S\big]\le1-\frac{\hat\theta\bar\xi}4+\frac32\hat\theta^2\delta_He^{\hat\theta\delta_H}\bar\xi\le1-\frac{\hat\theta\bar\xi}8\le1-\frac{c\hat\theta\sqrt s}n
    \end{equation*}
    for large $n$; this is the multiplicative drift of $\hat{\mathsf V}$.
    If $B(S)\le1$, then $B(S')\le1+\delta_H$, the function $b\mapsto e^{\hat\theta b_+}$ is $\hat\theta e^{\hat\theta(1+\delta_H)}$-Lipschitz on $(-\infty,1+\delta_H]$, and \eqref{eq:GaugeInjection} gives $\E[\hat{\mathsf V}(S')\mid S]\le\hat{\mathsf V}(S)+C\hat\theta pe^{\hat\theta(1+\delta_H)}$; to obtain \eqref{eq:HighDrift}, we add $\hat d\hat{\mathsf V}(S)\le\hat de^{\hat\theta}\le\hat\theta pe^{\hat\theta}$ to the right side.
    Finally, $\hat d\le\hat\theta\sqrt s/n\le n^{-1/4}$, and $\hat\beta/\hat d=(C/c)\sqrt se^{\hat\theta(1+\delta_H)}$.
\end{proof}

\begin{proposition}[Sharp radius with polynomially small failure probability]
    \label{prop:Bootstrap}
    For every fixed $k\ge1$ there are constants $C_k$ and $n_k$ such that the gauge rule satisfies, in both models, for all $n\ge n_k$, all $s$ in the range \eqref{eq:BootstrapRange}, and all integers $T\ge0$,
    \begin{equation}
        \label{eq:Bootstrap}
        \P\bc{\norm{S_T}_\infty>C_k\big(\sqrt s+\log\ell\big)}\le n^{-k}.
    \end{equation}
\end{proposition}

\begin{proof}
    \emph{Parameters.}
    Put $k_1\colonequals k+20$, let $C_{k_1}$ be the constant of \eqref{eq:CoarsePoly} for the exponent $k_1$, let $q_0\colonequals\floor{n/(D_0s\ell^8)}$ and $\vartheta_0\colonequals(C_{\mathrm{cl}}pq_0)^{-1/2}$ be as in \cref{cor:GaugeComposition}, and put
    \begin{equation}
        \label{eq:BootstrapLevels}
        \begin{gathered}
            H_0\colonequals\ceil{C_{k_1}\Big(\sqrt s+\log\ell+\frac\ell s\Big)},\qquad b_1\colonequals\ceil{H_0+2+h_*},\\
            u_1\colonequals\ceil{\frac{(k_1+10)\ell+\log(8e(q_0+1))}{\log\vartheta_0}},\qquad H\colonequals b_1+u_1.
        \end{gathered}
    \end{equation}
    The level $H_0+2$ is a barrier for the gauge, $b_1$ is the base of a local layer, and $H$ is a barrier for the heights.
    In the range \eqref{eq:BootstrapRange}, we have $\sqrt s\le\ell^{1/3}$ and $\ell/s<\ell/\log\ell$, so $1\le H_0\le3C_{k_1}\ell/\log\ell$ for large $n$.
    Moreover $h_*\le C(\sqrt s+\log\ell)\le\ell/\log\ell$ by \eqref{eq:GaugeEnvelope}, and $\log\vartheta_0\ge4\log\ell$ because $C_{\mathrm{cl}}pq_0\le C_{\mathrm{cl}}/(D_0\ell^8)\le\ell^{-8}$; hence $u_1\le(k_1+12)\ell/(4\log\ell)$ and $H\le(3C_{k_1}+k_1+5)\ell/\log\ell$.
    Since also $s^{3/2}\le\ell$ and $2\sqrt s\le2\ell^{1/3}$, we obtain $\delta_H\le n^{-1+o(1)}\le n^{-3/4}$ for large $n$, and \cref{lem:HighDrift} applies at every state with heights at most $H$.
    Finally, let $\bar w$ be the window of \cref{cor:GaugeComposition} for $k_1$ in place of $k$, $C_B=1$ and $B_0=2$, and put $w_g\colonequals\ceil{\hat\theta(H_0+1)/\hat d}$ and $w_1\colonequals w_g+\bar w$; then $w_g=O_k(n\ell)$ and $w_1\le n^3$.

    \emph{Coupled barriers.}
    Let $t_0\ge0$ be a deterministic time and $G\in\Fc_{t_0}$ an event on which $\norm{S_{t_0}}_\infty\le H_0$.
    Using times relative to $t_0$, let $\tau$ be the first time $t\ge0$ at which $B(S_{t_0+t})>H_0+2$ or $\norm{S_{t_0+t}}_\infty>H$, and put
    \begin{equation*}
        \hat F_t\colonequals\E\big[\hat{\mathsf V}(S_{t_0+t})\mathbf1_G\mathbf1\{\tau\ge t\}\big],\qquad\Lambda_t\colonequals\E\big[e_1(W(S_{t_0+t}))\mathbf1_G\mathbf1\{\tau\ge t\}\big],\qquad W_i\colonequals\vartheta_0^{(\abs{S_i}-b_1)_+}-1.
    \end{equation*}
    The event $\{\tau\ge t+1\}=\{\tau>t\}$ belongs to $\Fc_{t_0+t}$, and on it the state $S_{t_0+t}$ has heights at most $H$ and gauge at most $H_0+2$.
    Hence \eqref{eq:HighDrift} applies at this state, and so does \cref{lem:LocalClearance}\ref{item:LocalDrift} with $r=1$, $b=b_1$, $g=0$ and $q=q_0$, because the state is $(b_1,0)$-isolating with $N_{b_1}\le K_0n/Y_*\le q_0$ by \cref{lem:GaugeGeometry}.
    Enlarging the indicator $\mathbf1\{\tau>t\}$ to $\mathbf1\{\tau\ge t\}$ in the nonnegative leading terms gives
    \begin{equation*}
        \hat F_{t+1}\le(1-\hat d)\hat F_t+\hat\beta,\qquad\Lambda_{t+1}\le(1-\gamma)\Lambda_t+\mathsf a_1,
    \end{equation*}
    with $\gamma=p/4$ and $\mathsf a_1=2\vartheta_0^2(pq_0)^2$; these moments include the values at the exit time.
    On $G$, we have $B(S_{t_0})\le H_0$ and all heights are below $b_1$, so $\hat F_0\le e^{\hat\theta H_0}$ and $\Lambda_0=0$; therefore, for all $t\ge0$,
    \begin{equation}
        \label{eq:BootstrapMoment}
        \hat F_t\le e^{-\hat dt}e^{\hat\theta H_0}+C\sqrt se^{\hat\theta(1+\delta_H)},\qquad\Lambda_t\le\frac{\mathsf a_1}\gamma=\frac{8q_0}{C_{\mathrm{cl}}}.
    \end{equation}
    At the time $\tau$, either $\hat{\mathsf V}\ge e^{\hat\theta(H_0+2)}$, or some height is at least $b_1+u_1+1$ and then $e_1(W)\ge\vartheta_0^{u_1}$.
    Markov's inequality, a union bound over $t\le w_1$, $H_0\ge1$, $\delta_H\le1$ and the definition of $u_1$ give
    \begin{equation}
        \label{eq:BootstrapExit}
        \P\big(G\cap\{\tau\le w_1\}\big)\le(w_1+1)\Big[e^{-2\hat\theta}+C\sqrt se^{-\hat\theta}+\frac{8q_0}{C_{\mathrm{cl}}\vartheta_0^{u_1}}\Big]\le e^{-\hat\theta/2}+n^{-k_1-6}.
    \end{equation}
    Neither moment is conditioned on the future, and the support of every arrival keeps its uniform law.

    \emph{Clearance.}
    For $w_g\le t\le w_1$, \eqref{eq:BootstrapMoment} gives $\hat F_t\le e^{-\hat\theta}+C\sqrt se^{\hat\theta(1+\delta_H)}$, and hence
    \begin{equation*}
        \P\big(G\cap\{\tau>w_1\}\cap\{B(S_{t_0+t})>2\}\big)\le e^{-2\hat\theta}\hat F_t\le e^{-3\hat\theta}+C\sqrt se^{-\hat\theta(1-\delta_H)}.
    \end{equation*}
    Together with \eqref{eq:BootstrapExit} and a union bound over $t$,
    \begin{equation}
        \label{eq:BootstrapWindow}
        \P\Big(G\cap\Big\{\max_{w_g\le t\le w_1}B(S_{t_0+t})>2\Big\}\Big)\le n^{-k_1-6}+(w_1+2)e^{-\hat\theta/2}\le2n^{-k_1-6}.
    \end{equation}
    If $t_0=0$ and $G$ is the sure event, then $\hat F_0=1$ and $\hat F_t\le1+C\sqrt se^{\hat\theta(1+\delta_H)}$ for all $t\ge0$; the last display then holds for all $0\le t\le w_1$ with $e^{-3\hat\theta}$ replaced by $e^{-2\hat\theta}$, and \eqref{eq:BootstrapWindow} holds with $w_g$ replaced by zero.

    \emph{Conclusion.}
    Apply \cref{cor:GaugeComposition} with $k_1$ in place of $k$, with $C_B=1$, $B_0=2$, and with $u=u_*$ from \eqref{eq:FinalExcess}, again with $k_1$ in place of $k$.
    As in the proof of \cref{prop:CoarseTerminal}, the last two terms in \eqref{eq:GaugeComposition} are at most $n^{-k_1-10}+n^{-k_1-20}$, and the radius $b_J+u_*$ is at most $C_k(\sqrt s+\log\ell)$.
    If $T\ge w_1$, take $t_0\colonequals T-w_1$ and $G\colonequals\{\norm{S_{t_0}}_\infty\le H_0\}$; then $\P(G^c)\le n^{-k_1}$ by \eqref{eq:CoarsePoly}, and $[t_0+w_g,t_0+w_1]=[T-\bar w,T]$.
    If $T<w_1$, take $t_0=0$ and the sure event; the window $[(T-\bar w)_+,T]$ lies in $[0,w_1]$.
    In both cases \eqref{eq:BootstrapWindow} bounds the first term in \eqref{eq:GaugeComposition} by $n^{-k_1}+2n^{-k_1-6}$, and altogether $\P\{\norm{S_T}_\infty\ge b_J+u_*\}\le2n^{-k_1}\le n^{-k}$.
\end{proof}

The lower end of \eqref{eq:BootstrapRange} keeps $H$ of order at most $\ell/\log\ell$, so that $e^{\lambda H}=n^{o(1)}$ in \eqref{eq:SmallJumps}; the upper end is where \eqref{eq:CoarsePoly} becomes sharp.

\subsection{A hybrid rule for small sparsity}
\label{sec:CoshHybrid}

For bounded $s$, the coarse bound \eqref{eq:CoarsePoly} places the heights only below a level of order $\log n$, and for such a height region the jumps $\delta_H$ in \eqref{eq:SmallJumps} need not be small, so the bootstrap does not apply.
For $s\le\log\ell$, \cref{thm:Terminal} therefore uses a second rule, which combines a priority for high coordinates, as in \cref{def:GaugeRule}, with the gradient of the hyperbolic cosine potential of Spencer \cite{Spencer77}; its populations are controlled by moments of elementary symmetric polynomials, without a gauge.
Assume $n\ge4s$, and put
\begin{equation*}
    \omega\colonequals\frac1{1600\sqrt s},\qquad r_*\colonequals\floor{\frac n{2^{16}s^{3/2}}},
\end{equation*}
and suppose $r_*\ge1$.

\begin{definition}[The hybrid rule]
    \label{def:HybridRule}
    Let $b\ge1$ and $0\le q\le r_*$ be integers, let $V$ be an arrival with support $A$ and marks $\e_i\colonequals V(i)$, and put $\mathsf H\colonequals\{i:\abs{S_i}\ge b\}$.
    If $\abs{\mathsf H}\le q$ and $A\cap\mathsf H\ne\emptyset$, the \emph{hybrid rule with parameters $b,q$} moves the selected coordinate of $\mathsf H$ of largest height inward, with ties broken by the smallest index.
    Otherwise it chooses
    \begin{equation*}
        x\colonequals-\sgn\Big(\sum_{i\in A}\e_i\sinh(\omega S_i)\Big),
    \end{equation*}
    with $x\colonequals+1$ if the sum vanishes.
\end{definition}

If $N_b(S)\le q$, the state is $(b',0)$-isolating for every integer $b'\ge b$, since a selected coordinate that is the only selected one of height at least $b'$ is the highest selected coordinate of $\mathsf H$.

\begin{lemma}[Hyperbolic moments]
    \label{lem:CoshMoments}
    For the hybrid rule with parameters $b,q$, put $Y_i\colonequals\cosh(\omega S_i)$ and $F_r(S)\colonequals e_r(Y_1,\ldots,Y_n)$.
    In both models, at every state,
    \begin{equation}
        \label{eq:CoshDrift}
        \E\big[F_r(S')\bigm|S\big]\le\Big(1-\frac{\kappa_\omega r}2\Big)F_r(S)+\kappa_\omega(n-r+1)F_{r-1}(S),\qquad\kappa_\omega\colonequals\frac1{640000\,n},\quad1\le r\le r_*.
    \end{equation}
    Consequently, starting from $S_0=0$, for all $t\ge0$, $h\ge0$ and $1\le r\le r_*$,
    \begin{equation}
        \label{eq:CoshMarginals}
        \E F_r(S_t)\le2^r\binom nr,\qquad\P\big(N_h(S_t)\ge r\big)\le\Big(\frac{2en}{r\cosh(\omega h)}\Big)^r.
    \end{equation}
\end{lemma}

\begin{proof}
    We first consider binary inputs.
    Fix the state and a degree $r$, let $\chi_i,\zeta_i,v_i$ be the weights \eqref{eq:NewtonWeights} of $Y$, and put $\mathsf k(y)\colonequals(1-y^{-2})^{1/2}$, so that $\mathsf k(Y_i)=\tanh(\omega\abs{S_i})$.
    With $\pi_i\colonequals\mathsf k(Y_i)\chi_i$ and $\hat w_i\colonequals\mathsf k(Y_i)v_i=\sinh(\omega\abs{S_i})e_{r-1}(Y)/F_r$, the inequality $1-\tanh u\le1/\cosh u$ and \cref{lem:NewtonService}\ref{item:NewtonBasic} give
    \begin{equation}
        \label{eq:CoshWeights}
        0\le\chi_i-\pi_i\le\frac{\chi_i}{Y_i}=\frac{e_{r-1}(Y_{-i})}{F_r}\le\e_0\colonequals\frac rn\le\frac14.
    \end{equation}
    Let $\sigma_i\in\{\pm1\}$ with $\sigma_i=\sgn(S_i)$ if $S_i\ne0$; the choice at $S_i=0$ is immaterial, since $\pi_i=\hat w_i=0$ there.
    Since $\sum_{i\in A}\sinh(\omega S_i)=(F_r/e_{r-1}(Y))\sum_A\sigma_i\hat w_i$, the $\cosh$ choice is $x=-f$ for a sign $f$ of $\sum_A\sigma_i\hat w_i$, so \cref{lem:NewtonService}\ref{item:NewtonAveraged} applies to it whenever the support is a uniform $s$-subset of a set of at least $2s$ coordinates.

    We claim that the actual choice $x=-f_A$ satisfies
    \begin{equation}
        \label{eq:CoshService}
        \E\Big[f_A\sum_A\sigma_i\pi_i\Big]\ge\frac{\sqrt s}{400n}\sum_i\pi_i.
    \end{equation}
    For the pure $\cosh$ choice, with $\mathsf C=[n]$, the bracket in \eqref{eq:AveragedService} is at least $\sqrt s/(150n)$, since $r\le r_*$.
    For the hybrid, let $m\colonequals\abs{\mathsf H}$, $M_{\mathsf H}\colonequals\sum_{\mathsf H}\pi_i$ and $M_{\mathsf H^c}\colonequals\sum_{[n]\setminus\mathsf H}\pi_i$.
    If $m>q$, the rule is the pure $\cosh$ rule.
    Otherwise $n-m\ge3n/4$, and the support avoids $\mathsf H$ with probability at least $1-sm/n\ge1/2$.
    On supports avoiding $\mathsf H$ the rule is the $\cosh$ rule on a uniform $s$-subset of $[n]\setminus\mathsf H$, and \eqref{eq:AveragedService} bounds their contribution from below by $(\sqrt s/(288n)-18s^2r/n^2)M_{\mathsf H^c}$.
    On supports meeting $\mathsf H$, the selected coordinates outside $\mathsf H$ lose at most $s^2mM_{\mathsf H^c}/n^2$ in total, by conditioning on such a coordinate being selected and bounding the chance of also selecting a coordinate of $\mathsf H$.
    A selected coordinate of $\mathsf H$ moves inward unless another coordinate of $\mathsf H$ is selected; hence these coordinates contribute at least $(s/n)(1-2sm/n)M_{\mathsf H}\ge sM_{\mathsf H}/(2n)$.
    Since $r,m\le r_*$, these estimates prove \eqref{eq:CoshService}.

    For a fixed $r$-set $I$, let $\mathsf K_A\colonequals\abs{I\cap A}$ and $\mu\colonequals sr/n$.
    Exact hypergeometric factorial moments satisfy $\E(\mathsf K_A)_j\le\mu^j$; expanding in powers of $z-1$ and differentiating termwise yields $\E[\mathsf K_A^2z^{\mathsf K_A}]\le(\mu z+\mu^2z^2)\exp(\mu(z-1))\le4\mu$ for $z=e^\omega$.
    The second derivative of $u\mapsto\prod_{i\in I\cap A}\cosh(\omega(S_i+xu\e_i))/\cosh(\omega S_i)$ on $\abs u\le1$ is at most $\omega^2\mathsf K_A^2e^{\omega\mathsf K_A}$: relative to the product, it equals $\omega^2$ times the squared sum of the signed hyperbolic tangents plus the sum of the squared hyperbolic secants, which is the second moment of a sum of $\mathsf K_A$ random signs and hence at most $\mathsf K_A^2$.
    Taylor's inequality applies to either choice of $x$, including at coordinates with $S_i=0$, and the first-order terms sum to $-\omega F_rf_A\sum_A\sigma_i\pi_i$, since $\partial F_r/\partial S_i=\omega\sigma_i\pi_iF_r$.
    Summing over $I$ and using \eqref{eq:CoshService}, we obtain
    \begin{equation*}
        \E\big[F_r(S')-F_r(S)\bigm|S\big]\le-\frac{\omega\sqrt s}{400n}F_r\sum_i\pi_i+\frac{2\omega^2sr}nF_r.
    \end{equation*}
    By \eqref{eq:CoshWeights} and \eqref{eq:ElementaryIdentities}, $F_r\sum_i\pi_i\ge rF_r-\sum_ie_{r-1}(Y_{-i})=rF_r-(n-r+1)F_{r-1}$.
    Since $\omega\sqrt s/400=1/640000$ and $2\omega^2s=1/1280000$, this proves \eqref{eq:CoshDrift} for binary inputs.

    The hybrid rule depends on the state only through the magnitudes and the signs of the selected coordinates, and both sides of \eqref{eq:CoshDrift} depend only on the magnitudes; hence \cref{lem:MarkConjugation} transfers \eqref{eq:CoshDrift} to signed inputs.
    Finally, \cref{lem:TriangularRecursion} with $\mathsf u=\kappa_\omega/2$, $\mathsf v=\kappa_\omega$, $\mathsf M=2$ and $\varphi(h)=\cosh(\omega h)$ proves \eqref{eq:CoshMarginals}, since $F_r(S_0)=\binom nr$.
\end{proof}

\begin{proposition}[Terminal bound for the hybrid rule]
    \label{prop:CoshTerminal}
    Put $D_1\colonequals2^{40}$ and, whenever the integer $q_0$ below is positive,
    \begin{equation*}
        q_0\colonequals\floor{\frac n{D_1s^{3/2}}},\qquad b_0\colonequals\ceil{\omega^{-1}\log(4e^5n/q_0)}.
    \end{equation*}
    For every fixed $k\ge1$ there are constants $C_k$ and $n_k$ such that the hybrid rule with parameters $b_0,q_0$ satisfies, in both models, for all $n\ge n_k$, all $2\le s\le\sqrt n$ and all integers $T\ge0$,
    \begin{equation}
        \label{eq:CoshTerminal}
        \P\bc{\norm{S_T}_\infty>C_k\big(\sqrt s\log(s+1)+\log\ell\big)}\le n^{-k}.
    \end{equation}
\end{proposition}

\begin{proof}
    For large $n$, uniformly in $s\le\sqrt n$, we have $q_0\le\min\{r_*,n/(8A_{\mathrm{cl}}s)\}$ and $q_0\ge n^{1/4}/(2D_1)>2R$, where $R$ is as in \cref{prop:Composition}.
    Put
    \begin{equation*}
        \bar H\colonequals\ceil{\omega^{-1}\big((k+30)\log n+\log4\big)},\qquad\Gc_t\colonequals\big\{N_{b_0}(S_t)\le q_0,\ \norm{S_t}_\infty\le\bar H\big\}.
    \end{equation*}
    On $\Gc_t$ the state is $(b,0)$-isolating for every integer $b\ge b_0$, by the observation after \cref{def:HybridRule}, so \cref{prop:Composition} applies with these $b_0$, $q_0$ and $\bar H$.
    Its window satisfies $\bar w\le n^3$, and $q_J\le2R$ gives $\log\vartheta_J\ge(\log n)/8$; with $u_*$ as in \eqref{eq:FinalExcess}, we have $u_*=O(k+1)$ and $8eq_J/\vartheta_J^{u_*}\le n^{-k-20}$.
    By \eqref{eq:CoshMarginals}, at every time, $\P(N_{b_0}>q_0)\le e^{-4q_0}$, since $\cosh(\omega b_0)\ge2e^5n/q_0$, and $\P(\norm S_\infty>\bar H)\le2en/\cosh(\omega\bar H)\le en^{-k-29}$.
    Hence $\Gc_t$ fails for some $t\in[(T-\bar w)_+,T]$ with probability at most $(\bar w+1)(e^{-4q_0}+en^{-k-29})$, and \eqref{eq:Composition} gives
    \begin{equation*}
        \P\big(\norm{S_T}_\infty\ge b_J+u_*\big)\le(\bar w+1)\big(e^{-4q_0}+en^{-k-29}\big)+J(\bar w+1)e^{-c_*R}+n^{-k-20}\le n^{-k}.
    \end{equation*}
    Since $b_0=O(\sqrt s\log(s+1))$, $J=O(\log\ell)$ and $u_*=O(k+1)$, the radius $b_J+u_*=b_0+2J+u_*$ proves \eqref{eq:CoshTerminal}.
\end{proof}

\subsection{Proof of the terminal theorem}
\label{sec:ProofTerminal}

We combine \cref{prop:CoarseTerminal,prop:Bootstrap,prop:CoshTerminal} according to the ranges of $s$ in the table at the beginning of this section.

\begin{proof}[Proof of \cref{thm:Terminal}]
    Fix $k\ge1$, let $n$ be large, and recall that $\log\ell\le2\log\log n$ for $n\ge6$.
    If $s\le\log\ell$, then $s\le\sqrt n$ and $q_0\ge1$, so the rule is the hybrid rule, and \cref{prop:CoshTerminal} gives the radius $C_k(\sqrt s\log(s+1)+\log\ell)$ with failure probability $n^{-k}$; here $\sqrt s\log(s+1)\le\sqrt{\log\ell}\,\log(1+\log\ell)\le\log\ell$.
    If $\log\ell<s<\ell^{2/3}$, then \cref{prop:Bootstrap} applies.
    If $s\ge\ell^{2/3}$, then $\ell/s\le\sqrt s$, and \eqref{eq:CoarsePoly} gives the radius $C_k(2\sqrt s+\log\ell)$ with failure probability $n^{-k}$, in the binary model for $s\le n/2$ and in the signed model for $s\le n$.
    In each case the radius is at most $C_k'(\sqrt s+\log\log n)$, and the failure probability does not depend on $T$.
\end{proof}

\subsection{Quadratic Greedy for pairs}
\label{sec:GreedyTerminal}

For $s=2$, quadratic Greedy, the rule of \cref{thm:Prefix}\ref{item:PrefixPairs}, has the same terminal guarantee as the rule of \cref{thm:Terminal}.

\begin{corollary}[Quadratic Greedy at terminal times]
    \label{cor:GreedyTerminal}
    Let $s=2$, and let the signs be chosen by quadratic Greedy, with an arbitrary fixed rule at ties.
    For every fixed $k\ge1$ there are constants $C_k$ and $n_k$ such that, in both models, for all $n\ge n_k$ and every prescribed time $T\in\N$,
    \begin{equation}
        \label{eq:GreedyTerminal}
        \P\big\{\norm{S_T}_\infty>C_k\log\log n\big\}\le n^{-k}.
    \end{equation}
    Moreover, $\E\norm{S_T}_\infty\asymp\log\log n$ uniformly for $n\ge n_0$ and integers $T\ge n$, where $n_0$ is an absolute constant.
\end{corollary}

\begin{proof}
    Put $q_0\colonequals\floor{n/(64A_{\mathrm{cl}})}$, $b_0\colonequals\ceil{\log_2(32en/q_0)}+2$ and $\bar H\colonequals\ceil{(k+20)\log_2n}$.
    By \eqref{eq:GeometricMoments}, with $b=b_0$ and $r=q_0+1$, and with $b=\bar H+1$ and $r=1$, at every fixed time $t$,
    \begin{equation*}
        \P\{N_{b_0}(S_t)>q_0\}\le\Big(\frac{32en}{(q_0+1)2^{b_0}}\Big)^{q_0+1}\le4^{-q_0-1},\qquad\P\{\norm{S_t}_\infty>\bar H\}\le32en2^{-\bar H-1}\le16en^{-k-19}.
    \end{equation*}
    Every state is $(b,0)$-isolating for Greedy for every integer $b\ge1$, and \cref{prop:Composition} applies to Greedy with every rule at ties, as explained after \ref{item:OutcomePositive} and~\ref{item:OutcomeZero} in \cref{sec:Pairs}.
    We apply it with these $b_0$, $q_0$ and $\bar H$ and with $\Gc_t\colonequals\{N_{b_0}(S_t)\le q_0,\ \norm{S_t}_\infty\le\bar H\}$; for large $n$ we have $2R<q_0\le n/(8A_{\mathrm{cl}}s)$, $\bar w\le n^3$ and $\log\vartheta_J\ge(\log n)/4$.
    Hence $\Gc_t$ fails for some $t\in[(T-\bar w)_+,T]$ with probability at most $(n^3+1)(4^{-q_0-1}+16en^{-k-19})\le n^{-k-10}$.
    The second term in \eqref{eq:Composition} is at most $n^{-k-10}$, since $\bar w\le n^3$, $J\le n$ and $c_*R\ge R\ge100(k+10)\log n$; with $u_*$ from \eqref{eq:FinalExcess}, the third term is at most $n^{-k-20}$.
    Hence $\P(\norm{S_T}_\infty\ge b_J+u_*)\le n^{-k}$, and the radius $b_J+u_*=b_0+2J+u_*$ is $O_k(\log\log n)$, since $b_0=O(1)$, $J=O(\log\ell)$ and $u_*=O_k(1)$.
    For the expected value, combine \eqref{eq:GreedyTerminal} with $k=2$ and the bound $\norm{S_T}_\infty\le2n$ of \cref{lem:DoubleCover}\ref{item:DoubleCoverBound}, and use \cref{prop:Baseline}\ref{item:BaselineLogLog} for the lower bound.
\end{proof}

Thus, for pairs, a single parameter-free rule attains the optimal order of both objectives: of the expected prefix discrepancy for $n\le T\le e^{cn}$, by \cref{thm:Prefix}\ref{item:PrefixPairs} and \cref{cor:ExpectedPrefix}, and of the expected terminal discrepancy at every prescribed time $T\ge n$, by \cref{cor:GreedyTerminal}.
For the edge vectors $e_i-e_j$ of a uniformly random pair, Ajtai et al.\ showed that the expected unfairness of Greedy, which is half of the discrepancy, is of order $\log\log n$ at a typical time of a long execution \cite[Section~4]{Ajtai98}.
They also showed that, from every reachable state, the maximal unfairness falls to $O(\log n)$ within $n^4$ steps with high probability and stays there for a superpolynomial number of steps, and remarked that this implies fast convergence to the order $\log\log n$ \cite[Section~4.5]{Ajtai98}.
\Cref{cor:GreedyTerminal} gives the order $\log\log n$ at every prescribed time, with polynomially small failure probability, for our two models of pairs.

\section{Corollaries, benchmarks, and complements}
\label{sec:Consequences}

This section proves \cref{cor:ExpectedPrefix} in \cref{sec:ExpectedPrefix} and \cref{cor:ExpectedTerminal,cor:PolynomialPrefix} in \cref{sec:ExpectedTerminal}.
\Cref{sec:Restart} removes the dependence on the horizon from the rule of \cref{thm:Prefix}\ref{item:PrefixFixed} by restarts, and \cref{sec:BeyondExponential} determines the limit of the optimal prefix discrepancy beyond exponential horizons.
\Cref{sec:GrowingPrefix} extends the prefix bound of \cref{cor:PolynomialPrefix} beyond polynomial horizons for growing sparsity, and \cref{sec:Benchmarks} records benchmarks from known theorems and complements for binary inputs with large supports.

\subsection{Expected prefix discrepancy}
\label{sec:ExpectedPrefix}

\Cref{cor:ExpectedPrefix} determines the optimal expected prefix discrepancy for $s=2$ and $n\le T\le e^{cn}$, and for fixed $s\ge3$ it bounds it from below by $\log\log n+(\log T/\log n)^{\alpha_s}$ and from above by $\log\log n+(\log T/\log n)^{1/2}$, up to constants depending on $s$ and $c$.

\begin{proof}[Proof of \cref{cor:ExpectedPrefix}]
    For the lower bounds, recall that $\P\{D_T\ge h\}\ge1-\e$ implies $\E D_T\ge(1-\e)h$ for the nonnegative variable $D_T$.
    The constant $c'$ and the error term $\e_n$ in \cref{thm:Prefix}\ref{item:PrefixLower} do not depend on the algorithm, so the bound survives the infimum in \eqref{eq:Ondisc}.

    For the upper bounds, choose $\delta\colonequals T^{-2}$ in \cref{thm:Prefix}\ref{item:PrefixPairs} or \ref{item:PrefixFixed}.
    Every sample path satisfies $D_T\le T$, so a radius $r$ with failure probability $\delta$ gives $\E D_T\le r+T\delta=r+1/T$.
    Since $T\ge n$, we have $L=\log(16n(T+1)T^2)=O(\log T)$; for $T\le e^{cn}$ it satisfies $L\le C_0n$ for every fixed $C_0>3c$ and all large $n$, and also $L\le n^2$.
    For $s=2$, quadratic Greedy needs no such choice.
    For $s\ge3$, the rule of \cref{thm:Prefix}\ref{item:PrefixFixed} is selected in advance for this $L$; the restart rule of \cref{def:RestartRule} avoids this choice.
    Since $\log T\ge\log n$, the resulting radii are of the stated orders.
    For the finitely many small dimensions excluded in \cref{thm:Prefix}, both sides are bounded above and below by positive constants depending only on $s$ and $c$, since $1\le D_T\le T\le e^{cn}$ for $T\ge1$.
\end{proof}

For example, at stretched-exponential horizons $T=\ceil{e^{n^\vartheta}}$ with $0<\vartheta<1$ and fixed $s\ge3$, the lower and upper bounds of \cref{cor:ExpectedPrefix} are of order $(n^\vartheta/\log n)^{\alpha_s}$ and $(n^\vartheta/\log n)^{1/2}$, respectively; for $s=2$ both exponents equal $1/3$.
For fixed $s\ge3$, the horizon term of \cref{thm:Prefix}\ref{item:PrefixLower} exceeds the doubly logarithmic baseline precisely in the regime
\begin{equation}
    \label{eq:Transition}
    \log T\gg\log n\,(\log\log n)^{1/\alpha_s},
\end{equation}
whereas, as long as $\log T\lesssim\log n\,(\log\log n)^2$, the upper bound of \cref{thm:Prefix}\ref{item:PrefixFixed} is of order $\log\log n$.
Thus $\ondisc(\mu_s,T)\asymp\log\log n$ throughout the latter range, and this fails as soon as \eqref{eq:Transition} holds.
For $s\ge3$ these are comparisons of proved bounds; we do not claim to identify the exact transition.
For $s=2$ the transition occurs at $\log T\asymp\log n\,(\log\log n)^3$, and \cref{cor:ExpectedPrefix} identifies it exactly.

\subsection{A rule independent of the horizon}
\label{sec:Restart}

The rule of \cref{thm:Prefix}\ref{item:PrefixFixed} places its knee according to $L$.
Restarting it with geometrically increasing values of $L$ removes this dependence; this is the restart scheme with geometrically increasing guesses from online learning, often called the doubling trick, in which a restart is triggered once the current guess is exceeded \cite[Section~4.6]{CesaBianchi97}.
Let $s\ge3$ and $n\ge e^e$.
For $\lambda\ge\log(16n(n+1))$, let $\Ac_\lambda$ be the algorithm of \cref{thm:Prefix}\ref{item:PrefixFixed} for $L=\lambda$, that is, the rule \eqref{eq:WeightRule} with the knee that \cref{sec:ProofFixedSparsity} computes from $n$, $s$ and $\lambda$; it chooses each sign as a function of the current state and the arrival, and since it depends only on $n$, $s$ and $L$, for every $C_0>0$ and $n\ge n_0(s,C_0)$ it satisfies \eqref{eq:FixedSparsity} for every pair $(T,\delta)$ with $T\ge n$ and $\log(16n(T+1)/\delta)=\lambda\le C_0n$.
Let $A_s$ be the constant of \cref{thm:Prefix}\ref{item:PrefixFixed}, where we may assume $A_s\ge1$, and put, for $j\ge0$,
\begin{equation*}
    \lambda_j\colonequals4^j\log n\,(\log\log n)^2,\qquad B_j\colonequals A_s\Big(\log\log n+\Big(\frac{\lambda_j}{\log n}\Big)^{1/2}\Big)=A_s(1+2^j)\log\log n;
\end{equation*}
for large $n$, $\lambda_0\ge\log(16n(n+1))$.

\begin{definition}[Restart rule]
    \label{def:RestartRule}
    The \emph{restart rule} processes the arrivals in consecutive \emph{epochs} $j=0,1,\ldots$.
    In epoch $j$, it chooses each sign as $\Ac_{\lambda_j}$ would at the \emph{local state}, the signed sum of the arrivals of this epoch, which is zero at the beginning of the epoch.
    The epoch ends with the first arrival after which the local state has norm larger than $B_j$, and the next arrival begins epoch $j+1$.
\end{definition}

The state $S_t$ is never reset; it is the sum of the local states of all epochs.

\begin{proposition}[Restarts]
    \label{prop:Restart}
    Let $s\ge3$ and $C_0>0$.
    There is $n_1=n_1(s,C_0)$ such that, for all $n\ge n_1$, all integers $T\ge n$ and all $0<\delta<1$ with $L\le C_0n$, the restart rule satisfies, in both models,
    \begin{equation}
        \label{eq:Restart}
        \P\bc{D_T>8A_s\Big(\log\log n+\Big(\frac L{\log n}\Big)^{1/2}\Big)}\le\delta.
    \end{equation}
\end{proposition}

\begin{proof}
    \emph{Epoch sums.}
    Let $\tau_0\colonequals0$, and let $\tau_{i+1}$ be the time of the arrival that ends epoch $i$, with $\tau_{i+1}\colonequals\infty$ if epoch $i$ does not begin or does not end; these are stopping times, and epoch $i$ consists of the arrivals at the times in $(\tau_i,\tau_{i+1}]$.
    Every arrival changes every coordinate by at most one, so every value of the local state of epoch $i$, including its final value, has norm at most $B_i+1$.
    Hence, if $\tau_{j+1}\ge T$, then every $S_t$ with $t\le T$ is a sum of such values over the epochs $0,\ldots,j$, and
    \begin{equation*}
        D_T\le\sum_{i=0}^j(B_i+1)=A_s(j+2^{j+1})\log\log n+j+1\le4A_s2^j\log\log n,
    \end{equation*}
    since $A_s\log\log n\ge1$ and $2j+1\le2^{j+1}$.

    \emph{The critical epoch.}
    Given $T$ and $\delta$, let $j$ be the least index with $\lambda_j\ge L$, and put $\delta_j\colonequals16n(T+1)e^{-\lambda_j}$, so that $\delta_j\le\delta<1$ and $\log(16n(T+1)/\delta_j)=\lambda_j$.
    For large $n$, depending on $s$ and $C_0$, we have $\lambda_0\le C_0n$, hence $\lambda_j\le\max\{\lambda_0,4L\}\le4C_0n$.
    On the event $\{\tau_j=k\}\in\Fc_k$ with $k<T$, the local run of epoch $j$ agrees, up to the end of the epoch, with the run of $\Ac_{\lambda_j}$ from zero on the arrivals $V_{k+1},V_{k+2},\ldots$, which are independent of $\Fc_k$ and have the original law, and $\tau_{j+1}<T$ requires this run to exceed $B_j$ within its first $T$ arrivals.
    Since $B_j=A_s(\log\log n+(\lambda_j/\log n)^{1/2})$, the guarantee of $\Ac_{\lambda_j}$, with $4C_0$ in place of $C_0$, horizon $T$ and failure probability $\delta_j$, bounds the conditional probability of this event by $\delta_j$ for $n\ge n_0(s,4C_0)$.
    If $\tau_j\ge T$, then $\tau_{j+1}\ge T$.
    Summing over $k<T$ gives $\P(\tau_{j+1}<T)\le\delta_j\le\delta$.

    \emph{Conclusion.}
    Outside this event, $D_T\le4A_s2^j\log\log n$.
    If $j=0$, this is $4A_s\log\log n$; if $j\ge1$, then $\lambda_{j-1}<L$, so that $2^j\log\log n=(\lambda_j/\log n)^{1/2}<2(L/\log n)^{1/2}$.
    In both cases $D_T\le8A_s(\log\log n+(L/\log n)^{1/2})$.
\end{proof}

The restart rule is deterministic and uses no information about $T$ or $\delta$, so the bounds \eqref{eq:Restart} for all $T$ and $\delta$ concern one sequence of signs; in particular, for $s\ge3$ it attains the upper bound of \cref{cor:ExpectedPrefix} with one rule for all horizons $n\le T\le e^{cn}$.
Unlike the rules of \cref{sec:FixedSparsity,sec:Terminal}, it does not choose the sign as a function of the current state and the arrival: it keeps an epoch counter and a local state, and at each restart it disregards the global state $S_t$.

\subsection{Beyond exponential horizons}
\label{sec:BeyondExponential}

The restriction $T\le e^{cn}$ in \cref{cor:ExpectedPrefix} concerns only the bounds, not the problem.
As $T\to\infty$, the optimal prefix discrepancy converges, and its limit is governed by the width bound \eqref{eq:BaranyWidth}.
For dense Rademacher inputs, Bansal and Spencer \cite[Section~3]{Bansal20a} observed that the exponential lower bound of B\'ar\'any \cite{Barany79} for adaptive adversaries persists for random inputs if the horizon may be chosen adaptively, since a worst-case adaptive sequence eventually occurs, after about $\exp(\exp(n))$ arrivals.
The next corollary makes this precise for every $2\le s\le n$ and adds a matching upper bound.

\begin{corollary}[Beyond exponential horizons]
    \label{cor:BeyondExponential}
    Let $2\le s\le n$, and put $\Delta\colonequals2^{s-1}\binom{n-1}{s-1}$ in the signed model and $\Delta\colonequals\binom{n-1}{s-1}$ in the binary model.
    Then $\ondisc(\mu_s,T)$ is nondecreasing in $T$, and $\sup\P\{D_T<\Delta/2\}\to0$ as $T\to\infty$, where the supremum is taken over all online algorithms.
    Conversely, there is a rule that satisfies, for all $T$, $D_T\le\Delta/2$ in the signed model with $s\ge3$, $D_T\le\Delta/2+1$ in the signed model with $s=2$, and $D_T<\Delta/2+s$ in the binary model.
    In particular, $\lim_{T\to\infty}\ondisc(\mu_s^\pm,T)=2^{s-2}\binom{n-1}{s-1}$ for $s\ge3$, and for fixed $s$ the limit of $\ondisc(\mu_s,T)$ is of order $n^{s-1}$ in both models.
\end{corollary}

\begin{proof}
    Every online algorithm for horizon $T+1$ is also one for horizon $T$, and $D_T\le D_{T+1}$; hence $\ondisc(\mu_s,T)$ is nondecreasing in $T$.
    Let $\Vc$ be the set of possible arrivals and their negatives, that is, $\Vc=\Vc^{=}_{n,s}$ in the signed model and $\Vc=\{\pm\mathbf1_A:A\subseteq[n],\ \abs A=s\}$ in the binary model, and let $\Vc^+$ contain one vector from each pair $\pm v$.
    No two vectors of $\Vc^+$ are collinear, and each coordinate lies in the supports of exactly $\Delta$ of them.

    Let $R$ be the largest integer with $2R<\Delta$.
    As in the proof of \cref{prop:InfiniteValue}, \eqref{eq:BaranyWidth} shows that the safety sets of $\Vc$ on all $n$ coordinates, with cube radius $R$, satisfy $Q_J=\emptyset$ for some $J$, which depends only on $n$ and $s$.
    The next arrival lies in a given pair $\pm v$ with probability $p_*\colonequals1/\abs{\Vc^+}$, independently of the past, and then the state moves to $S+v$ or to $S-v$, whatever sign the algorithm chooses.
    The proof of \cref{prop:BlockBound}, with the tracked state $Y\colonequals S$ and the sets $E(v)\colonequals\{\pm v\}$, therefore gives $\P\{D_T\le R\}\le(1-p_*^J)^{\floor{T/J}}$ for every online algorithm, and $D_T>R$ means $D_T\ge\Delta/2$, since $D_T$ is an integer.

    For the upper bounds, let the algorithm follow the parity rule of \cref{sec:SafetyGame}, which moves the state by $+v$ at the odd-numbered arrivals from a pair $\pm v$ with $v\in\Vc^+$ and by $-v$ at the even-numbered ones.
    Then the state is a subset sum of $\Vc^+$ at all times, so $D_T\le\max_i\max\{N_i^+,N_i^-\}$, where $N_i^\pm$ is the number of $v\in\Vc^+$ with $v_i=\pm1$, and $N_i^++N_i^-=\Delta$.
    In the signed model with $s\ge3$, the choice of $\Vc^+$ in the proof of \cref{prop:InfiniteValue} gives $N_i^+=N_i^-=\Delta/2$.
    For $s=2$, the vectors $(-1)^a(e_b\pm e_a)$ with $a<b$ from that proof give $N_i^+-N_i^-=2\sum_{a<i}(-1)^a\in\{-2,0\}$.
    In the binary model, $\Vc^+$ consists of vectors $\e_A\mathbf1_A$ with signs $\e_A$, and $N_i^+-N_i^-=\sum_{A\ni i}\e_A$.
    Each set $A$ lies in exactly $s$ of the families $F_i\colonequals\{A':i\in A'\}$, $i\in[n]$, so by the Beck--Fiala theorem \cite{BeckFiala81} the signs can be chosen such that $\abs{N_i^+-N_i^-}=\abs{\sum_{A\in F_i}\e_A}\le2s-1$ for every $i$.
    The last assertion follows, since $\binom{n-1}{s-1}\asymp_sn^{s-1}$.
\end{proof}

In the signed model with $s\ge3$, the parity rule is thus optimal in the limit, and the limit is the infinite-horizon value of the adaptive game of \cref{prop:InfiniteValue} with $m=n$ and $r=s$.
In both models, the limit is approached only far beyond the horizons of \cref{cor:ExpectedPrefix}, which bounds $\ondisc(\mu_s,T)$ by $O_{s,c}((n/\log n)^{1/2})$ for $T\le e^{cn}$.
For $s=2$, the order $n$ of the limit is reached once $\log T\ge Cn^3\log n$, by \cref{rem:PairsBeyondExponential}; for fixed $s\ge3$, it is open at which horizons the order $n^{s-1}$ is reached.
The forcing strategy of \cref{lem:AdmissibleHeights}, with $k=s-1$ and the largest admissible height $H\asymp_sn^{1/\gamma_{s-1}}$, reaches only heights of order $n^{2-2^{2-s}}$, since $1/\gamma_{s-1}=2-2^{2-s}$, and the width bound \eqref{eq:BaranyWidth} behind the limit does not bound the forcing time.

\subsection{Consequences of the terminal bound}
\label{sec:ExpectedTerminal}

\Cref{cor:ExpectedTerminal} states that, in the ranges of \cref{thm:Terminal}, the optimal expected terminal discrepancy is of order $\sqrt s+\log\log n$ at every time $T\ge n$, and that the rule of \cref{thm:Terminal} attains this order.

\begin{proof}[Proof of \cref{cor:ExpectedTerminal}]
    By \cref{prop:Baseline}\ref{item:BaselineCombined}, every online algorithm satisfies $\P\{\norm{S_T}_\infty\ge c(\sqrt s+\log\log n)\}\ge1-\e_n$ for every $T\ge n$, where $\e_n\to0$ does not depend on the algorithm or on $T$; hence $\E\norm{S_T}_\infty\ge(1-\e_n)c(\sqrt s+\log\log n)$, which proves the lower bound for large $n$.

    For the upper bound, consider the rule of \cref{thm:Terminal}, and put $\ell\colonequals\log(en)$.
    We first bound the second moment uniformly in time.
    If the rule is the gauge rule, \cref{lem:GaugeGeometry} and \cref{cor:GaugeMoments} give $\norm S_\infty\le B(S)_++C(\sqrt s+\ell)$ at every state and $\sup_t\E e^{\theta_0B(S_t)_+}\le C\sqrt s$; since $x^2\le2\theta_0^{-2}e^{\theta_0x}$ for $x\ge0$, this gives $\E\norm{S_T}_\infty^2\le C(s+\ell^2)$.
    If it is the hybrid rule, \eqref{eq:CoshMarginals} with $r=1$ gives $\P\{\norm{S_T}_\infty\ge h\}\le4ene^{-\omega h}$ for all $h\ge0$, with $\omega^{-1}=1600\sqrt s$, and integration gives $\E\norm{S_T}_\infty^2\le Cs\ell^2$.
    Let $R_2\colonequals C_2(\sqrt s+\log\log n)$ be the radius of \eqref{eq:Terminal} with $k=2$.
    By the Cauchy--Schwarz inequality,
    \begin{equation*}
        \E\norm{S_T}_\infty\le R_2+\big(\E\norm{S_T}_\infty^2\big)^{1/2}\P\{\norm{S_T}_\infty>R_2\}^{1/2}\le R_2+\frac{C\sqrt s\,\ell}n\le C'(\sqrt s+\log\log n)
    \end{equation*}
    for $n\ge n_2$, uniformly in $T$.
\end{proof}

\Cref{cor:PolynomialPrefix} states that, in the same ranges, the rule of \cref{thm:Terminal} also controls the prefix discrepancy at every polynomial horizon $T\le n^\beta$, with radius $O_{\beta,k}(\sqrt s+\log\log n)$ and failure probability $n^{-k}$.

\begin{proof}[Proof of \cref{cor:PolynomialPrefix}]
    Apply \cref{thm:Terminal} with the exponent $k'\colonequals\ceil{k+\beta}+1$, and let $n\ge n_{k'}$.
    Since the algorithm does not depend on $k'$, the bounds at the times $1\le t\le T$ concern one sequence of signs, and a union bound over these at most $n^\beta$ times gives
    \begin{equation*}
        \P\bc{D_T>C_{k'}\big(\sqrt s+\log\log n\big)}\le n^\beta\cdot n^{-k-\beta-1}\le n^{-k};
    \end{equation*}
    the time $t=0$ contributes nothing, since $S_0=0$.
    This proves \eqref{eq:PolynomialPrefix}.
    For the expected value, apply \eqref{eq:PolynomialPrefix} with $\ceil\beta+1$ in place of $k$, and let $R_\beta$ be the resulting radius.
    Every sample path satisfies $D_T\le T\le n^\beta$, so $\E D_T\le R_\beta+n^\beta\cdot n^{-\beta-1}\le R_\beta+1$ for $n\ge n_{\beta,\ceil\beta+1}$.
    For $n\le T$, the lower bound $\E D_T\ge\E\norm{S_T}_\infty\ge c(\sqrt s+\log\log n)$ for $n\ge n_0$ follows from \cref{prop:Baseline}\ref{item:BaselineCombined}.
    For the finitely many remaining dimensions, both sides are bounded above and below by positive constants depending only on $\beta$, since $1\le D_T\le T\le n^\beta$ for $T\ge1$ and $\sqrt2\le\sqrt s+\log\log n\le\sqrt n+\log\log n$ for $n\ge3$.
\end{proof}

The constants in \cref{cor:PolynomialPrefix} depend on $\beta$ and $k$, and neither parameter may grow with $n$ without a further quantitative argument.

\subsection{Prefix bounds for growing sparsity}
\label{sec:GrowingPrefix}

Letting $\beta$ grow with $n$ in \cref{cor:PolynomialPrefix} requires bounds that hold up to $T$ simultaneously.
For $2\le s\le\log^2(en)$, the moderate moments of \cref{cor:ModerateMoments} give such bounds for the gauge rule, with absolute constants; for $s>\log\log(en)$, this is the rule of \cref{thm:Terminal}.

\begin{proposition}[Prefix bounds for growing sparsity]
    \label{prop:GrowingPrefix}
    Put $\ell\colonequals\log(en)$.
    There are absolute constants $C$ and $n_0$ such that the gauge rule satisfies, in both models, for all $n\ge n_0$, all $2\le s\le\ell^2$ and all integers $n\le T\le e^{\ell^2}$,
    \begin{equation}
        \label{eq:GrowingPrefix}
        \E D_T\le C\Big(\sqrt s+\log\ell+\frac{\log T}{\min\{s,\ell\}}\Big).
    \end{equation}
    In particular, $\ondisc(\mu_s,T)\asymp\sqrt s+\log\log n$ uniformly for $n\ge n_0$, $2\le s\le\ell^2$ and integers $T\ge n$ with $\log T\le\sqrt s\min\{s,\ell\}$.
\end{proposition}

\begin{proof}
    \emph{The gauge.}
    Let $\theta$ be as in \eqref{eq:ModerateExponent}.
    For large $n$, uniformly for $s\le\ell^2$, we have $\log(n/s)\ge\ell/2$, hence $\theta\ge c_0\min\{s,\ell\}/2\ge c_0$.
    Fix $0\le y\le2\ell^2$, and put
    \begin{equation*}
        \beta_y\colonequals2+\frac{\log T+4\log\ell+y}\theta,\qquad b_0\colonequals\ceil{\beta_y+h_*},\qquad q_0\colonequals\floor{\frac n{D_0s\ell^8}},\qquad R\colonequals\ceil{\log T+y}.
    \end{equation*}
    Let $C_{\mathrm{win}}$ denote the constant $C$ in \eqref{eq:ModerateWindow}.
    By \eqref{eq:ModerateWindow} with $t_0=0$ and $w=T$, and since $\max\{\sqrt s,\theta\}+\theta pT\le2\ell^3T/n$, the event $\Vc$ that $B(S_t)\le\beta_y$ for all $t\le T$ fails with probability at most $2C_{\mathrm{win}}\ell^3(T/n)e^{2\theta-\theta\beta_y}=2C_{\mathrm{win}}e^{-y}/(\ell n)\le e^{-y}$.

    \emph{Populations and the last layer.}
    By \cref{lem:GaugeGeometry}, on $\Gc_t\colonequals\{B(S_t)\le\beta_y\}$ we have $N_{b_0}(S_t)\le K_0n/Y_*\le q_0$, and $S_t$ is $(b,0)$-isolating for every integer $b\ge b_0$.
    For large $n$, $q_0\le n/(8A_{\mathrm{cl}}s)$ and $q_0\ge n/(2D_0\ell^{10})>2R$, since $R\le3\ell^2+1$.
    Since we need all times up to $T$, we apply \cref{lem:LayerIteration} directly, rather than \cref{prop:Composition}, with $g=0$, these $b_0$, $q_0$ and $R$, the events $\Gc_t$, $t_0=\cdots=t_{J+1}=0$ and $M=0$; its hypotheses hold, since all heights vanish at time zero.
    By \eqref{eq:LayerFailure}, the probability that $\Vc$ holds and a population condition $N_{b_{j+1}}(S_t)\le q_{j+1}$ with $j<J$ and $t\le T$ fails is at most $J(T+1)e^{-c_*R}\le J(T+1)T^{-c_*}e^{-y}\le e^{-y}$, since $c_*>1$, $T\ge n$ and $J\le\log_2\log_2n+3$.
    Moreover, $q_J\le2R$ gives $\vartheta_J^2=n/(C_{\mathrm{cl}}sq_J)\ge\sqrt n$, so $u\colonequals\ceil{4(\log(2T)+\log(8eq_J)+y)/\log n}$ satisfies $(T+1)\cdot8eq_J/\vartheta_J^u\le e^{-y}$.
    Hence \eqref{eq:LayerFinal} with $r=1$ and $d=u$, together with a union bound over $t\le T$, shows that the event that $\Vc$ holds, no population condition fails and some height up to time $T$ reaches $b_J+u$ has probability at most $e^{-y}$; together with the two failure bounds above, $\P(D_T\ge b_J+u)\le3e^{-y}$.

    \emph{Radius and expectation.}
    By \eqref{eq:GaugeEnvelope}, $h_*=O(\sqrt s+\log\ell)$; moreover $J=O(\log\ell)$, $\theta\ge c_0\min\{s,\ell\}/2$ and $\log n\ge\ell/2\ge\min\{s,\ell\}/2$.
    Hence $b_J+u\le\beta_y+h_*+1+2J+u\le r_y\colonequals C'(\sqrt s+\log\ell+(\log T+y)/\min\{s,\ell\})$ for an absolute constant $C'$.
    Since $r_y$ is affine in $y$ with slope at most $C'$ and $D_T\le T\le e^{\ell^2}$,
    \begin{equation*}
        \E D_T\le r_0+C'\int_0^{2\ell^2}3e^{-y}\,dy+T\cdot3e^{-2\ell^2}\le r_0+3C'+3,
    \end{equation*}
    which proves \eqref{eq:GrowingPrefix}.
    The range $\log T\le\sqrt s\min\{s,\ell\}$ lies in $\log T\le\ell^2$, and there \eqref{eq:GrowingPrefix} is $O(\sqrt s+\log\log n)$; \cref{prop:Baseline}\ref{item:BaselineCombined} gives the matching lower bound for every online algorithm.
\end{proof}

The range $\log T\le\sqrt s\min\{s,\ell\}$ goes beyond every polynomial horizon as soon as $s/\ell^{2/3}\to\infty$; for instance, for $s=\ceil\ell$ it contains all $T$ with $\log T\le\ell^{3/2}$.

\subsection{Benchmarks and complements}
\label{sec:Benchmarks}

\begin{remark}[Generic benchmarks]
    \label{rem:GenericBenchmarks}
    Known theorems for fixed input sequences give the following benchmarks for our inputs; they are not new results about random supports.
    Let $L_0\colonequals\log(3T/\delta)$.
    Aden-Ali's algorithm \cite[Theorem~2]{AdenAli26}, applied to $V_t/\sqrt s$, has prefix discrepancy $O(\sqrt{sL_0})$ with probability at least $1-\delta$, in both models and for all $1\le s\le n$, since its private randomness is independent of the inputs; for fixed $s$, $T\ge n$ and polynomially small $\delta$, this is $O(\sqrt{\log T})$, which \cref{thm:Prefix}\ref{item:PrefixFixed} improves at $T=e^{cn}$ by a factor of order $\sqrt{\log n}$, and \cref{thm:Prefix}\ref{item:PrefixPairs} by a factor of order $n^{1/6}(\log n)^{1/3}$.
    Hmadi's theorem \cite[Corollary~3.5]{Hmadi26}, applied with sparsity parameter $\max\{s,CL_0\log^2(e+L_0)\}$, gives prefix discrepancy $O(\sqrt s+\sqrt{L_0}\log(e+L_0))$ with probability at least $1-\delta$; hence, if $s\ge CL_0\log^2(e+L_0)$, it is $O(\sqrt s)$, which matches \cref{prop:Baseline} for $T\ge n$ in the signed model and, for $s\le n/2$, in the binary model.
    For $s=n$, this gives prefix discrepancy $O(\sqrt n)$ up to horizons $T\le\exp(cn/\log^2(en))$, with polynomially small failure probability \cite[Corollary~3.6]{Hmadi26}; the earlier theorem of Altschuler and Tikhomirov gives the analogous statements with $\log^{2+\eta}$ in place of $\log^2$, for a fixed $\eta>0$ \cite[Theorem~1.1 and Corollary~1.3]{Altschuler26}.
\end{remark}

\begin{remark}[Binary complements]
    \label{rem:BinaryComplements}
    In the binary model, the complement $\bar A_t\colonequals[n]\setminus A_t$ of a uniform $s$-subset is a uniform $(n-s)$-subset, and with $Q_t\colonequals\sum_{u\le t}x_u$ and $\bar S_t\colonequals\sum_{u\le t}x_u\mathbf1_{\bar A_u}$ we have, pathwise, $S_t=Q_t\mathbf1-\bar S_t$ and $\sum_iS_i(t)=sQ_t$.
    For $s\ge n/2$, this gives $\abs{Q_t}\le2\norm{S_t}_\infty$ and $\norm{\bar S_t}_\infty\le3\norm{S_t}_\infty$.
    Since complementation is a bijection at each arrival, every online algorithm for $s$-supports induces one for the fresh uniform $(n-s)$-supports $\bar A_t$, with signed sums $\bar S_t$.
    Hence, for $r\le n/2$, every lower bound for sparsity $r$ transfers to sparsity $s=n-r$ with a loss of a factor at most three.
    In particular, in the binary model with $2\le s\le n-2$ and $T\ge n$, every online algorithm satisfies $\norm{S_T}_\infty\ge c(\sqrt{\min\{s,n-s\}}+\log\log n)$ with probability $1-o(1)$, and for every fixed $r\ge2$, the lower bound of \cref{thm:Prefix}\ref{item:PrefixLower} for sparsity $r$, with the exponent $\alpha_r$ and with $c'/3$ in place of $c'$, also holds for $s=n-r$.
    Conversely, $\norm{S_t}_\infty\le2\norm{(Q_t,\bar S_t)}_\infty$, so upper bounds for the augmented inputs $(1,\mathbf1_{\bar A_t})$ transfer to $S_t$; in this way, $s$ may be replaced by $\min\{s,n-s+1\}$ in the benchmarks of \cref{rem:GenericBenchmarks}.
    The augmented inputs always contain their first coordinate, however, so they are not uniform sparse inputs, and \cref{thm:Prefix,thm:Terminal} do not apply to them.
    We therefore do not determine the long-horizon order for binary inputs with $s=n-2$, although its lower bound is the same as for $s=2$, and \cref{thm:Terminal} does not cover binary inputs with $n/2<s<n$.
    By \cref{rem:SizeOne}, the optimal prefix discrepancy is one for $s=1$ in both models and for $s=n$ in the binary model.
\end{remark}

\section{Open questions and a conjecture}
\label{sec:OpenQuestions}

We collect the main questions left open by our results.
Throughout, $c>0$ is fixed and $n\le T\le e^{cn}$, unless stated otherwise.
By \cref{cor:PolynomialPrefix}, the order of the prefix discrepancy is settled at all polynomial horizons, for every sparsity in the range of \cref{thm:Terminal}; the questions on prefixes therefore concern horizons with $\log T\gg\log n$.
We first discuss prefixes, then terminal times, and finally random edge orientations.

\paragraph{The exponent for larger supports.}
For pairs, \cref{cor:ExpectedPrefix} determines the optimal order.
For fixed $s\ge3$, it determines the order up to the exponent of $\log T/\log n$, which lies in $[\alpha_s,\frac12]$; the gap $\frac12-\alpha_s=1/(2(2^s-1))$ decreases rapidly, but it is positive for every $s$.
Several facts point towards the upper exponent.
First, the lower-bound exponents $\alpha_s$ are the best possible for our width certificate with static coefficients, evaluated as in \cref{lem:WidthCertificate} (\cref{prop:StaticBarrier}), so the gap to $1/2$ may reflect the method rather than the game.
Second, for the game itself we only know a weaker limit, at exponent $1/2$: Greedy survives at least a quadratic number of requests (\cref{prop:QuadraticSurvival}), so for fixed $s$ no forcing argument based on the game, with the block prices of \cref{prop:BlockBound}, can give an exponent larger than $1/2$.
Third, the most natural candidate rule does not beat the square-root law: in the signed model with every fixed $s\ge3$, quadratic Greedy with ties broken towards $+1$ can be forced to every height $H$ within $O(H^2)$ adaptive arrivals and has $D_T=\Omega_{s,c}(\sqrt{n/\log n})$ at $T=\floor{e^{cn}}$ with high probability (\cref{prop:GreedyThreeTouches}), which is also the order achieved by the rule of \cref{thm:Prefix}\ref{item:PrefixFixed}.
This evidence is indirect: the first two observations concern limitations of the forcing method, and the third concerns one particular rule rather than all online algorithms.

\begin{conjecture}[Square-root law]
    \label{conj:SquareRoot}
    For every fixed $s\ge3$, uniformly for $n\le T\le e^{cn}$,
    \begin{equation*}
        \ondisc(\mu_s,T)\asymp_{s,c}\log\log n+\Big(\frac{\log T}{\log n}\Big)^{1/2}.
    \end{equation*}
\end{conjecture}

By \cref{lem:Realization,prop:BlockBound}, the lower bound in the conjecture would follow from a quadratic forcing time with a moderate pool: an absolute constant $C$ such that, for every $R\ge1$, the safety sets of $\Vc_{m,r}$ with cube radius $R$ and some pool size $m\le CR^2$ satisfy $Q_J=\emptyset$ for $J=CR^2$.
For $r=3$, this would give the lower bound for every $s\ge3$, including $s=3$, and for $r=4$ for every $s\ge4$; by \cref{prop:QuadraticSurvival}, such a forcing time would be optimal up to the constant.
\Cref{prop:GreedyThreeTouches} gives such forcing times, already for three touches and with pools of size $O(\log R)$, against quadratic Greedy with ties broken towards $+1$, but not against every controller.

\begin{question}[Quadratic forcing]
    \label{q:QuadraticForcing}
    Can the adversary force exit from the cube of radius $R$ in the three-touch safety game within $O(R^2)$ requests, using a pool of $O(R^2)$ coordinates?
    If not, can it do so in the four-touch game?
\end{question}

For three touches, the smallest forcing time with such pools lies between order $R^2$, by \cref{prop:QuadraticSurvival}, and $O(R^{7/3})$, by \cref{prop:OptimizedRadices} with $k=2$.
By \cref{prop:StaticBarrier}, a positive answer cannot come from \cref{lem:WidthCertificate} with fixed coefficients alone and requires further information about the safety sets.

\paragraph{Joint dependence on the sparsity.}
\Cref{rem:GrowingSparsity} shows that the horizon term persists for growing $s$, up to a factor $\sqrt{\log\log(\log T/\log(en/s))}$, as long as the pool of its forcing strategy fits into the $n$ coordinates.
On the other hand, the generic bounds of \cref{rem:GenericBenchmarks} give $O(\sqrt s)$ once $s\gtrsim\log T\,(\log\log T)^2$.

\begin{question}[Joint dependence]
    \label{q:Joint}
    Is it true that
    \begin{equation*}
        \ondisc(\mu_s,T)\asymp\sqrt s+\log\log n+\Big(\frac{\log T}{\log(en/s)}\Big)^{1/2}
    \end{equation*}
    uniformly for $3\le s\le n/2$ and $n\le T\le e^{cn}$?
\end{question}

The proposed order holds (i) at all polynomial horizons, by \cref{cor:PolynomialPrefix}, since its last term is at most $\sqrt{2\beta s}$ for $T\le n^\beta$; (ii) for $s\gtrsim\log T\,(\log\log T)^2$, by \cref{rem:GenericBenchmarks}; and (iii) for $s\le\log^2(en)$ and $\log T\le\sqrt s\min\{s,\log(en)\}$, by \cref{prop:GrowingPrefix}, since its last term is then $O(s^{1/4})$.
For fixed $s$, it reduces to \cref{conj:SquareRoot}, and it is consistent with \cref{rem:GrowingSparsity}.
Even for dense Rademacher inputs, $s=n$, we do not know whether the prefix discrepancy remains $O(\sqrt n)$ up to $T=e^{cn}$; the bounds of Altschuler and Tikhomirov \cite[Corollary~1.3]{Altschuler26} and of Hmadi \cite[Corollary~3.6]{Hmadi26} cover horizons up to $\exp(c_\eta n/\log^{2+\eta}(en))$ and $\exp(cn/\log^2(en))$, respectively (\cref{rem:GenericBenchmarks}).

\paragraph{Terminal discrepancy at small failure probabilities.}
\Cref{thm:Terminal} controls the terminal discrepancy with failure probability $n^{-k}$ for each fixed $k$, and its constants depend on $k$.
For $0<\delta<1$, let $r_s(n,\delta)$ be the smallest $r$ such that some online algorithm satisfies $\P\{\norm{S_T}_\infty>r\}\le\delta$ at every prescribed time $T\ge0$.
A terminal bound at all times $t\le T$ gives a prefix bound up to $T$ by a union bound; this is how \cref{cor:PolynomialPrefix} follows from \cref{thm:Terminal}.
In particular, for $T=\floor{\delta^{-1/2}}$, the union bound and $D_T\le T$ give $\ondisc(\mu_s,T)\le r_s(n,\delta)+1$.
For fixed $s$ and $e^{-n}\le\delta\le n^{-C_s}$, \cref{rem:TerminalSmallConfidence} shows directly that $r_s(n,\delta)$ is at least of order $(\log(1/\delta)/\log n)^{\alpha_s}$, whereas the subgaussian guarantee of Kulkarni, Reis and Rothvoss \cite[Theorem~1 and Corollary~3]{Kulkarni24} gives $r_s(n,\delta)=O(\sqrt{s\log(n/\delta)})$.

\begin{question}[Terminal radius at small failure probabilities]
    \label{q:TerminalConfidence}
    For fixed $s$, is
    \begin{equation*}
        r_s(n,\delta)\lesssim_s\ondisc(\mu_s,\ceil{1/\delta})
    \end{equation*}
    uniformly for $e^{-n}\le\delta\le n^{-C_s}$, with $C_s$ as in \cref{rem:TerminalSmallConfidence}?
\end{question}

For $s=2$, the right side is of order $\log\log n+(\log(1/\delta)/\log n)^{1/3}$, by \cref{cor:ExpectedPrefix}; for $s\ge3$, \cref{conj:SquareRoot} predicts the exponent $1/2$.
For polynomially small $\delta$, both sides are of order $\log\log n$, by \cref{thm:Terminal,cor:PolynomialPrefix,prop:Baseline}.

\paragraph{Related input laws.}
The lower-bound method is not specific to our two models; \cref{rem:EdgeOrientations} shows that random edge orientations on the complete graph also have prefix discrepancy $\Omega((\log T/\log n)^{1/3})$, which makes the dependence on $n$ explicit in the bound $\Omega((\log T)^{1/3})$ stated in \cite[Section~1]{BansalCarpool26}.

\begin{question}[Random edge orientations]
    \label{q:Edges}
    For arrivals $e_i-e_j$ with a uniformly random pair $\{i,j\}$, is the optimal prefix discrepancy of order $\log\log n+(\log T/\log n)^{1/3}$ for $n\le T\le e^{cn}$?
\end{question}

A positive answer would be consistent with the question of Bansal, Prabhu, Singla and Sundaram whether random edges of an arbitrary graph admit prefix discrepancy $O((\log T)^{1/3})$ \cite[Section~1.1]{BansalCarpool26}; Kulkarni, Reis and Rothvoss ask for the same order for every oblivious sequence of edge vectors \cite[Conjecture~2]{Kulkarni24}.
Our proof for pairs does not apply directly: in the edge model, the deviation of the deterministic outcome from the signed average has the sign $+\sigma_i\sigma_j$, opposite to the binary model, so that \eqref{eq:PSD} gives no upper bound for the correction in \cref{lem:GeometricMoments,lem:GlobalQuadratic}.

\appendix
\numberwithin{equation}{section}
\section{Limits of the lower-bound method}
\label[appendix]{sec:MethodLimits}

This appendix explains the exponents in \cref{thm:Prefix}\ref{item:PrefixLower} from the point of view of the finite safety game of \cref{sec:SafetyGame}.
We show that, for a fixed number of touches and with the block success probabilities $p_*^J$ of \cref{prop:BlockBound}, no forcing argument based on this game can give an exponent larger than $1/2$, and that, for two touches, Greedy survives a cubic number of requests, so that the cubic forcing time is the true order in the game (\cref{rem:CubicForcing}).
We also show that the radix construction has the optimal exponent among width certificates with static coefficients, as long as they are evaluated with the budget of \cref{lem:WidthCertificate}.

\subsection{Quadratic survival}
\label{sec:QuadraticSurvival}

For integers $r\ge2$ and $R\ge1$ put
\begin{equation*}
    F_{r-1}(R)\colonequals\min\Big\{\sum_{i=1}^{r-1}y_i^2:\ y\in\Z^{r-1},\ \sum_{i=1}^{r-1}y_i\ge R\Big\}\ge\frac{R^2}{r-1},
\end{equation*}
where the inequality is the Cauchy--Schwarz inequality.

\begin{proposition}[Quadratic survival]
    \label{prop:QuadraticSurvival}
    Let $m\ge1$ and $r\ge2$, and consider the game with alphabet $\Vc_{m,r}$ and cube radius $R$.
    Suppose that the controller plays Greedy, the game analogue of quadratic Greedy: it moves the state to $z+xv$ with $x\inp zv\le0$, breaking ties arbitrarily.
    If the state starts at $z_0\in Q_0$ and leaves $Q_0$ for the first time at the $t$-th request, then
    \begin{equation*}
        \norm{z_0}_2^2+r(t-1)\ge R^2+F_{r-1}(R).
    \end{equation*}
    In particular, $0\in Q_j$ for every $j\le(R^2+F_{r-1}(R))/r$, and every forcing time $J$ satisfies $J>R^2/(r-1)$.
\end{proposition}

\begin{proof}
    Since $x\inp zv\le0$ and $\norm v_2^2\le r$, every Greedy step increases $\norm z_2^2$ by at most $r$; hence the state $z$ before the exiting request satisfies $\norm z_2^2\le\norm{z_0}_2^2+r(t-1)$.
    In the exiting request, some coordinate $i$ with $\abs{z_i}=R$ moves outward, so $xv_iz_i=R$.
    Since $x\inp zv\le0$, the other at most $r-1$ selected coordinates satisfy $\sum_{j\ne i}(-xv_jz_j)\ge R$, and therefore $\sum_{j\ne i}z_j^2\ge F_{r-1}(R)$.
    Hence $\norm z_2^2\ge R^2+F_{r-1}(R)$.
    For $z_0=0$, Greedy survives at least $(R^2+F_{r-1}(R))/r\ge R^2/(r-1)$ requests, whatever the adversary does.
\end{proof}

Thus, even with an unlimited supply of coordinates, every forcing time in \cref{prop:BlockBound} is at least $R^2/(r-1)$.
Since $p_*\le p_1\le s/n$, the right-hand side of \eqref{eq:BlockBound} is at most $1/2$ only if $J\log(n/s)\le\log T+1$, so the lower bound of \cref{prop:BlockBound} can give at most heights of order $\sqrt{r\log T/\log(n/s)}$, that is, of order $\sqrt{\log T/\log n}$ for fixed $s$.
For three or more touches, the bound of \cref{prop:QuadraticSurvival} is sharp up to a constant factor for quadratic Greedy with ties broken towards $+1$, which can be forced to every height $H$ within $O(H^2)$ requests (\cref{prop:GreedyThreeTouches}).
Starting from the axis state $(R,0,\ldots,0)$, the same argument shows that $Q_j$ contains a point with a coordinate of absolute value $R$ for at least $F_{r-1}(R)/r$ steps; hence no argument can reduce the largest coordinate in $Q_j$ by one every $O(R)$ steps.

\begin{remark}[Two touches]
    \label{rem:CubicForcing}
    For two touches, the cubic forcing time of \cref{prop:Radix} with $k=1$ and $q_1=4R$ is the true order: for all $m,R\ge1$, the safety sets of $\Vc_{m,2}$ with cube radius $R$ satisfy $0\in Q_J$, and in particular $Q_J\ne\emptyset$, for every integer $0\le J\le R^3/128$.
    Indeed, the double cover in the proof of \cref{lem:DoubleCover} also allows requests with one nonzero entry: a request $\alpha e_i$ is realized by a single Greedy move between the chips $(i,\alpha)$ and $(i,-\alpha)$, whose positions differ by $2\inp zv$ for $v=\alpha e_i$, whereas a request with two nonzero entries uses two moves.
    If the controller plays Greedy from $0$, at most $2J$ expanding moves occur during $J$ requests, so \cref{thm:ChipGame}\ref{item:ChipBound} gives $\norm z_\infty\le4(2J)^{1/3}\le R$ throughout.
\end{remark}

\subsection{A barrier for static width certificates}
\label{sec:StaticBarrier}

Together with \cref{prop:OptimizedRadices}, the next proposition shows that the exponents $\alpha_s=1/\beta_{s-1}$ are exactly the limit of the width certificate of \cref{lem:WidthCertificate} with static coefficients, evaluated with its budget and the guaranteed unit decrease at all but $2R$ roots.
This is a statement about this evaluation of the certificate, not about the game.

\begin{proposition}[Static certificate barrier]
    \label{prop:StaticBarrier}
    Let $k\ge1$, let $a\in\Z^m$ with $a_1=0$, and put $\Lambda\colonequals\norm a_1\ge1$.
    The directions $v\in\Vc_{m,k+1}$ with $v_1=1$ have at most $K_k\Lambda^{1-2^{-k}}$ distinct roots $-\inp av$, where $K_0\colonequals1$ and $K_k\colonequals2k+1+2K_{k-1}$.
    Consequently, whenever \cref{lem:WidthCertificate} is applied with the alphabet $\Vc_{m,k+1}$ and $N>2R$ distinct roots, its bound on the forcing time satisfies
    \begin{equation*}
        \frac{2RW}{N-2R}\ge2R\Lambda \ge2R\Big(\frac{2R}{K_k}\Big)^{\eta_k}\ge c_kR^{\beta_k},\qquad c_k\colonequals2^{1+\eta_k}K_k^{-\eta_k}.
    \end{equation*}
\end{proposition}

\begin{proof}
    Let $\Ac$ be the set of distinct positive values $\abs{a_i}$, so that $\sum_{c\in\Ac}c\le\Lambda$, and let $\Sigma_j(\Ac)$ be the set of sums of at most $j$ signed elements of $\Ac$, repetitions allowed.
    Every root is an element of $\Sigma_k(\Ac)$.
    We show $\abs{\Sigma_j(\Ac)}\le K_j\Lambda^{1-2^{-j}}$ by induction on $j$; the case $j=0$ is trivial.
    Put $\tau\colonequals \Lambda^{1-2^{-j}}\ge1$.
    Sums that use only values at most $\tau$ are integers in $[-j\tau,j\tau]$, so there are at most $(2j+1)\tau$ of them.
    A sum that uses a value larger than $\tau$ is determined by the choice of one such value, of which there are at most $\Lambda/\tau=\Lambda^{2^{-j}}$, its sign, and a sum in $\Sigma_{j-1}(\Ac)$.
    Hence there are at most $2\Lambda^{2^{-j}}K_{j-1}\Lambda^{1-2^{-(j-1)}}=2K_{j-1}\tau$ such sums, which completes the induction.

    In \cref{lem:WidthCertificate}, each normal has $\ell_1$-norm $\Lambda+\abs{t_\nu}\ge\Lambda$, so $W\ge N\Lambda$ and $2RW/(N-2R)\ge2R\Lambda$.
    If $N>2R$, then $2R<N\le K_k\Lambda^{1-2^{-k}}$, that is, $\Lambda>(2R/K_k)^{\eta_k}$.
\end{proof}

The same limitation holds if the scores are combined with nonnegative weights $\omega_\nu$ and evaluated in the same way.
The guaranteed decrease per step is then $d_\omega\colonequals\sum_\nu\omega_\nu$ minus the sum of the $2R$ largest weights, while the budget, the length of the a priori interval for the weighted sum of the scores, becomes $2R\sum_\nu\omega_\nu(\Lambda+\abs{t_\nu})\ge2R\Lambda\sum_\nu\omega_\nu\ge2R\Lambda d_\omega$.
Thus the resulting bound on the forcing time is again at least $2R\Lambda$ whenever $d_\omega>0$, and $d_\omega>0$ requires more than $2R$ roots with positive weight, so that the counting argument applies as before.

The proposition does not rule out smaller forcing times, and hence larger exponents, obtained from larger actual decrements of the scores, from actual ranges of smaller order than the budget, from adaptive coefficients, or from different adversaries; the symmetry of the safety sets alone makes the scores nonnegative, but this only halves the range.

\section{Forcing quadratic Greedy}
\label[appendix]{sec:SpecificRules}

This appendix analyzes the most natural alternative to the rule of \cref{sec:FixedSparsity} in the signed model: for every fixed $s\ge3$, quadratic Greedy with ties broken towards $+1$ can be forced to height $H$ within $O(H^2)$ adaptive requests, so it does not beat the square-root law.
The forcing strategy is adapted to this rule and gives no lower bound against other rules.

\begin{proposition}[Quadratic forcing of Greedy]
    \label{prop:GreedyThreeTouches}
    Consider the signed model with $3\le s\le n$ and quadratic Greedy, $x_t=-\sgn\inp{S_{t-1}}{V_t}$, with $x_t\colonequals+1$ if $\inp{S_{t-1}}{V_t}=0$; in this appendix we call it simply Greedy.
    For every integer $H\ge1$, put $R_H\colonequals48H^2+1$; there is an integer $m(H)\le2\ceil{\log_2(2H)}+2$ with the following property.
    If $m(H)+s-3\le n$, then for any $m(H)$ \emph{main} and $s-3$ \emph{auxiliary} coordinates, all distinct, and from every state in which the main heights are smaller than $H$, an adaptive sequence of at most $R_H$ arrivals supported on these coordinates, each with prescribed support and marks, forces one of the main coordinates to height $H$.
    Consequently, if $m(H)+s-3\le n$, then for every $T\ge1$,
    \begin{equation}
        \label{eq:GreedyForcing}
        \P(D_T<H)\le\exp\Big\{-\floor{\frac T{R_H}}\Big(2^s\binom ns\Big)^{-R_H}\Big\}.
    \end{equation}
    In particular, for fixed $s\ge3$ and $k>0$ there are $c_s>0$ and $C_{s,k}$ such that, for all $n\ge n_0(s,k)$ and all integers $T$ with $C_{s,k}\log n\le\log T\le n^2$, Greedy has $D_T\ge c_s(\log T/\log n)^{1/2}$ with probability at least $1-n^{-k}$; at $T=\floor{e^{cn}}$ with fixed $c>0$, this is of order $\sqrt{n/\log n}$.
\end{proposition}

\begin{proof}
    \emph{Directions and two certificates.}
    For $m\ge4$, identify $m$ main coordinates with $[m]$; we regard $w$, the certificate $y$ below and the directions $u_j$ as vectors on the main coordinates, extended by zero.
    Define $w_1\colonequals0$, $w_2=w_3\colonequals2$ and $w_j\colonequals w_{j-1}+w_{j-2}$ for $j\ge4$, also for $j>m$, and put $w\colonequals(w_1,\ldots,w_m)$ and $W\colonequals w_m$.
    Consider the directions
    \begin{gather*}
        u_0=(1,1,1,0,\ldots,0),\qquad u_1=(-1,-1,1,0,\ldots,0),\qquad u_2=(-1,1,-1,0,\ldots,0),\\
        u_j=-e_{j-2}-e_{j-1}+e_j\qquad(4\le j\le m),
    \end{gather*}
    indexed by $I\colonequals\{0,1,2\}\cup\{4,\ldots,m\}$; the index $3$ is not used, so that $u_j$ has its positive entry at coordinate $j$ for $j\ge4$.
    Each of them has exactly three nonzero entries, all in $\{\pm1\}$.

    The forcing strategy presents a direction $u_j$ with $\inp S{u_j}\le0$, which Greedy then adds on the main coordinates.
    The first certificate below shows that such a direction exists while all main heights are below $H$, and the second shows that each such step increases the bounded quantity $\inp yS$ by one.

    First, $w=\sum_{j\in I}\lambda_ju_j$ with nonnegative coefficients.
    Put $\lambda_j\colonequals0$ for $j>m$ and $\lambda_j\colonequals w_j+\lambda_{j+1}+\lambda_{j+2}$ for $m\ge j\ge4$, so that the $i$-th coordinate of $\sum_{j\ge4}\lambda_ju_j$ equals $\lambda_i-\lambda_{i+1}-\lambda_{i+2}=w_i$ for every $i\ge4$.
    On the first three coordinates, $\sum_{j\ge4}\lambda_ju_j$ equals $(0,-\lambda_4,-\lambda_4-\lambda_5)$, and the coefficients
    \begin{equation*}
        \lambda_0\colonequals2+\lambda_4+\tfrac12\lambda_5,\qquad\lambda_1\colonequals1+\tfrac12(\lambda_4+\lambda_5),\qquad\lambda_2\colonequals1+\tfrac12\lambda_4
    \end{equation*}
    make these coordinates equal to $w_1=0$ and $w_2=w_3=2$; all coefficients are nonnegative.
    Second, the integer vector $y$ with $y_i\colonequals w_i-1$, that is, $y_1=-1$, $y_2=y_3=1$ and $y_i=w_i-1$ for $4\le i\le m$, satisfies $\inp y{u_j}=1$ for every $j\in I$, as a direct check shows; for $j\ge4$ this is the recursion for $w_j$.
    Hence
    \begin{equation}
        \label{eq:GreedyCertificate}
        \sum_{j\in I}\lambda_j=\sum_{j\in I}\lambda_j\inp y{u_j}=\inp yw=\norm w_2^2-\norm w_1,\qquad\norm y_1=\norm w_1-m+2\le\norm w_1.
    \end{equation}

    \emph{Size of the pool.}
    For $j\ge2$ the recursion gives $w_j\le w_{j+1}\le2w_j$, and hence $w_{j+2}\ge2w_j$.
    By induction on $m$, $\norm w_1=w_{m+2}-2=w_{m+1}+w_m-2\le3W$.
    Let $m=m(H)$ be the least integer $m\ge4$ with $W\ge4H$.
    Then $4H\le W<8H$: for $m=4$ we have $W=4$ and $H=1$, and for $m>4$ we have $W\le2w_{m-1}<8H$.
    Since $w_{2r+2}\ge2^rw_2=2^{r+1}$ for $r\ge0$, we also have $m\le2\ceil{\log_2(2H)}+2$.

    \emph{Forcing.}
    Let $S$ be an integer state in which the $m$ main coordinates have heights at most $H-1$.
    Some direction satisfies $\inp S{u_j}\le0$.
    Otherwise all these integers are at least one, and \eqref{eq:GreedyCertificate} gives $\inp wS=\sum_{j\in I}\lambda_j\inp{u_j}S\ge\norm w_2^2-\norm w_1>W^2-3W$, whereas $\inp wS\le(H-1)\norm w_1\le3(H-1)W\le W^2-3W$, since $W\ge4H$; this is impossible.
    Present such a direction $u_j$, chosen by a fixed rule, with exactly these marks, as the next arrival; if $s>3$, add the $s-3$ auxiliary coordinates, with mark $-\sgn(S_i)$ at a coordinate with $S_i\ne0$ and mark $+1$ otherwise.
    For this arrival $V$, the auxiliary coordinates contribute $-\sum\abs{S_i}\le0$ to $\inp SV$, so $\inp SV\le\inp S{u_j}\le0$.
    Greedy then chooses $x=+1$: if $\inp SV<0$ by the sign rule, and if $\inp SV=0$ by the tie rule.
    Thus it adds exactly $u_j$ on the main coordinates, and $\inp yS$, which involves only the main coordinates, increases by exactly one.
    As long as the $m$ main heights are at most $H-1$, we have $\abs{\inp yS}\le(H-1)\norm y_1$.
    Therefore, if $N$ consecutive arrivals follow this strategy and all $m$ main heights are at most $H-1$ before and after each of them, then $N\le2(H-1)\norm y_1$.
    Consequently, some main height reaches $H$ within
    \begin{equation*}
        2(H-1)\norm y_1+1\le6(H-1)W+1\le48H^2+1=R_H
    \end{equation*}
    arrivals.
    No reachability of the initial state is needed, the auxiliary coordinates may have arbitrary heights, and the coordinates outside the $m+s-3$ chosen ones play no role.

    \emph{Probability.}
    Fix $m(H)$ main and $s-3$ auxiliary coordinates, which is possible since $m(H)+s-3\le n$, divide the time axis into consecutive blocks of $R_H$ arrivals, and argue as in the proof of \cref{prop:BlockBound}.
    For a block, consider the event that every arrival $V_u$ of the block such that all $m$ main heights are smaller than $H$ at time $u-1$ equals the arrival prescribed by the strategy at the state $S_{u-1}$.
    This arrival is $\Fc_{u-1}$-measurable and $V_u$ is independent of $\Fc_{u-1}$, so each requirement holds with conditional probability $p_s\colonequals(2^s\binom ns)^{-1}$; since the tie rule is not symmetric, we count only the prescribed orientation, not its negative.
    Hence the event has conditional probability at least $p_s^{R_H}$ given the past before the block, and on this event some main height reaches $H$ within the block if all $m$ main heights are smaller than $H$ at its beginning.
    Iterating over the $\floor{T/R_H}$ complete blocks proves \eqref{eq:GreedyForcing}.

    For the last claim, let $a\colonequals\max\{1,k\log n\}$, $\Lambda_n\colonequals\log(2^s\binom ns)$, and
    \begin{equation*}
        H\colonequals\floor{\sqrt{\frac{\log(T/(2a))}{49(1+\Lambda_n)}}}.
    \end{equation*}
    Since $1+\Lambda_n\le(s+1)\log n$ for large $n$, and $\log(T/(2a))\ge\frac12\log T$ because $\log T\ge C_{s,k}\log n$, we have
    \begin{equation*}
        H\ge\sqrt{\frac{\log T}{98(s+1)\log n}}-1\ge c_s\Big(\frac{\log T}{\log n}\Big)^{1/2},\qquad c_s\colonequals\frac1{\sqrt{392(s+1)}},
    \end{equation*}
    once $C_{s,k}\ge392(s+1)$; in particular $H\ge1$.
    Since $\log T\le n^2$, we also have $H\le\sqrt{\log T}/7\le n$, so that $m(H)+s-3=O_s(\log n)\le n$ for large $n$.
    Since $R_H\le49H^2$, we get $\log R_H+R_H\Lambda_n\le49H^2(1+\Lambda_n)\le\log(T/(2a))$; in particular $R_H\le T/2$, so $\floor{T/R_H}\ge T/(2R_H)$ and
    \begin{equation*}
        \floor{\frac T{R_H}}p_s^{R_H}\ge\exp\Big\{\log\frac T2-\log R_H-R_H\Lambda_n\Big\}\ge a.
    \end{equation*}
    Thus $\P(D_T<H)\le e^{-a}\le n^{-k}$.
    For $T=\floor{e^{cn}}$, the bound $H\ge c_s(\log T/\log n)^{1/2}$ is of order $\sqrt{n/\log n}$.
\end{proof}

By \cref{thm:Prefix}\ref{item:PrefixFixed}, the order $\sqrt{n/\log n}$ at $T=e^{cn}$ is also achievable for every fixed $s\ge3$, so the proposition does not show that Greedy with this tie rule is worse than the rule of \cref{sec:FixedSparsity}.
The strategy relies on the decisions of Greedy: against the directions $u_j$, the parity rule of \cref{sec:SafetyGame}, which moves the state by $+u_j$ at the odd-numbered requests of $u_j$ and by $-u_j$ at the even-numbered ones, keeps the state on the main coordinates equal to a sum of distinct directions $u_j$; since every main coordinate lies in the supports of at most five of them, all main heights stay at most five, whatever the order of the requests.
Thus the construction gives no forcing bound that holds against every controller (\cref{q:QuadraticForcing}).

\section*{Acknowledgments}
The author used Anthropic's Claude and OpenAI's Codex to assist with mathematical exploration, literature research, manuscript organization, English exposition, and the checking of proofs.
The author retains full responsibility for the content of the manuscript, including all mathematical claims and references.

\bibliographystyle{alpha}
\bibliography{References}
\end{document}